\documentclass[twocolumn,usenatbib,twocolappendix]{aastex63}
\usepackage{lipsum}
\usepackage{totcount}
\usepackage{comment}
\usepackage{amsmath}

\newtotcounter{citnum} %From the package documentation
\def\oldbibitem{} \let\oldbibitem=\bibitem
\def\bibitem{\stepcounter{citnum}\oldbibitem}
\usepackage{array}
\usepackage{floatrow}
\newcommand{\degree}{\mbox{$^{\circ}$}}

\newcommand{\edits}[1]{#1} 	% Revision 1 modifications
\newcommand{\editstwo}[1]{#1} 	% Revision 2 modifications

\graphicspath{{./}{}}

\submitjournal{AJ}

\shorttitle{Characterization of the WASP-4 system}
\shortauthors{J.D. Turner, L. Flagg, A. Ridden-Harper, R. Jayawardhana}
\newcommand\NewPeriod{1.338231587$\pm$0.000000022 days}
\newcommand\Newdecayrate{-7.33$\pm$0.71 msec year$^{-1}$}

\begin{document}

\title{Characterizing the WASP-4 system with TESS and radial velocity data:\\ Constraints on the cause of the hot Jupiter's changing orbit and evidence of an outer planet }

%Investigating the changing orbit of WASP-4b with TESS observations}
%%%%%%%%%%%%%%Authors%%%%%%%%%%%%%%
\correspondingauthor{Jake D. Turner}
\email{astrojaketurner@gmail.com, jaketurner@cornell.edu}

\author[0000-0001-7836-1787]{Jake D. Turner}
\affil{Department of Astronomy and Carl Sagan Institute, Cornell University, Ithaca, New York 14853, USA}
\affil{NHFP Sagan Fellow}

\author[0000-0001-6362-0571]{Laura Flagg}
%\affil{Department of Physics and Astronomy, Rice University, 6100 Main St. MS-108, Houston, TX 77005, USA}
\affil{Department of Astronomy, Cornell University, Ithaca, New York 14853, USA}

\author[0000-0002-5425-2655]{Andrew Ridden-Harper}
\affil{Department of Astronomy and Carl Sagan Institute, Cornell University, Ithaca, New York 14853, USA}

\author[0000-0001-5349-6853]{Ray Jayawardhana}
\affil{Department of Astronomy, Cornell University, Ithaca, New York 14853, USA}



\begin{abstract}
%250 work limit
Orbital dynamics provide valuable insights into the evolution and diversity of exoplanetary systems. Currently, only one hot Jupiter, WASP-12b, is confirmed to have a decaying orbit. Another, WASP-4b, exhibits hints of a changing orbital period \edits{that could be caused by orbital decay, apsidal precession, or the acceleration of the system towards the Earth.} We have analyzed all data sectors from NASA's Transiting Exoplanet Survey Satellite together with all radial velocity (RV) and transit data in the literature to characterize WASP-4b's orbit. Our analysis shows that the full RV data set is consistent with no acceleration towards the Earth. \edits{Instead, we find evidence of a possible additional planet in the WASP-4 system, with an orbital period of $\sim$7000 days and $M_{c} sin(i)$ of $5.47^{+0.44}_{-0.43} M_{Jup}$.} Additionally, we find that the \edits{transit timing variations of all of the WASP-4b transits cannot be explained by the second planet but} can be explained with either a decaying orbit or apsidal precession, with a slight preference for orbital decay. Assuming the decay model is correct, we find an updated period of 1.338231587$\pm$0.000000022 days, a decay rate of -7.33$\pm$0.71 msec year$^{-1}$, and an orbital decay timescale of $\tau = P/\lvert\dot{P}\rvert$ = 15.77$\pm$1.57 Myr. If the observed decay results from tidal dissipation, we derive a modified tidal quality factor of $Q^{'}_{\star}$ = 5.1$\pm$0.9 $\times 10^4$, which is an order of magnitude lower than values derived for other hot Jupiter systems. However, more observations are needed to determine conclusively the cause of WASP-4b's changing orbit and confirm the existence of an outer companion. 

%Some studies have proposed orbital decay or apsidal precession as the cause, but a recent radial-velocity (RV) study suggests that the period change results from the system accelerating towards the Earth. 

%Some studies have proposed orbital decay, apsidal precession, or the system accelerating towards the Earth as the cause.

\end{abstract}

\keywords{planets and satellites: gaseous planets --  planet-star interactions }


\section{Introduction} \label{sec:intro}

One of the most powerful tools used to study exoplanets is observing them while they transit across the disk of their stars. The transit method can be used to search for temporal variations in the planetary orbital parameters and allows for a direct measurement of the planetary radius and thus its atmosphere (e.g. \citealt{Charbonneau2000, Charbonneau2007}). Specifically, light curves of transiting planets can be used to search for transit timing variations (TTVs; \citealt{Agol2005,2018haexAgol}), transit duration variations (\citealt{2018haexAgol}), and impact parameter variations (\citealt{Herman2018,Szabo2020MNRAS}). The presence of TTVs may indicate additional bodies in the system, a decaying planetary orbit, precession in the orbit, or a variety of other effects (e.g., \citealt{Applegate1992}, \citealt{Dobrovolskis1996}, \citealt{Miralda2002}; \citealt{Schneider2003}; \citealt{Agol2005};  \citealt{Mazeh2013}; \citealt{2018haexAgol}). 

%Some exoplanetary systems exhibit temporal variations in their orbital parameters. Such systems provide valuable insights into their physical properties and the processes that cause the variations. Light curves of transiting planets can be used to search for transit timing variations (TTVs; \citealt{Agol2005,2018haexAgol}) and, less frequently, for variations in transit duration (\citealt{2018haexAgol}) and the impact parameter (\citealt{Herman2018,Szabo2020MNRAS}). The presence of TTVs can indicate additional bodies in the system or an unstable orbit resulting from tidal forces of the star (e.g., \citealt{Miralda2002}; \citealt{Mazeh2013}). Also, TTVs in tightly packed multi-planetary systems have been detected and used to constrain the masses of planets in the system (e.g., \citealt{2018haexAgol}).

It is theorized that orbital decay might occur on short-period massive planets orbiting stars with surface convective zones due to exchange of energy with their host stars through tidal interactions (e.g., \citealt{Rasio1996,Lin1996,Chambers2009,Lai2012,Penev2014,Barker2020}). Measurements of orbital decay would expand our understanding of the hot Jupiter population and its evolution (e.g. \citealt{Jackson2008,Hamer2019}). Even though such decay may occur over millions of years, it is possible to search for small changes in the orbital period of hot Jupiter systems since many of these planets have been monitored for decades. 

%\textbf{[reword]} Theory suggests that short-period massive planets orbiting stars with surface convective zones may exchange energy with their host stars through tidal interactions, causing the host star to spin faster and the planet’s orbit to decay (e.g., \citealt{Lin1996,Chambers2009,Lai2012,Penev2014,Barker2020}). As the planet’s orbit decays over millions of years, its orbital period will change by a small, yet potentially detectable amount. Observations of this decay are now possible because some hot Jupiter systems have been monitored for decades. Such measurements enhance our understanding of the hot Jupiter population (e.g. \citealt{Jackson2008,Hamer2019}).

Currently, WASP-12b is one of the few hot Jupiters confirmed to have a varying period. It is an ultra-hot planet around a G0 star with an orbital period of 1.09 days (\citealt{Hebb2009}). WASP-12b is believed to have an escaping atmosphere (e.g., \citealt{Lai2010,Bisikalo2013B,Turner2016a}) as suggested by Hubble Space Telescope near-ultraviolet observations (\citealt{Fossati2010b,Haswell2012,Nichols2015}). \citet{Maciejewski2016} were the first to detect its decreasing orbital period, and subsequent studies have confirmed the period change \citep{Patra2017,Maciejewski2018,Bailey2019,Baluev2019} and established orbital decay as its cause \citep{Yee2020}. The decaying orbit of WASP-12b was confirmed also using transit and occultation observations with NASA's Transiting Exoplanet Survey Satellite (TESS; \citealt{Ricker2015}) \citep[see also \citealt{Owens2021}]{Turner2021}. The decay rate of WASP-12b was found to be 32.53$\pm$1.62 msec yr$^{-1}$ corresponding to an orbital decay timescale of $\tau = P/|\dot P| = 2.90\pm0.14$ Myr \citep{Turner2021}, shorter than the estimated mass-loss timescale of 300 Myr \citep{Lai2010,Jackson2017}. Assuming the observed decay results from tidal dissipation, \citet{Turner2021} derived a modified tidal quality factor, a 
dimensionless quantity that describes the efficiency of tidal dissipation, of $Q'_{\star}$ = 1.39$\pm$0.15 $\times 10^5$, which falls at the lower end of values derived for binary star systems \citep{Meibom2015} and hot Jupiters \citep{Jackson2008,Husnoo2012,Barker2020}.

%Since its discovery in 2009, WASP-12b's orbital parameters (e.g., \citealt{Campo2011,Maciejewski2016}) and atmosphere (e.g., \citealt{Croll2011,Stevenson2014,Sing2016}) have been studied extensively. Observations with the Cosmic  Origins  Spectrograph  (COS) on the Hubble Space Telescope reveal an early ingress in the near-ultraviolet (\citealt{Fossati2010a,Haswell2012,Nichols2015}). An escaping atmosphere was suggested as the cause of the early ingress (e.g., \citealt{Lai2010,Bisikalo2013B,Turner2016a}). \citet{Maciejewski2016} were the first to report evidence of a decreasing orbital period for WASP-12b. At the time, the cause of the changing orbit was uncertain, and could be ascribed to orbital decay, apsidal precession, or the Romer effect. Follow-up observations confirmed the changing orbital period, with models slightly favoring orbital decay as the cause  (\citealt{Patra2017,Maciejewski2018,Bailey2019}). Recently, using new transit, occultation and radial velocity observations, \citet{Yee2020} presented strong evidence that WASP-12b's period variation is caused by orbital decay. The decay rate has been constrained to between 20-30 milliseconds/yr. 

Besides WASP-12b, WASP-4b is a well-studied system with hints of a changing period. WASP-4b is a typical hot Jupiter with a short orbital period of 1.34 days and orbits a G7V star \citep{Wilson2008}. The planet's atmosphere (e.g. \citealt{Caceres2011,Ranjan2014,Bixel2019}) and orbital parameters (e.g. \citealt{Southworth2009,Winn2009,Hoyer2013,Baluev2020}) have been studied extensively since its discovery in 2008. \citet{Bouma2019} first reported an orbital period variation of WASP-4b using TESS and ground-based observations. \citet{Southworth2019} confirmed that the period was decaying with additional ground-based observations and found a decay rate of 9.2 msec yr$^{-1}$. They found that orbital decay and apsidal precession could explain the TTVs after ruling out instrumental issues, stellar activity, the Applegate mechanism, and light-time effect. \citet{Bouma2020} obtained additional radial-velocity (RV) data on WASP-4b using HIRES on Keck and found that the observed orbital period variation could be explained by the system accelerating toward the Sun at a rate of -0.0422 m s$^{-1}$ day$^{-1}$. Recently, \citet{Baluev2020} analyzed a comprehensive set of 129 transits and additional RV data (presented in \citealt{Baluev2019}) from 2007-2014 (mostly in years not covered by the RV data presented in \citealt{Bouma2020}). They also confirmed a period change in WASP-4b's orbit but do not confirm the RV acceleration found by \citet{Bouma2020}. However, \citet{Baluev2020} did not include the new RV HIRES data from \citet{Bouma2020} in their analysis. Therefore, the cause of the period variation in WASP-4b is still an open question.

\begin{figure}[bth]
 \includegraphics[width=\textwidth]{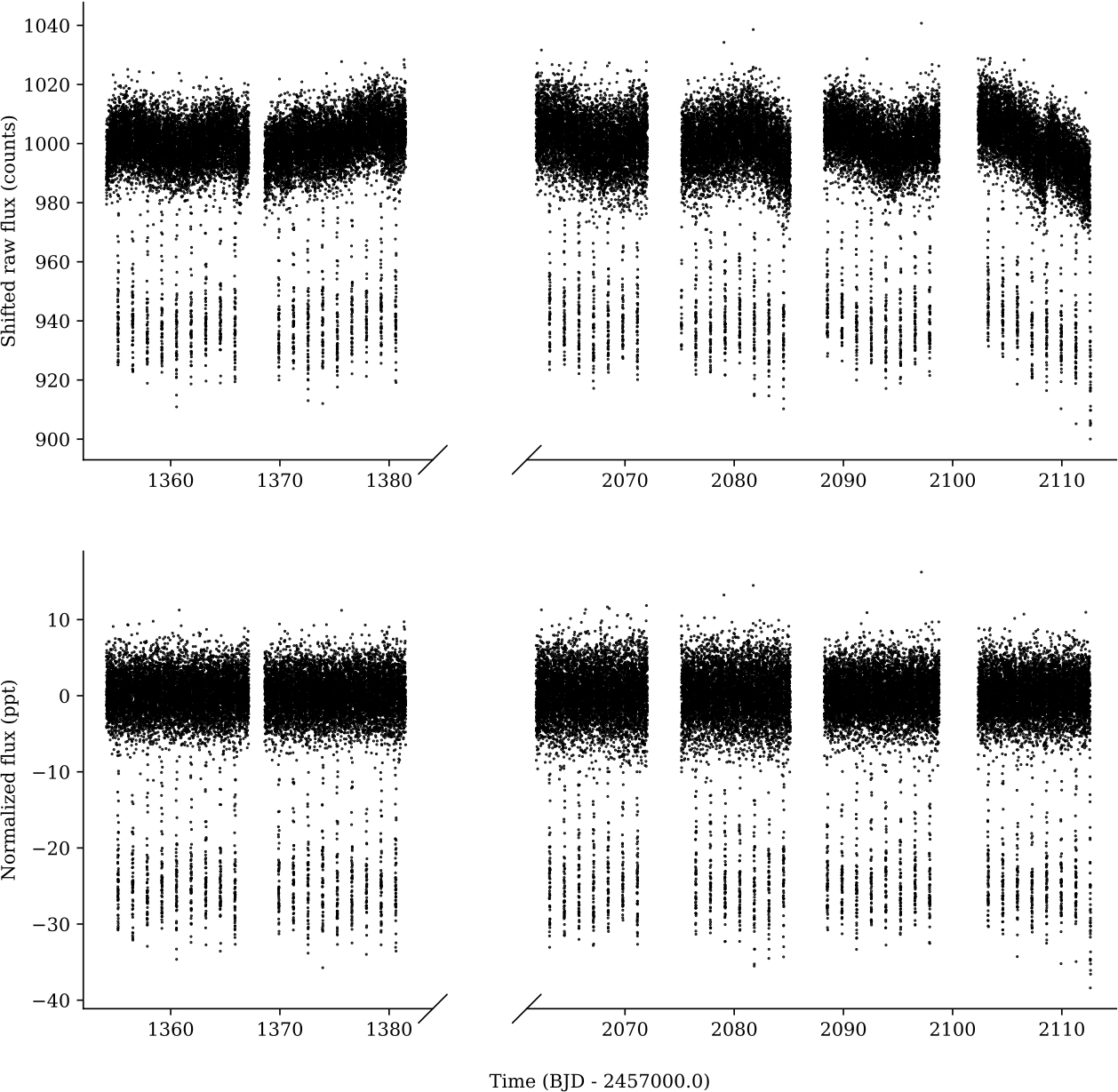}
\caption{TESS light curve of WASP-4b in Sectors 2, 28, and 29. Top: Raw simple aperture photometry light curves. Bottom: Detrended Data Validation Timeseries (DVT). }
\label{fig:lightcurves}
\end{figure}

Motivated by the possible changing period of WASP-4b, we analyze all three sectors (Figure \ref{fig:lightcurves}) from TESS and combine the results with all transit, occultation, and RV measurements from the literature to verify its changing period and derive updated planetary properties. TESS is well suited for our study because it provides high-precision time-series data, ideal for searching for TTVs (e.g. \citealt{Hadden2019,Pearson2019}; \citealt{vonEssen2020}; \citealt{RiddenHarper2020,Turner2021}). Altogether, the transit data span 13 years from 2007-2020 and the RV data span 12 years from 2007-2019. Using the combined data set we hope to shed light on the cause of WASP-4b's period variation.

\section{Observations and Data Reduction}
TESS observed WASP-4b in Sector 2 (2018-Aug-22 to 2018-Sep-20), Sector 28 (2020-Jul-30 to 2020-Aug-26), and Sector 29 (2020-Aug-26 to 2020-Sep-22). The TESS observations were processed by the Science Processing Operations Center (SPOC) pipeline\footnote{All of the SPOC data products are publicly accessible from the Mikulski Archive for Space Telescopes at \url{https://archive.stsci.edu/}} \citep{Jenkins2016}. The SPOC pipeline produces light curves ideal for characterizing transiting planets since they are corrected for systematics. SPOC produces Presearch Data Conditioning (PDC) and Data Validation Timeseries (DVT) light curves. The PDC light curves are corrected for instrumental systematics (pointing or focus related), discontinuities resulting from radiation events in the CCD detectors, outliers, and flux contamination. The DVT light curves are created by using a running median filter on the PDC light curves to remove any long-period systematics. We use the DVT light curves (Figure \ref{fig:lightcurves}) for our analysis because they have less scatter in their out-of-transit (OoT) baseline. As shown in \citet{RiddenHarper2020} for XO-6b TESS data, the DVT and PDC light curves produce similar results on the timing of the transits. For the Sector 2 data, the light curves produced from the SPOC \edits{pipeline} have a known issue\footnote{The issue is related to inaccurate uncertainties in the 2D black model, which represents the fixed pattern that is visible in the black level for a sum of many exposures. See TESS Data Release Notes: Sector 27, DR38.} that overestimates their uncertainties. Therefore, we estimated the uncertainties using the scatter in the OoT baseline as we did in our previous study of WASP-12b \citep{Turner2021} and as recommended (Barclay, T., private communication). The Sector 28 and 29 data are unaffected by this problem. Therefore, we used the uncertainties provided by the SPOC pipeline.

\section{Data Analysis}
\subsection{Transit Modeling}
All the TESS transits of WASP-4b were modeled with the EXOplanet MOdeling Package (\texttt{EXOMOP}; \citealt{Pearson2014,Turner2016b,Turner2017})\footnote{\texttt{EXOMOPv7.0}; \href{https://github.com/astrojake/EXOMOP}{https://github.com/astrojake/EXOMOP} } to find a best-fit. \texttt{EXOMOP} creates a model transit using the analytic equations of \cite{Mandel2002} and the data are modeled using a Differential Evolution Markov Chain Monte Carlo (DE-MCMC; \citealt{Eastman2013}) analysis. The residual permutation, time-averaging, and wavelet methods are incorporated into  \texttt{EXOMOP} to account for red noise in the light curve. See \citet{Pearson2014} and \citet{Turner2016b} for a more detailed descriptions of \texttt{EXOMOP}. 

Each TESS transit (Figure \ref{fig:lightcurves}) was modeled with \texttt{EXOMOP} independently. We used 20$^{6}$ links and 20 chains for the DE-MCMC model and use the Gelman-Rubin statistic (\citealt{Gelman1992}) to ensure chain convergence (\citealt{Ford2006}). The mid-transit time ($T_{c}$), scaled semi-major axis ($a/R_{*}$), planet-to-star radius ($R_{p }/R_{*}$), and inclination ($i$) are set as free parameters for every transit. The linear and quadratic limb darkening coefficients and period ($P_{orb}$) are fixed during the analysis. The linear and quadratic limb darkening coefficients are taken from \citet{Claret2017} and are set to 0.382 and 0.210, respectively.

The parameters derived for every TESS transit event can be found in Tables \ref{tb:lighcurve_model_TESS}-\ref{tb:lighcurve_model_TESS3} in Appendix \ref{app:individual_transits}. The modeled light curves for each individual transit can be found in Figures \ref{fig:ind_transits_sec2_1}--\ref{fig:ind_transits_sec29_2} in Appendix \ref{app:individual_transits}. All parameters for each transit event are consistent within 2$\sigma$ of every other transit. The Sector 2 data for WASP-4b was analyzed in \citet{Bouma2019} and our timing analysis for each individual transit is consistent within 1$\sigma$ of their findings (see Figure \ref{fig:LC_S2_comp} in Appendix \ref{sec:diffTESS}). 

\begin{figure}
 \includegraphics[width=\textwidth]{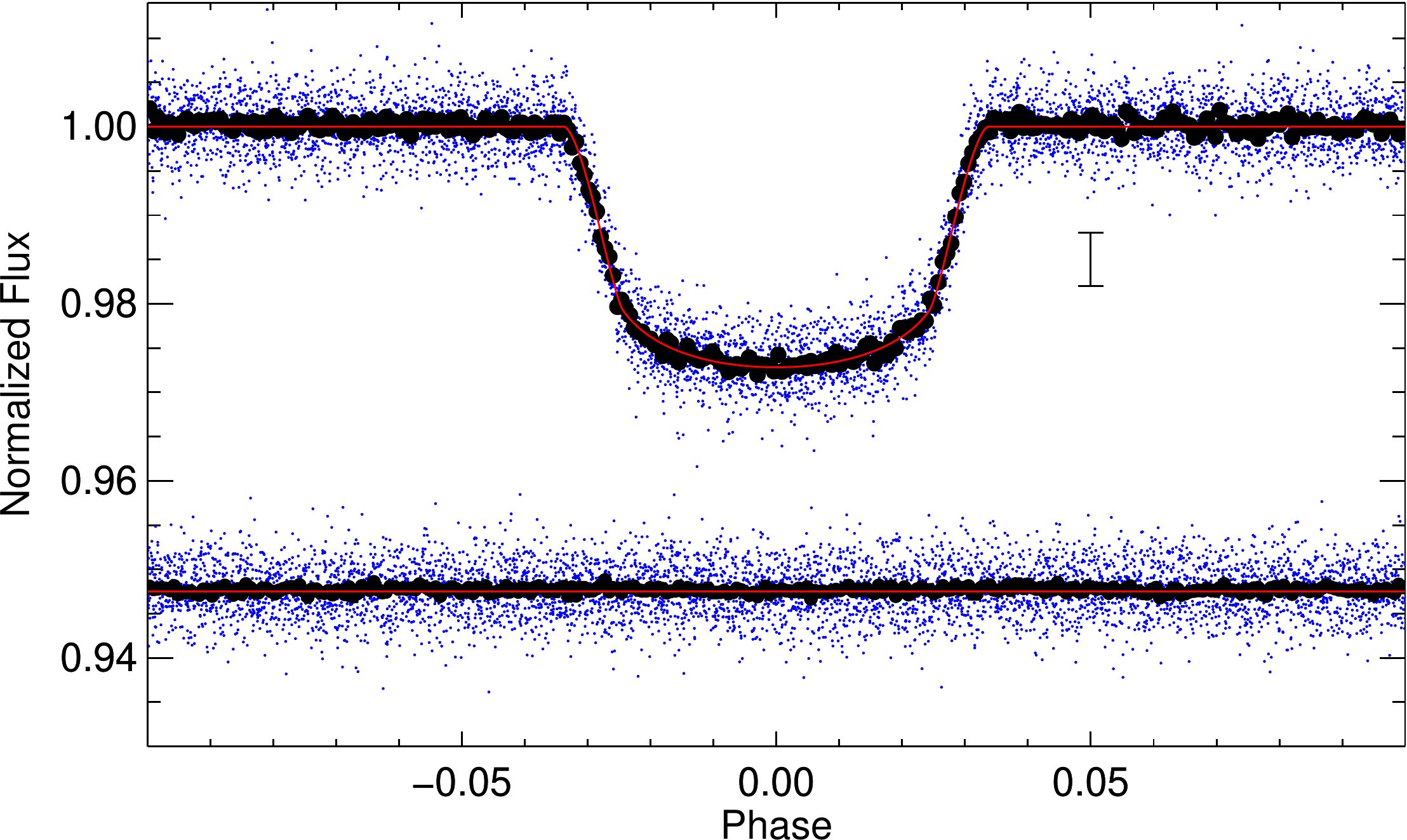}
\caption{Phase-folded transit light curve of WASP-4b from TESS. The unbinnned and binned data are shown in blue and black, respectively. The binned transit has been binned by 5 minutes. The best-fitting model obtained from the EXOplanet MOdeling Package (\texttt{EXOMOP}) is shown as a solid red line. The residuals (light curve - model) are shown below the light curve.} 
\label{fig:modelfit}
\end{figure}

The light curve of WASP-4b was phase-folded at each derived mid-transit time and modeled with \texttt{EXOMOP} to find the final fitted parameters. The phase-folded light curve and model fit can be found in Figure \ref{fig:modelfit}. We use the light curve model results combined with literature values to calculate the planetary mass (M$_{p}$;  \citealt{Winn2010b}), radius (R$_{p}$), density ($\rho_{p}$), equilibrium temperature (T$_{eq}$), surface gravity ($\log{g_{p}}$; \citealt{Southworth2007a}), orbital distance ($a$), inclination ($i$), Safronov number\footnote{The Safronov number is a qualitative measure of the potential of a planet to capture or gravitationally scatter other objects in close by orbits} ($\Theta$; \citealt{Safronov1972}; \citealt{Southworth2010a}), and stellar density ($\rho_{\ast a}$; \citealt{Seager2003}). The planet properties we derived for WASP-4b are shown in Table \ref{tb:planet_parameters}. All the planetary parameters are consistent with their discovery values \citep{Wilson2008} but their precision is greatly improved.

\begin{table}
\caption{Physical  properties  of  WASP-4b and WASP-4c}
    \centering
    \begin{tabular}{lcccc}
     \hline
     Parameter              &  units    & value  & 1 $\sigma$ uncertainty  \\
     \hline
  \multicolumn{4}{c}{\textbf{WASP-4b}}\\
  \hline
  P                         &  days              &   1.338231587      & 0.000000022            \\
  R$_p$/R$_\ast$            &               &  0.15158      & 0.00057   \\
  a/R$_\ast$                &               &  5.410        & 0.088        \\
  Inclination               & \degree       & 88.02         & 0.69 \\
  Duration                  & mins          & 129.594       & 0.097      \\
  b                         &               & 0.187   &0.065   \\
  R$_{p}$                   & R$_{Jup}$     & 1.312       & 0.045 \\
  M$_{p}$                   & M$_{Jup}$     &  1.164 &  0.082                 \\
  $\rho_{p}$                & g cm$^{-3}$   &   0.639 & 0.079 \\
  $\log{g_{p}}$   	        &   cgs         &   3.216        & 0.035\              \\
  $\rho_{\ast}$           & g cm$^{-3}$   &   1.67 & 0.16     \\
  T$_{eq}$                  &   K           &  1641.65      & 27.36  \\
  $\Theta$                  &               & 0.046872      & 0.006376              \\ 
  a                         &  AU           &   0.02239  & 0.000 84\\
     \hline
  \multicolumn{4}{c}{\textbf{WASP-4c}}\\
 P              & days       &  7001.0  & 6.6 \\ 
 M$_{p}$sin(i)   & M$_{Jup}$   & 5.47  & 0.44\\
  a              & AU        &     6.82  & 0.25  \\
  \hline
    \end{tabular}
    \tablecomments{
    \edits{To calculate the planetary mass (M$_{p}$) of WASP-4b we used a stellar RV amplitude (K$_{b}$) of 237.3$\pm$2.2 m $s^{-1}$ as listed in Table \ref{tb:RV_results_All} from the best-fit 2-planet RV model (Model $\#$3). The period of WASP-4b was taken from the orbital decay model in Table \ref{tb:timing_models}. All the physical parameters for WASP-4c were taken from the best-fit 2-planet RV model (Model $\#$3).} 
    }
    \label{tb:planet_parameters}
\end{table}

%%%%%%%%%%%%%%%%%%%%%%%%%%%% occultation %%%%%%%%%%%%%%%%%%
\subsection{Occultation}
\begin{figure}[th!]
\centering
 \begin{tabular}{c}
 \includegraphics[width=0.85\textwidth,page=1]{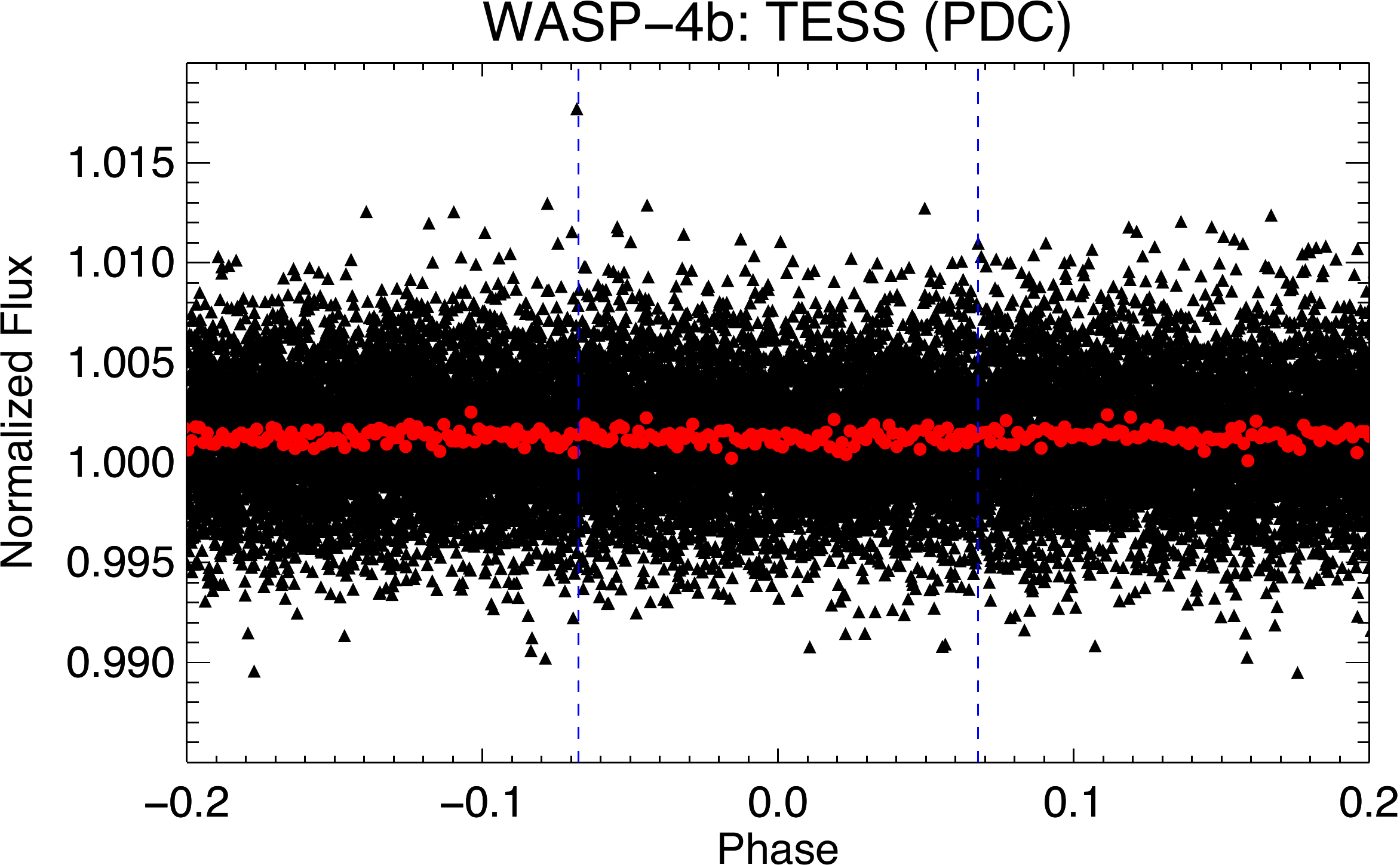}
  %\sidesubfloat[\textbf{(b.)}]{    \includegraphics[width=0.85\textwidth,page=2]{WASP4b_Occ.pdf}} \\
 \end{tabular}
\caption{Phased-folded occulation PDC light curve of WASP-4b from TESS. The unbinnned and data binned by 18 minutes are shown in black and red, respectively. \edits{The vertical dashed lines show the time range of the expected secondary eclipse.}
}
\label{fig:tess_occ}
\end{figure}

We created an occultation light curve by phase-folding all the data about the secondary eclipse using the first TESS transit as the reference transit time.  As shown in Figure \ref{fig:tess_occ}, we do not see an occultation of WASP-4b in the TESS data. We only show the PDC light curve but this is also the case for the DVT light curve. We find a 3-$\sigma$ upper limit on the occultation depth ($\delta$$_{occ}$) to be 1.34 $\times 10^{-5}$. The geometric albedo (A$_{g,occ}$) of WASP-4b can be calculated assuming no thermal contribution
\begin{equation}
 \delta_{occ} = A_{g,occ} \left(  \frac{R_{p}}{a}\right)^2. \label{occ_eq}
\end{equation}
We find a 3-$\sigma$ upper limit on A$_{g,occ}$ of 0.017 using Equation $\eqref{occ_eq}$. Our upper limit on A$_{g,occ}$ is consistent with the overall trend that hot Jupiters are very dark (e.g. \citealt{Kipping2011,Movcnik2018,Kane2020}).

%%%%%%%%%%%% RV Modeling %%%%%%%%%%%%%%%%
\subsection{Modeling Radial Velocities}
We combined all available radial velocity (RV) data of WASP-4 in the literature in our analysis. The complete RV data set is given in Table \ref{tab:all_RV}. The set includes CORALIE and HARPS data from \citet{Baluev2019}, HARPS data from \citet{Pont2011}, and HIRES data from \citet{Bouma2020}. We did not include the RV data from \citet{Triaud2010} as these data were taken for Rossiter–McLaughlin measurements and were reduced with a nonstandard pipeline. 

%%%%%%%% RV Obs Table %%%%%%%
\begin{table}[!htb]
    \centering
    \begin{tabular}{ccccc}
    \hline
    Time          & RV          & $\sigma_{RV}$ & Instrument  & Source \\
    (BJD$_{TDB}$) & (m s$^{-1}$)  & (m s$^{-1}$)\\
    \hline
    2454321.12345 & 42          & 0.42         & HIRES          & 5 \\
    \hline     
    \end{tabular}
    \caption{All WASP-4b Radial Velocity Measurements}
    \label{tab:all_RV}
    \tablecomments{
    This table is available in its entirety in machine-readable form.
    \tablerefs{
    1.) \citealt{Wilson2008}; 
    2.) \citealt{Triaud2010}; 
    3.) \citealt{Pont2011}; 
    4.) \citealt{Husnoo2012}; 
    5.) \citealt{Knutson2014_RV}; 
    6.) \citealt{Baluev2019}
    }
    }
\end{table}

\begin{table}[phtb]
 \caption{Comparison of all the RV models performed with \texttt{RadVel}. \edits{ Column 1: Model number. Column 2: Main free parameters. Column 3: Number of free parameters. Column 4: RMS of the residuals. Column 5. Natural log of the likelihood. Column 6. BIC value. }}
    \centering
    \begin{tabular}{llllll}
    \hline
Model &  Free Parameters & $N_{\rm free}$  & RMS & $\ln{L}$ & BIC  \\
        &               &                   & (m s$^{-1}$) \\
  \hline
  \hline 
  \multicolumn{6}{c}{\textbf{One-Planet Model}}\\
  1 & $K_{b}$, {$\sigma$}, {$\gamma$} & 6 & 29.85 & -325.78 & 675.08  \\
  2 &  $K_{b}$, $\sigma$, $\gamma$, $\dot{\gamma}$ & 7  & 29.78 & -651.38 & 682.76  \\
 %  $K_{b}$, $\sigma$, $\gamma$, $\dot{\gamma}$, $\ddot{\gamma}$ & 8 & 74 &19.33 & -297.58 & 626.94 & 610.72\\ 
 \multicolumn{6}{c}{\textbf{Two-Planet Model}} \\
3   & $K_{b}$, $K_{c}$, {$\sigma$}, {$\gamma$}  & 9 & 18.91& -304.40 & 628.34    \\
%4   & $K_{b}$, $K_{c}$, {$\sigma$},  {$\gamma$}, $\dot{\gamma}$  & 10  & 74 & 18.97 & -304.11 & 632.08 & 612.53 \\
4   & $K_{b}$, $K_{c}$, {$\sigma$}, {$\gamma$}, e$_{c}$  & 11 & 19.28 & -629.22 & 637.74  \\
%6   & $K_{b}$, $K_{c}$, {$\sigma$}, {$\gamma$}, $\dot{\gamma}$, e$_{c}$ & 12  & 74 & 18.99 & -304.12 & 640.72 & 618.19 \\
  \hline
    \end{tabular}
   \tablecomments{The number of data points ($N_{\rm data}$) for all the models was 74.}
    \label{tb:RV_compare}
\end{table}

The RV data were modeled with \texttt{RadVel} \citep{Fulton2018}. \edits{We ran four models in total for the RV analysis (Table \ref{tb:RV_compare}). The priors used in the analysis are summarized in Table \ref{tb:priors_RV} in Appendix \ref{app:RV}. We used Gaussian priors for the orbital period and time of inferior conjunction that were set to the values derived by \citet{Bouma2019}. We set the eccentricity of the orbit to zero as indicated by previous upper limit studies (\citealt{Beerer2011,Knutson2014_RV,Bonomo2017}. }

\begin{figure*}[pthb!]
\centering
 \begin{tabular}{ccc}
 \textbf{I. Model $\#$1} & & \textbf{II. Model $\#$2}\\
 \includegraphics[width=0.39\textwidth,page=1]{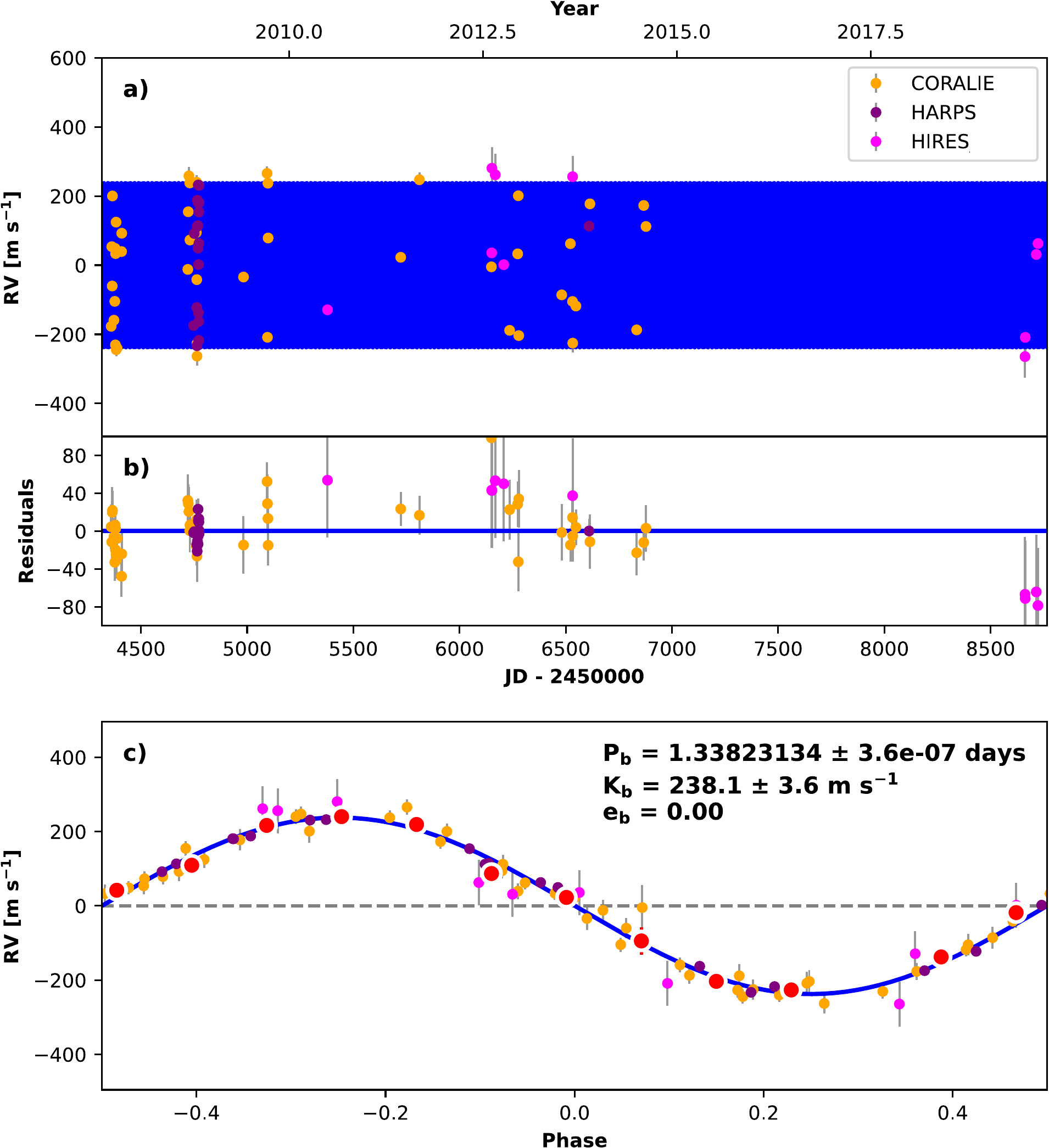}  & &  \includegraphics[width=0.39\textwidth,page=1]{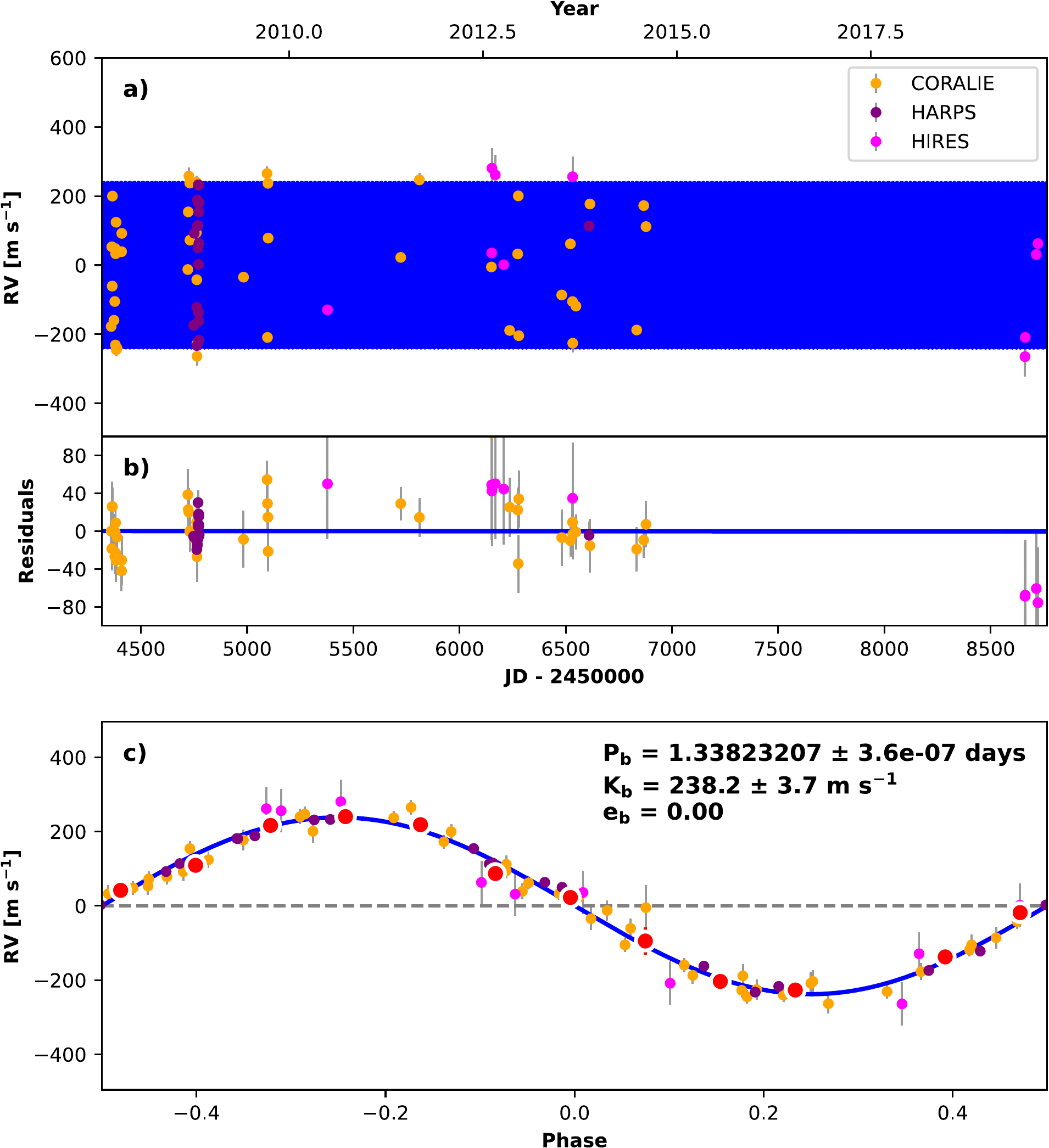} \\
 &&\\ 
  \textbf{III. Model $\#$3} &  &\textbf{IV. Model $\#$4}\\
  \includegraphics[width=0.39\textwidth,page=1]{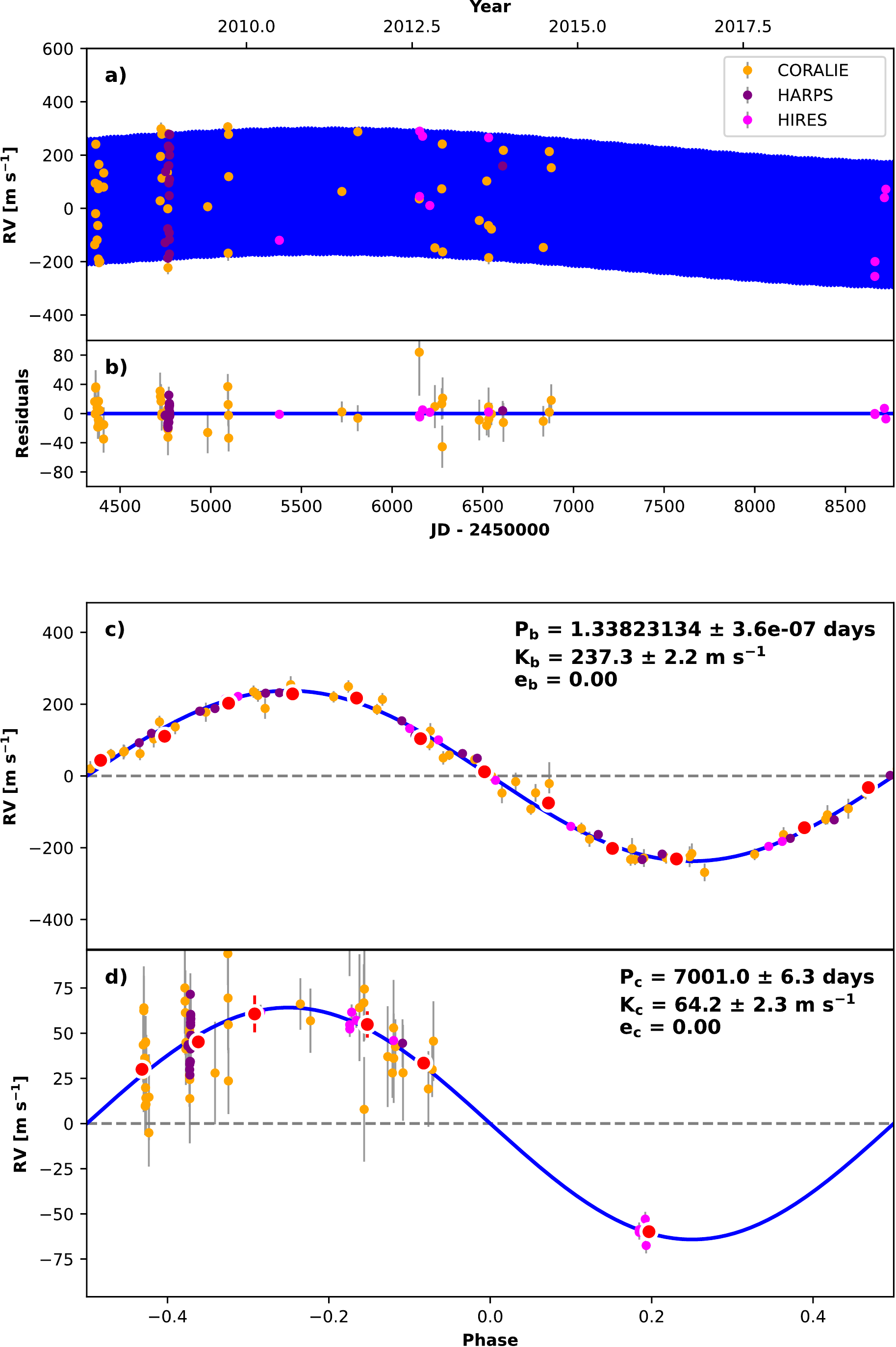}  & & 
 \includegraphics[width=0.39\textwidth,page=1]{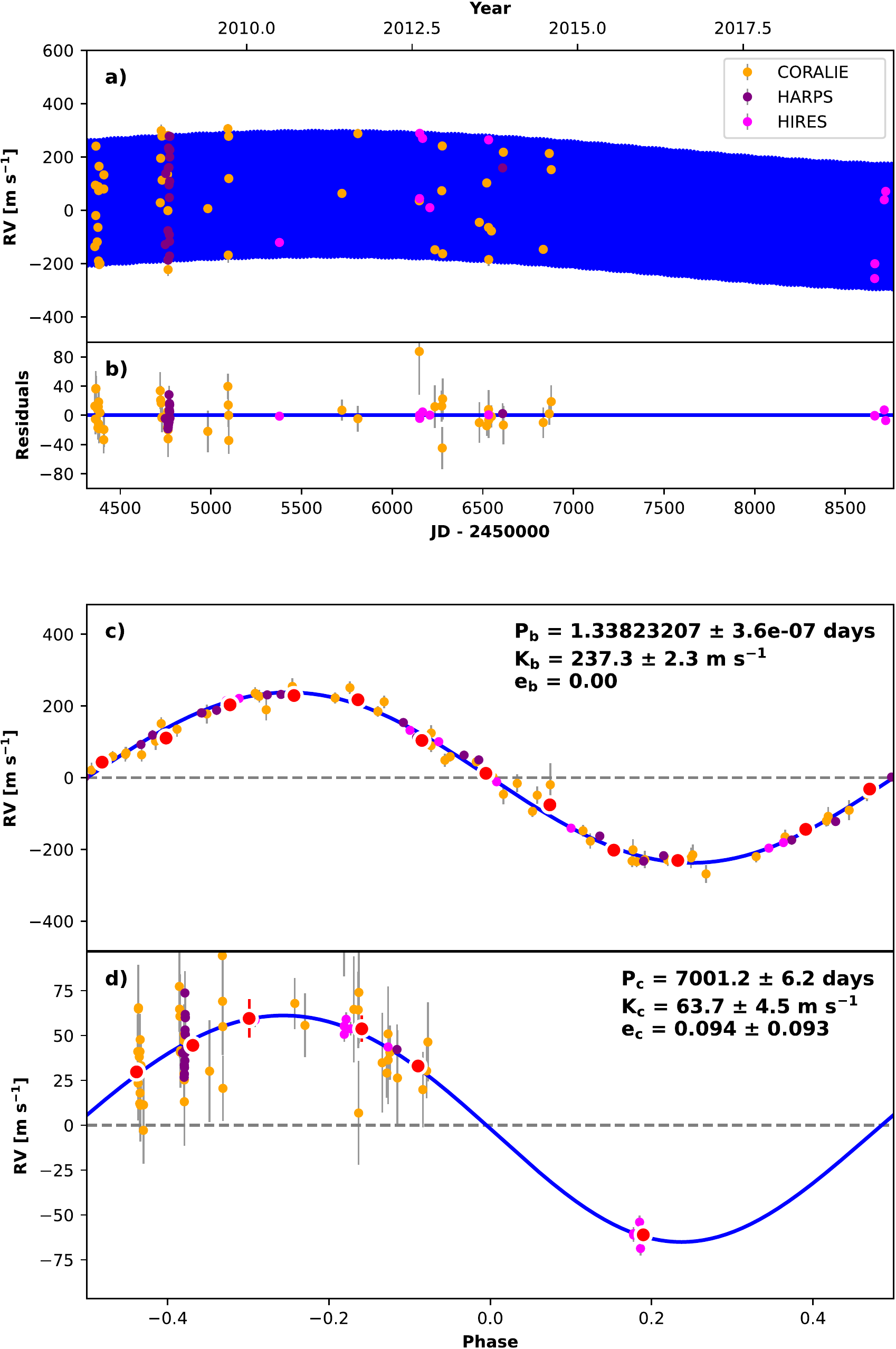} \\
 \end{tabular}
\caption{Best-fit Keplerian orbital model
  for the WASP-4 system for Models $\#$1 (top-left), $\#$2 (top-right), $\#$3 (bottom-left), and $\#$4 (bottom-right) using \texttt{RadVel}. {\bf a panels)} The maximum likelihood model is plotted while the orbital parameters listed in Table \ref{tb:RV_results_All} are the median values of the posterior distributions.  The thin blue line is the best fit model. We add in quadrature
  the RV jitter terms listed in Table \ref{tb:RV_results_All} with the
  measurement uncertainties for all RVs.  {\bf b panels)} Residuals to the
  best fit model. The error bars of the residuals reflect both the measurement error and the jitter from the MCMC fit. The larger error bars for the 1 planet fits reflect the larger amounts of jitter needed to fit the data with only 1 planet. {\bf c panels)} RVs phase-folded to the ephemeris of WASP-4b. The phase-folded model for WASP-4b is shown as the blue line. The Keplerian orbital models for all other planets (if modeled) have been subtracted. {\bf d panels)} RVs phase-folded to the ephemeris of the second planet WASP-4c. The Keplerian orbital model for the other planet (panels c, WASP-4b) has been subtracted. Red circles are the velocities binned in 0.08 units of orbital phase. }
  \label{fig:RVModels_all}
\end{figure*}

As done in previous studies, we first modeled the data with only one planet. The free parameters in the model were the orbital velocity semi-amplitude (K$_{b}$), the instrument zero-points, white noise instrument jitter for each instrument ($\sigma$, added in quadrature to its uncertainties), and the linear acceleration ($\dot{\gamma}$). We ran models with and without $\dot{\gamma}$ to check if an acceleration is needed to fit the data (\edits{Models $\#$1 and $\#$2}). The results of the 1-planet models of WASP-4b can be found in Table \ref{tb:RV_results_All} and Figure \ref{fig:RVModels_all}. We use the Bayesian Information Criterion (BIC) to assess the preferred model. The BIC is defined as 
\begin{equation}
BIC =  \chi^{2} + k \ln{(N_{pts})}, \label{eq:BIC}
\end{equation}
where $k$ is the number of free parameters in the model fit and $N_{pts}$ is the number of data points. The power of the BIC is the penalty for a higher number of fitted model parameters, making it a robust way to compare different best-fit models. The preferred model is the one that produces the lowest BIC value. We find a BIC value of 675.08 and 682.76 for the model without \edits{(Model $\#$1)} and with fitting $\dot{\gamma}$ \edits{(Model $\#$2)}, respectively. \edits{For two generic models $i$ and $j$, we can relate the difference in the BIC values between models, $\Delta$(BIC$_{j,i}$) = BIC$_{j}$ - BIC$_{i}$, and the Bayes factor $B_{i,j}$, the ratio of the likelihood between models $i$ and $j$, assuming a Gaussian distribution for the posteriors (e.g. \citealt{Faulkenberry2018})}: 
\begin{align}
    B_{ij} =& \exp[\Delta(BIC_{ji})/2], \nonumber \\
    B_{ij} =& \exp[-\Delta(BIC_{ij})/2].
    \label{eq:Bayes_BIC}
\end{align}
 \edits{Therefore, Model $\#1$ without fitting for $\dot{\gamma}$ is the preferred model with a $\Delta$(BIC$_{1,2}$) = BIC$_{1}$ - BIC$_{2}$ = -7.68 and Bayes factor, B$_{1,2}$, of 46.5.} When fitting for $\dot{\gamma}$, we find a linear acceleration term that is positive but is consistent with zero within the uncertainties ($\dot{\gamma} = 0.0001^{+0.0034}_{-0.0036}$ m s$^{-1}$ day$^{-1}$). Our results are in conflict with the findings by \citet{Bouma2020} that find an acceleration along our line of sight at a rate of $\dot{\gamma} = -0.0422^{+0.0028}_{-0.0027}$ m s$^{-1}$ day$^{-1}$. We can reproduce the results of \citet{Bouma2020} by modeling only their data (See Figure \ref{fig:RVfit_Bouma} and Table \ref{tb:BoumaRV} in Appendix \ref{app:RV}). Based on this test, we conclude that the acceleration found by \citet{Bouma2020} was caused by modeling only part of the full RV data set. Therefore, we conclude that the changing orbital period detected using the transit data is not caused by the WASP-4 system accelerating towards Earth.    

\begin{table*}[p]
\caption{\edits{Results of the one-planet and two-planet radial-velocity fitting of the WASP-4 system with \texttt{RadVel}}}
\begin{tabular}{llllll}
\hline
Parameter	&	Model $\#$1	&	Model $\#$2	&	Model $\#$3	&	Model $\#$4	&	Units\\																					
%	&	1 planet	&	1-planet 	&	2 planet	&	2 planet	&	\\																					
%	&	\bf{Fit without $\ddot{\gamma}$}	&	\bf{Fit with $\ddot{\gamma}$} 	&	\bf{Fit without e$_{c}$}	&	\bf{Fit with e$_{c}$}	&	\\																					
 \hline																															
\multicolumn{6}{c}{\textbf{Modified MCMC Step Parameters}}\\																						
$P_{b}$	&	$1.338231343^{+7.2e-07}_{-1.8e-09}$	&	$1.338232067^{+3e-09}_{-7.2e-07}$	&	$1.3382313419^{+7.3e-07}_{-1.2e-09}$	&	$1.3382320688^{+1.5e-09}_{-7.3e-07}$	&	days \\																					
$T\rm{conj}_{b}$	&	$745.515752^{+1.9e-05}_{-1.9e-05}$	&	$745.515752^{+1.9e-05}_{-1.9e-05}$	&	$745.515752^{+1.9e-05}_{-1.9e-05}$	&	$745.515752^{+1.9e-05}_{-1.9e-05}$	&	JD \\																					
$e_{b}$	&	$\equiv0.0$	&	$\equiv0.0$	&	$\equiv0.0$	&	$\equiv0.0$	&	\\																					
$\omega_{b}$	&	$\equiv0.0$	&	$\equiv0.0$	&	$\equiv0.0$	&	$\equiv0.0$	&	radians \\																					
$K_{b}$	&	$238.1^{+3.7}_{-3.6}$	&	$238.2^{+3.8}_{-3.7}$	&	$237.3\pm 2.2$	&	$237.3^{+2.2}_{-2.3}$	&	m s$^{-1}$ \\																					
$P_{c}$	&	---	&	---	&	$7001.0^{+6.0}_{-6.6}$	&	$7001.2^{+6.2}_{-6.3}$	&	days \\																					
$T\rm{conj}_{c}$	&	---	&	---	&	$0^{+2300}_{-2100}$	&	$1^{+2300}_{-2100}$	&	JD \\																					
$T\rm{peri}_{c}$	&	---	&	---	&	$-1750^{+2300}_{-2100}$	&	$801\pm 2400$	&	JD \\																					
$e_{c}$	&	---	&	---	&	$\equiv0.0$	&	$0.094^{+0.12}_{-0.067}$	&	\\																					
$\omega_{c}$	&	---	&	---	&	$\equiv0.0$	&	$2.7^{+1.3}_{-1.2}$	&	radians \\																					
$K_{c}$	&	---	&	---	&	$64.2^{+2.3}_{-2.2}$	&	$63.7^{+5.7}_{-3.3}$	&	m s$^{-1}$ \\																					
 \hline																															
\multicolumn{6}{c}{\bf{Orbital Parameters}} \\																															
$P_{b}$	&	$1.338231343^{+7.2e-07}_{-1.8e-09}$	&	$1.338232067^{+3e-09}_{-7.2e-07}$	&	$1.3382313419^{+7.3e-07}_{-1.2e-09}$	&	$1.3382320688^{+1.5e-09}_{-7.3e-07}$	&	days \\																					
$T\rm{conj}_{b}$	&	745.515752$^{+1.9e-05}_{-1.9e-05}$	&	745.515752$^{+1.9e-05}_{-1.9e-05}$	&	745.515752$^{+1.9e-05}_{-1.9e-05}$	&	745.515752$^{+1.9e-05}_{-1.9e-05}$	&	JD \\																					
$e_{b}$	&	$\equiv0.0$	&	$\equiv0.0$	&	$\equiv0.0$	&	$\equiv0.0$	&	\\																					
$\omega_{b}$	&	$\equiv0.0$	&	$\equiv0.0$	&	$\equiv0.0$	&	$\equiv0.0$	&	radians \\																					
$K_{b}$	&	$238.1^{+3.7}_{-3.6}$	&	$238.2^{+3.8}_{-3.7}$	&	$237.3\pm 2.2$	&	$237.3^{+2.2}_{-2.3}$	&	m s$^{-1}$ \\																					
$P_{c}$	&	---	&	---	&	$7001.0^{+6.0}_{-6.6}$	&	$7001.2^{+6.2}_{-6.3}$	&	days \\																					
$a_{c}$	&	---	&	---	&	$6.82^{+0.23}_{-0.25}$	&6.83$^{+0.23}_{-0.25}$		&	AU \\																					
$T\rm{conj}_{c}$	&	---	&	---	&	$0^{+2300}_{-2100}$	&	$1^{+2300}_{-2100}$	&	JD \\																					
$T\rm{peri}_{c}$	&	---	&	---	&	$-1750^{+2300}_{-2100}$	&	$801\pm 2400$	&	JD \\																					
$e_{c}$	&		&		&	$\equiv0.0$	&	$0.094^{+0.12}_{-0.067}$	&	\\																					
$\omega_{c}$	&	---	&	---	&	$\equiv0.0$	&	$2.7^{+1.3}_{-1.2}$	&	radians \\																					
$K_{c}$	&	---	&	---	&	$64.2^{+2.3}_{-2.2}$	&	$63.7^{+5.7}_{-3.3}$	&	m s$^{-1}$ \\																					
$M_{c} sin(i)$	&	---	&	---	&	$5.47^{+0.44}_{-0.43}$	& 5.40$^{+0.59}_{-0.50}$		&	M$_{Jup}$ \\																					
 \hline																															
\multicolumn{6}{c}{\bf{Other Parameters}}\\																															
$\gamma_{\rm HIRES}$	&	$\equiv-33.5423$	&	$\equiv-33.4244$	&	$\equiv-42.8019$	&	$\equiv-41.3003$	&	m s$-1$ \\																					
$\gamma_{\rm HARPS}$	&	$\equiv-36.1144$	&	$\equiv-36.6269$	&	$\equiv-82.2669$	&	$\equiv-81.6365$	&	m s$-1$ \\																					
$\gamma_{\rm CORALIE}$	&	$\equiv57750.7122$	&	$\equiv57751.1433$	&	$\equiv57710.1977$	&	$\equiv57710.7494$	&	m s$-1$ \\																					
$\dot{\gamma}$	&	$\equiv0.0$	&	0.0001$^{+0.0034}_{-0.0036}$	&	$\equiv0.0$	&	$\equiv0.0$	&	m s$^{-1}$ d$^{-1}$ \\																					
$\ddot{\gamma}$	&	$\equiv0.0$	&	$\equiv0.0$	&	$\equiv0.0$	&	$\equiv0.0$	&	m s$^{-1}$ d$^{-2}$ \\																					
$\sigma_{\rm HIRES}$	&	$65^{+17}_{-13}$	&	$64^{+18}_{-14}$	&	$4.5^{+2.5}_{-1.8}$	&	$4.4^{+2.5}_{-1.8}$	&	$\rm m\ s^{-1}$ \\																					
$\sigma_{\rm HARPS}$	&	$12.7^{+3.1}_{-2.3}$	&	$13.5^{+3.5}_{-2.6}$	&	$12.3^{+2.8}_{-2.0}$	&	$12.7^{+2.9}_{-2.2}$	&	$\rm m\ s^{-1}$ \\																					
$\sigma_{\rm CORALIE}$	&	$10.6^{+4.9}_{-5.6}$	&	$11.0^{+5.1}_{-5.6}$	&	$5.2^{+4.9}_{-3.5}$	&	$5.3^{+5.0}_{-3.8}$	&	$\rm m\ s^{-1}$ \\																					
\hline																															
\end{tabular}
%\tablecomments{ 1940000 links saved}
\tablecomments{
  Reference epoch for $\gamma$,$\dot{\gamma}$,$\ddot{\gamma}$: 2455059. The HARPS and CORALIE data was taken from \citealt{Baluev2019} and the HIRES data was taken from \citealt{Bouma2020}. 
  \vspace{2em}
  }
\label{tb:RV_results_All}
\end{table*}

%%%%%%%% Timing Obs Table %%%%%%%
\begin{table*}[tbhp!]
    \centering
    \begin{tabular}{cccccc}
    \hline
    Event  & Midtime     & Error & Epoch & Timing Source  & Transit Source \\
    Type   & BJD$_{TDB}$ & days  &  \\
    \hline
tra	&	2458843.00493	&	0.00054	&	2325	&	This Paper & This Paper	\\
    \hline     
    \end{tabular}
    \caption{All WASP-4b Transit and Occultation Times}
    \label{tab:all_times}
    \tablecomments{
    This table is available in its entirety in machine-readable form.
    \tablerefs{
    \citealt{Wilson2008}; 
    \citealt{Winn2009_WASP4b};
    \citealt{Dragomir2011}; 
    \citealt{Caceres2011}; 
    \citealt{Beerer2011}; 
    \citealt{SanchisOjeda2011};
    \citealt{Nikolov2012}; 
    \citealt{Hoyer2013}; 
    \citealt{Zhou2015};
    \citealt{Huitson2017};
    \citealt{Southworth2019}; 
    \citealt{Baluev2019};
    \citealt{Bouma2019}; 
    \citealt{Baluev2020}   
    }
    }
\end{table*}

%%%%% Comment out%%%%%%% 
\begin{comment}
%%% 2 planet, no acc
\begin{figure}[thb!]
\centering
 \begin{tabular}{c}
 \includegraphics[width=1\textwidth,page=1]{wasp4_config_july30_2pv5_rv_multipanel_2planet_new.pdf}}
 \end{tabular}
\caption{Best-fit 2-planet Keplerian orbital model
  for WASP-4b using \texttt{RadVel}. In this model, we do not fit for the linear acceleration ($\dot{\gamma}$) and fix the eccentricity of both bodies to zero. {\bf a)} The maximum likelihood model is plotted while the orbital parameters listed in Table \ref{tb:RV_results_twoplanet} are the
  median values of the posterior distributions.  The thin blue line is
  the best fit 2-planet model. We add in quadrature
  the RV jitter terms listed in Table \ref{tb:RV_results_twoplanet} with the
  measurement uncertainties for all RVs.  {\bf b)} Residuals to the
  best fit 2-planet model. {\bf c)} RVs phase-folded
  to the ephemeris of WASP-4b. The Keplerian orbital models for all other planets have been subtracted. The small point colors
  and symbols are the same as in panel {\bf a}.  Red circles are the same velocities binned in 0.08 units of orbital phase. The phase-folded model for WASP-4b is shown as the blue line. {\bf d)} RVs phase-folded
  to the ephemeris of WASP-4c. The Keplerian orbital models for all other planets have been subtracted. The phase-folded model for WASP-4c is shown as the blue line.}
\label{fig:RVfits_2planet}
\end{figure}
\end{comment}

Next, we fit the RV data with a 2-planet model because the residuals of the one-planet fit showed some sinusoidal structure \edits{(panel b in Figures \ref{fig:RVModels_all}I-II)}. This sinusoidal trend is not caused by stellar activity as the S-index time series from the HIRES data shows no signs of secular or sinusoidal trends (\citealt{Bouma2020}). We performed several different models where we fit for K$_{b}$, $\sigma$, $\dot{\gamma}$, the orbital velocity semi-amplitude of the 2nd body (K$_{c}$). and an eccentricity of the 2nd planet (e$_{c}$). \edits{The results of Models $\#$3 and $\#$4 are summarized in Table \ref{tb:RV_compare}. We find that Model $\#$3} with K$_{b}$ and K$_{c}$ set as free parameters and e$_{b}$, e$_{c}$, and $\dot{\gamma}$ fixed to zero finds the best fit with a BIC of 628.34. \edits{The derived orbital parameters of this model can be found in Table \ref{tb:RV_results_All} and the two-planet RV fit can be found in Figures \ref{fig:RVModels_all}III-IV.} The two-planet model \edits{(Model $\#3$) is highly preferred over the one-planet model (Model $\#1$) with \edits{a $\Delta(BIC_{3,1})$ = -46.74 and a Bayes factor, B$_{3,1}$, of 1.41$\times10^{10}$}.} For the second body, we find a period (P$_{c}$) of 7001.0$^{+6.0}_{-6.6}$ days, a semi-major axis of $6.82^{+0.23}_{-0.25}$ AU, and a $M_{c} sin(i)$ of $5.47^{+0.44}_{-0.43} M_{Jup}$ \edits{(Table \ref{tb:planet_parameters}, Table \ref{tb:RV_results_All})}. \editstwo{The companion is expected to be much fainter than the planet host star because \citet{Wilson2008} found no evidence for changing spectral line bisectors in their spectroscopic observation.} \citet{Becker2017} found that distant exterior companions to hot Jupiters around cool stars ($T_{star} <$ 6200 K) are typically coplanar within 20-30 degrees. Therefore, we find that the mass of WASP-4c is between 5.47--6.50 $M_{Jup}$ assuming that its inclination is within 30 degrees of WASP-4b's inclination. However, more RV measurements are needed to verify the existence of this second planet around WASP-4.

\vspace{-0.22em}
\subsection{Transit Timing Variations}

\begin{table*}[ptbh!]
\caption{Best-Fit parameters for the timing models}
    \centering
    \begin{tabular}{lllll}
     \hline
     Parameter                          &  Symbol       &   units       & value  & 1 $\sigma$ uncertainty  \\
     \hline
    \multicolumn{5}{c}{\textbf{Constant Period Model}} \\
        Period                          & P$_{orb}$         & days          & 1.338231392         & 0.000000014\\
        Mid-transit time                & T$_{c,0}$       &BJD$_{TDB}$      & 2455804.515677      & 0.000015\\
        \hline                                                              
        N$_{dof}$                         &               &               & 180+2          & \\
        $\chi_{min}^2$                    &               &               &  386.40         &       \\
        BIC     &               &               &  396.78         &\\
     \hline
     \hline 
   \multicolumn{5}{c}{ \textbf{Orbital Decay Model}}\\ 
        Period                          & P$_{orb}$ &  days         &  1.338231587      & 0.000000022\\
        Mid-transit time                & T$_{c,0}$ &BJD$_{TDB}$      &  2455804.515781   & 0.000018 \\
        Decay Rate                      & dP/dE     &days/orbit     & -0.000000000311   & 0.000000000030  \\
        Decay Rate                      & dP/dt     & msec/yr       &  -7.33            & 0.71    \\
        \hline
        N$_{dof}$                         &           &               &  180+3   \\ 
        $\chi_{min}^2$                    &           &               &  276.35      \\
        BIC \                           &           &               &  291.93         \\
        \hline 
        \hline
      \multicolumn{5}{c}{  \textbf{Apsidal Precession Model}} \\
        Sidereal Period                  & P$_{s}$    & days          & 1.338231448   &    0.000000098       \\
        Mid-transit time                 & T$_{c,0} $ & BJD$_{TDB}$     & 2455804.51545 &    0.00022                 \\
        Eccentricity                     & e         &               & 0.00090      &    0.00051\\
        Argument of Periastron           & $\omega_{0}$& rad           &  2.70        &    0.36        \\
        Precession Rate                  & $d\omega/dN$& rad/orbit     & 0.0011       &    0.00037               \\
       \hline                                                          
        N$_{dof}$                          &           &               &   180+4                \\ 
        $\chi_{min}^2$                     &           &               &   270.42              \\
        BIC \                            &           &               &   296.39              \\
        \hline 
        \hline
    \end{tabular}
    \label{tb:timing_models}
\end{table*}

For the timing analysis, we combined the TESS transit data with all the prior transit and occultation times. All the transit and occultation times used in this analysis can be found in Table \ref{tab:all_times}. In the table we give the original reference in which the data were reported and the reference for the timing if different from the original source. We combine transit data as tabulated by \citet{Bouma2020} and additional transits reported by amateur observers from \citet{Baluev2020}. This table is available in its entirety in machine-readable form online.

Similar to what was done in \citet{Yee2020}, \citet{Patra2017}, and \citet{Turner2021}, we fit the timing data to three different models. The first model is the standard constant period formalization:
\begin{align}
    t_{\text{tra}}(E) =& T_{c,0} + P_{orb} \times E,  \\
   t_{\text{occ}}(E)  =&    T_{c,0} +\frac{P_{orb}}{2}  + P_{orb} \times E,
\end{align}
\noindent where $T_{0}$ is the reference transit time, $P_{orb}$ is the orbital period, $E$ is the transit epoch, and $T_{tra}(E)$ is the calculated transit time at epoch $E$. 

\begin{figure*}[htb!]
\centering
 \begin{tabular}{ll}
  \textbf{(a.)} & \textbf{(b.)}\\ 
  \includegraphics[width=0.45\textwidth,page=1]{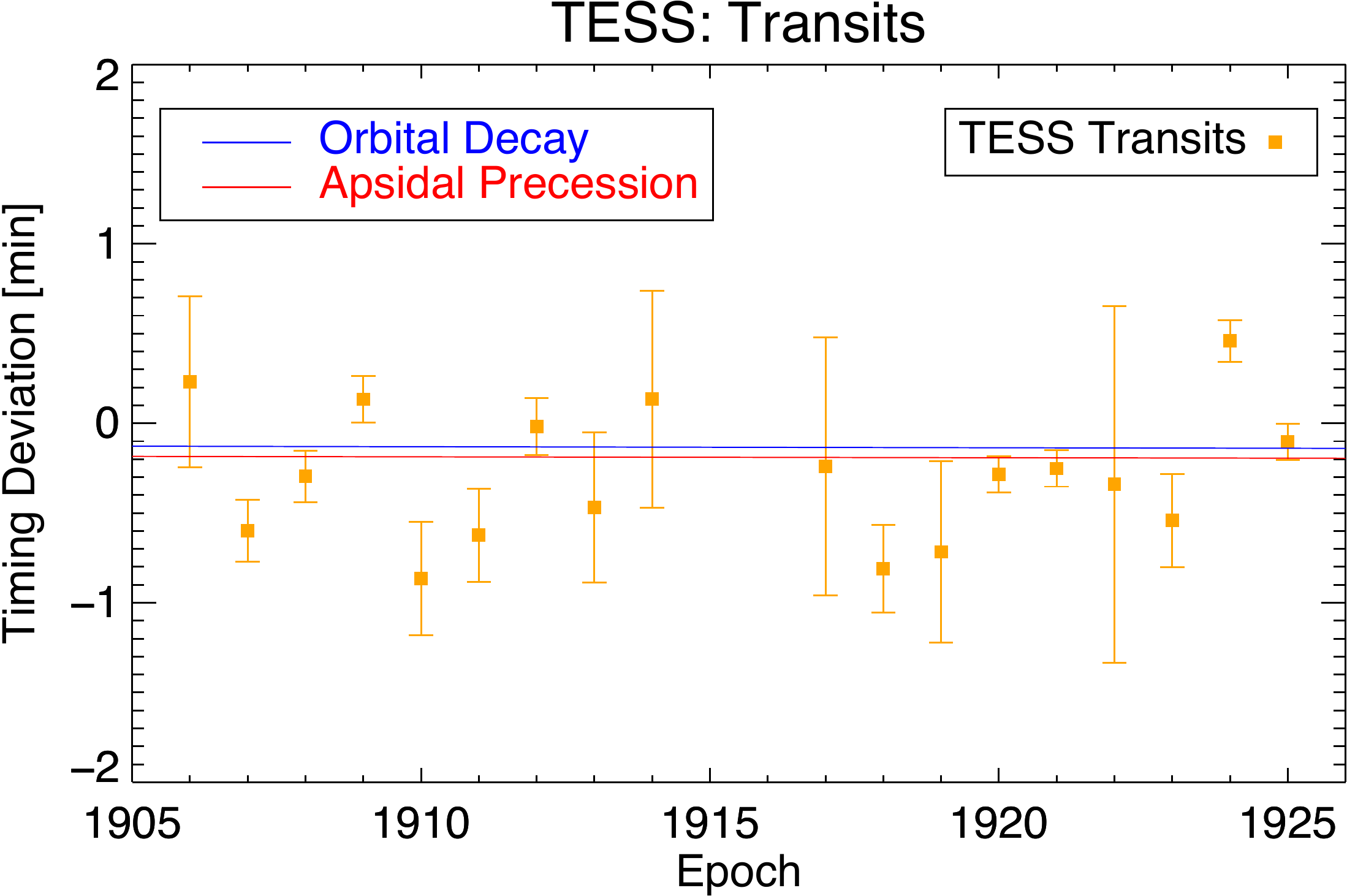} & \includegraphics[width=0.45\textwidth,page=2]{WASP4b_TTVS_data_with_models.pdf} \\
 \textbf{(c.)} & \textbf{(d.)}\\ 
\includegraphics[width=0.45\textwidth,page=4]{WASP4b_TTVS_data_with_models.pdf} &  \includegraphics[width=0.45\textwidth,page=5]{WASP4b_TTVS_data_with_models.pdf} \\
  % \sidesubfloat[\textbf{a.}]{   \includegraphics[width=0.45\textwidth,page=1]{WASP4b_TTVS_data_with_models.pdf}} & 
  %\sidesubfloat[\textbf{b.}]{    \includegraphics[width=0.45\textwidth,page=2]{WASP4b_TTVS_data_with_models.pdf}} \\
  % \sidesubfloat[\textbf{c.}]{ \includegraphics[width=0.45\textwidth,page=4]{WASP4b_TTVS_data_with_models.pdf}} &
   %   \sidesubfloat[\textbf{d.}]{ \includegraphics[width=0.45\textwidth,page=5]{WASP4b_TTVS_data_with_models.pdf}}\

 \end{tabular}
\caption{WASP-4b transit (panels a, b, and c) and occultation (panel d) timing variations after subtracting the data with a constant-period model. The filled black triangles are the data points from the literature and the square orange points are from the TESS data in this paper. All the transit and occultation times can be found in Table \ref{tab:all_times}. The orbital decay and apsidal precession models are shown as the blue and red lines, respectively.
}
\label{fig:timing_fits}
\end{figure*}

The second model assumes that the orbital period is changing uniformly over time:
\begin{align}
    t_{\text{tra}}(E) =& T_{c,0} + P_{orb} \times E  + \frac{1}{2} \frac{\mathrm{d}P_{orb}}{\mathrm{d}E} E^2, \\
    t_{\text{occ}}(E) =&  T_{c,0} +\frac{P_{orb}}{2}  + P_{orb} \times E + \frac{1}{2} \frac{\mathrm{d}P_{orb}}{\mathrm{d}E} E^2,
\end{align}
where $\mathrm{d}P_{orb}/\mathrm{d}E$ is the decay rate. 

The third model assumes the planet is precessing uniformly (\citealt{Gimenez1995}):
\begin{align}
     t_{\text{tra}}(E) =&   T_{c,0} + P_{s} \times E  - \frac{e P_{a}}{\pi} \cos{\omega(N)},\\
      t_{\text{occ}}(E) =&  T_{c,0} + \frac{P_{orb}}{2} + P_{s} \times E  + \frac{e P_{a}}{\pi} \cos{\omega(N)},\\
      \omega(N) =& \omega_{0} + \frac{\mathrm{d} \omega}{\mathrm{d}E} E,\\
      P_{s} =& P_{a} \left( 1 - \frac{1}{2\pi} \frac{\mathrm{d} \omega}{\mathrm{d}E} \right),
\end{align}
where $e$ is a nonzero eccentricity, $\omega$ is the argument of pericenter, P$_{s}$ is the sidereal period and P$_{a}$ is the anomalistic period.

For all three models, we found the best-fitting model parameters using a DE-MCMC analysis. We used 20 chains and 20$^{6}$ links in the model and again we ensure chain convergence using the Gelman-Rubin  statistic. 

The results of timing model fits can be found in Table \ref{tb:timing_models}. Figure \ref{fig:timing_fits} shows the transit and occultation timing data fit with the orbital decay and apsidal precession models. In this figure, the best-fit constant-period model has been subtracted from the timing data. The orbital decay model fits the transit and occultation data slightly better than the precession model (Table \ref{tb:timing_models}, Figure \ref{fig:timing_fits}).

Our finding that the constant-period model does not fit the data well is consistent with previous studies (\citealt{Bouma2019,Southworth2019,Baluev2020}). The orbital decay and apsidal precession models fit the data with a minimum chi-squared ($\chi_{min}^{2}$) of 276.35 and 270.42, respectively. We find that the orbital decay model is the preferred model with a \edits{$\Delta$(BIC) = -5.93 and a Bayes factor of 9.3.} Therefore, based on our analysis of the observed timing residuals, the orbital decay model is only slightly preferred over apsidal precession.

\begin{figure}[]
\centering
 \begin{tabular}{c}
   \includegraphics[width=\textwidth]{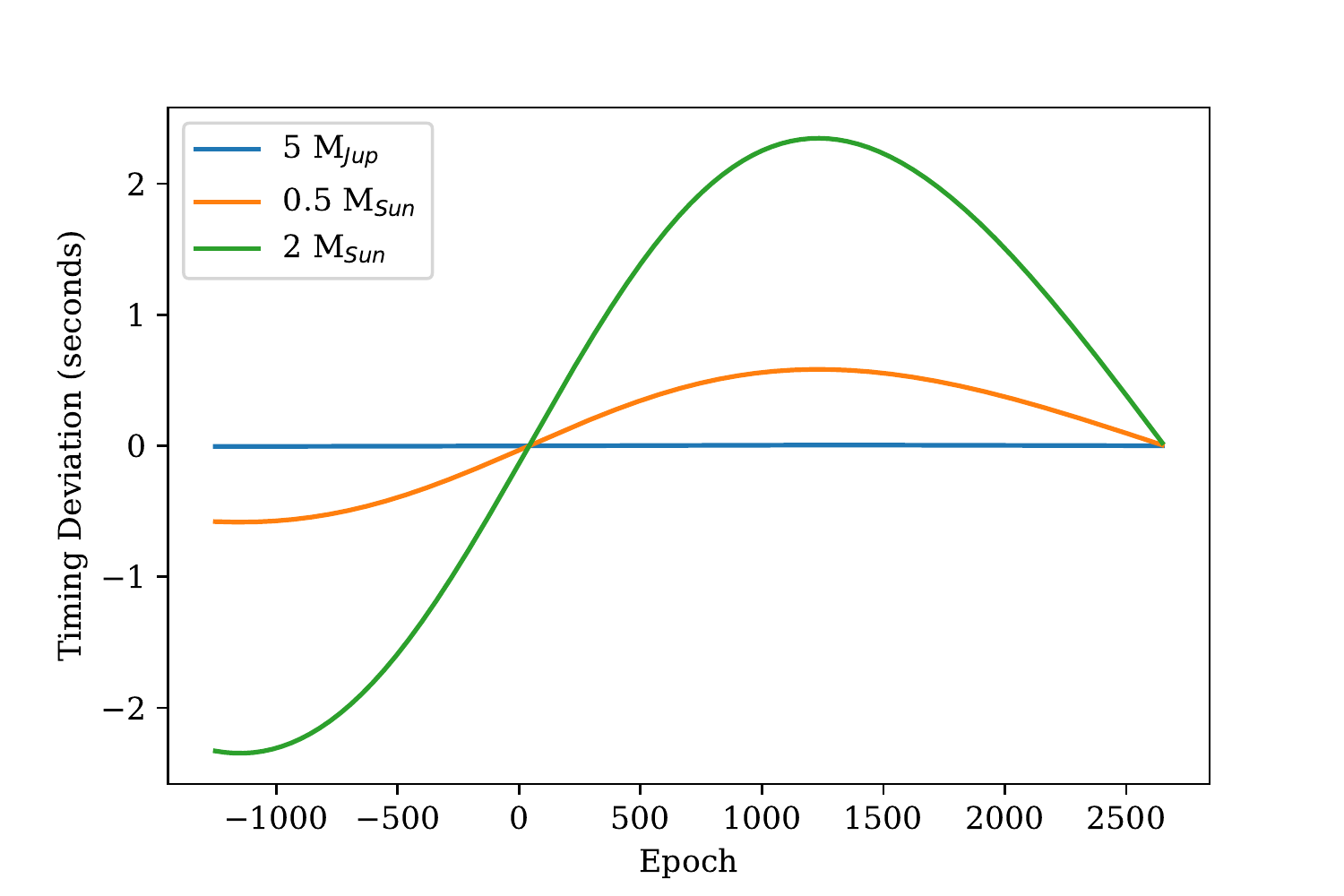} 
 \end{tabular}
\caption{Forward models of the TTV signal of WASP-4b produced by the possible outer companion. These models were created using \texttt{OCFit}. The models used the priors as listed in Table \ref{tb:RV_results_All} where the outer planet has an eccentricity of 0.094 (Model $\#4$). The two objects are assumed to be co-planar. }
\label{fig:TTVs_predict_2ndplanet}
\end{figure}

Due to the RV measurements showing evidence of a possible second planet, we modeled the two-planet system to see if they could reproduce the TTVs. For this analysis, we used the publicly available TTV analysis package, \texttt{OCFit}\footnote{\url{https://github.com/pavolgaj/OCFit}} \citep{Gajdos2019OCFitRef}. Specifically, we used the \texttt{AgolExPlanet} function which is an implementation of Equation 25 in \citet{Agol2005}.  The priors in \texttt{OCFit} were set to the values given for Model $\#4$  in Table \ref{tb:RV_results_All} for both planets where the outer planet has an eccentricity of 0.094. We were not able to fit the TTVs well with an outer planet consistent with the RV constraints. We also produced several forward models with \texttt{OCFit} that show that the expected TTV signal from the outer body is less than 2 seconds (dependent on the real mass of the body) over the full observational period (Figure \ref{fig:TTVs_predict_2ndplanet}). \edits{We did not use the preferred two-planet model (Model $\#$3) because this model did not produce \editstwo{a detectable signal.}} The two objects are assumed to be co-planar but relaxing this condition will only decrease the TTV signal. 

We can also analytically calculate the expected TTV signal on WASP-4b ($\delta t_{b}$) using the following equation from \citet{Agol2005} assuming non-resonant planets with large period-ratios on circular orbits: 
\begin{equation}
    \delta t_{b} = \frac{M_{c}}{M_{star}} \left( \frac{P_{b}}{P_{c}}\right)^2 P_{b},  \label{eq:TTV_signal}
\end{equation}
where $M_{c}$ and $P_{c}$ are the mass and orbital period of the outer companion, $M_{star}$ is the mass of WASP-4, and $P_{b}$ is the period of WASP-4b. The assumptions of equation \eqref{eq:TTV_signal} are all satisfied within the constraints of the best-fit RV model parameters found by \texttt{RadVel} (Table \ref{tb:RV_results_All}). For a $M_{c}$ between 5.47 $M_{Jup}$ and 2 M$_{sun}$, we find a $\delta t_{b}$ using equation \eqref{eq:TTV_signal} to be between $1.9\times10^{-10}$ secs to $7.4\times10^{-8}$ secs. Hence, the expected TTV signal is many orders magnitude below the observed TTVs regardless of the mass of the outer companion. Our results are expected as resonant perturbations between close planets is the main cause of large ($\sim>$mins) TTVs (e.g. \citealt{Agol2005,Steffen2012,Nesvorn2013,Dawson2019}). Therefore, we conclude that the observed TTVs are caused by orbital decay or apsidal precession and not gravitational perturbations from the outer body.

\begin{figure}[htb!]
\centering
 \begin{tabular}{l}
 \textbf{(a.)}\\
 \includegraphics[width=0.9\textwidth,page=2]{MCMC_model_prediction.pdf}\\
  \textbf{(b.)}\\
 \includegraphics[width=0.9\textwidth,page=1]{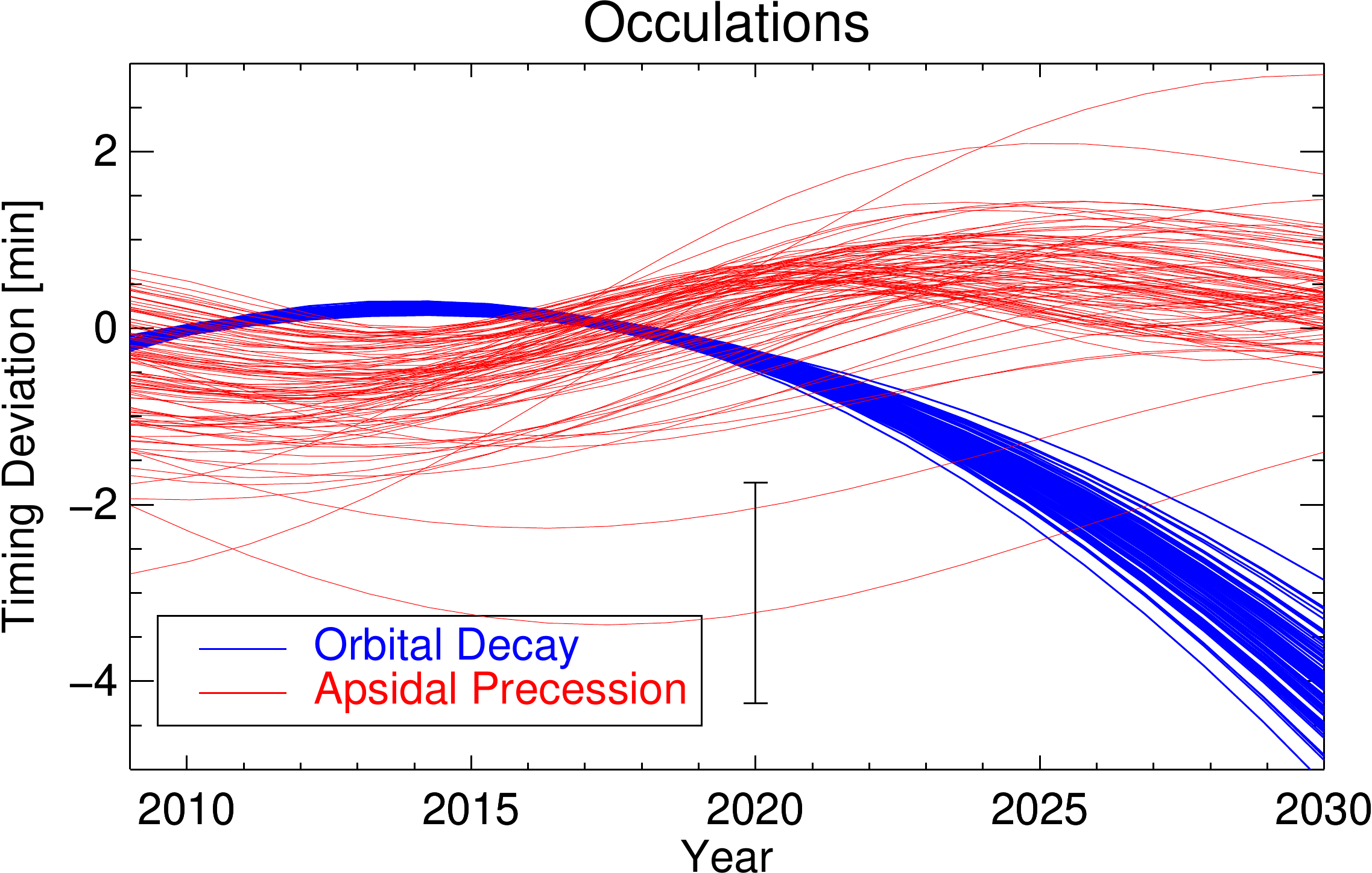}
%  \sidesubfloat[\textbf{(a.)}]{   \includegraphics[width=0.9\textwidth,page=1]{MCMC_model_prediction.pdf}} \\
 \end{tabular}
\caption{WASP-4b transit (panels a) and occultation (panel b) timing variations predictions using the best-fit models to the timing data presented in this paper. The lines are 100 random draws from the posteriors of the orbital decay (blue) and apsidal precession (red) models. On both panels, we show the average error of the previous observations. 
}
\label{fig:timing_fits_future}
\end{figure}

\section{Discussion}
From our analysis, we derived updated planetary parameters and find that the orbital decay model is slightly preferred to explain the observed TTVs. We conclusively rule out linear acceleration as the cause of the period change. We also show that TTVs can not be caused by the second body orbiting WASP-4 regardless of its mass. However, apsidal precession is not ruled out. Further transit and occultation measurements of WASP-4b are needed to disentangle the cause of the variation. The orbital decay and apsidal precession models exhibit very different timing variations in the mid-2020s (Figure \ref{fig:timing_fits_future}). Therefore, we predict that it will be possible to conclusively determine the cause of WASP-4b's changing orbit by then.

%We predict that by the mid-2020s that the cause of WASP-4b's period change can be conclusively determined because the orbital decay and apsidal precession models exhibit very different timing variations over that time range (Figure \ref{fig:timing_fits_future}).  

%We predict that by the mid-2020s that the orbital decay and apsidal precession models exhibit different timing variations that cause of the period change on WASP-4b can conclusively be determined (Figure \ref{fig:timing_fits_future}).  

Assuming the TTVs can be explained entirely by orbital decay, we derive an updated period of \NewPeriod{} and a decay rate of \Newdecayrate{}. Our results indicate an orbital decay timescale of $\tau = P/|\dot P| = 15.77\pm1.57$ Myr, slightly longer than the value derived by \citet{Bouma2019} of 9.2 Myr. Assuming that the planet mass is constant, the rate of change of WASP-4b's orbital period, $\dot P$, can be related to its host star's modified tidal quality factor by the constant-phase-lag model of \cite{Goldreich1966}, defined as 
\begin{align}
    \dot P = - \frac{27\pi}{2 Q'_\star} \left ( \frac{M_p}{M_\star} \right ) \left ( \frac{R_\star}{a} \right )^5,
\end{align} 
where $M_p$ is the mass of the planet, $M_\star$ is the mass of the host star, $R_\star$ is the radius of the host star and $a$ is the semi-major axis of the planet. Using our derived value of $\dot P$ and planetary parameters from Table \ref{tb:planet_parameters}, we find a modified tidal quality factor of $Q'_\star$ = 5.1$\pm$0.9 $\times 10^4$. This value is higher than the value found by \cite{Bouma2019} of $Q'_\star$ = 2.9$\pm$0.3 $\times 10^4$ and by \cite{Southworth2019} of $Q'_\star$ = 3.8$\pm$0.3 $\times 10^4$. The cause of this small discrepancy (all $Q'_\star$ values are within 2$\sigma$) is due to our including more transit data in our analysis than previous studies.  

It is currently not clear how to account theoretically for all the observed low values of $Q'_\star$.  Our value is an order of magnitude lower than the observed values for binary star systems \citep[10$^5$ - 10$^7$;][]{Meibom2015} and hot Jupiters (10$^{5} -10^{6.5}$; e.g., \citealt{Jackson2008,Husnoo2012,Barker2020}) and theoretically predicted values ($10^5 - 10^{10}$; e.g.,  \citealt{Ogilvie2014,Essick2016}; \edits{\citealt{CollierCameron2018,Linhao2021}}). However, our value is consistent with \citet{Hamer2019} who found that hot Jupiter host stars tend to be young requiring $Q'_\star \lesssim 10^7$. Some possible causes of a large tidal dissipation rate might be that WASP-4 is turning off the main sequence (as suspected for WASP-12; \citealt{Weinberg2017}) or that an exterior planet could be trapping WASP-4b's spin vector in a high-obliquity state (also predicted for WASP-12b; \citealt{Millholland2018}). The latter theory is intriguing as we now have evidence for an additional body in the system (Figure \ref{fig:RVModels_all}). More investigation is needed to understand the low value of $Q'_\star$. 

%Previous theoretical and observational studies have found a large range of $Q'_\star$ values.
%observe: 10^5.5 sample of extrasolar planets Jackson2008 
% theory: theoretical study 

If confirmed, WASP-4b would be the second exoplanet after WASP-12b (\citealt{Yee2020,Turner2021}) to show evidence of tidal orbital decay. Future observations of WASP-4b are needed to verify this possibility. Other planets such as WASP-103b, KELT-16b, and WASP-18b are also predicted to exhibit large rates of tidal decay \citep{Patra2020}. \edits{Additionally, the planets HATS-2b and WASP-64b are also ideal candidates to search for orbital decay because they are in systems similar to WASP-4 (\citealt{Southworth2019}).} To understand the formation and evolution of the hot Jupiter population, timing observations of additional systems are needed.

The discovery of WASP-4c makes the WASP-4 system unique, as it would be the most widely separated companion of a transiting hot-Jupiter known to date. However, our discovery of WASP-4c is not surprising because \citet{Knutson2014_RV} estimated that 27$\%\pm6\%$ of hot Jupiters have a planetary companion at a distance of 1-10 AU and a mass of 1-13 M$_{Jup}$. Similar planets may be common in other systems but could only be detected with long time-baseline RV data sets. Therefore, we expect that the unique status of the WASP-4 system is a result of observational bias rather than an intrinsic rarity of such systems. 

Future observations of the WASP-4 system can help us put the system in context with the overall hot Jupiter population and shed light on the possible formation scenarios of the system. Specifically, stronger constraints on the obliquity, mutual inclinations, and full orbital parameters of WASP-4c will help us understand planetary migration in this system.

% one paragraph on how WASP-4 system is not special but WASP-4c is a unique discovery. WASP-4c has larger orbit of companion of a hot Jupiter. More to come!   
% one paragraph with speculation of formation. Not really have a conclusion but that we add to evidence of how these can form. a/a_roche is useful but our planet is decaying so comparison is hard to understand. obliquity is interesting but inconlusive  

\section{Conclusions}
We analyzed all available sectors of TESS data of WASP-4b to investigate its possible changing orbit. Our TESS transit timing investigations confirm that the planet's orbit is changing (Figure \ref{fig:timing_fits}). We conclude that the acceleration of the WASP-4 system towards Earth is not the cause of the period variation after analyzing all available RV data (Figure \ref{fig:RVModels_all}; Table \ref{tb:RV_results_All}). From the RV analysis, we also find evidence of a possible second planet orbiting WASP-4 with a period (P$_{c}$) of 7001.0$^{+6.0}_{-6.6}$ days, semi-major axis of $6.82^{+0.23}_{-0.25}$ AU, and a $M_{c} sin(i)$ of $5.47^{+0.44}_{-0.43} M_{Jup}$ (Figure \ref{fig:RVModels_all}, Table \ref{tb:RV_results_All}). WASP-4c is the most widely separated companion of a transiting hot-Jupiter discovered to date. This outer planet is not the cause of the observed TTVs (Figure \ref{fig:TTVs_predict_2ndplanet}). Our timing analysis slightly favors the orbital decay scenario over apsidal precession as the cause of the TTVs (Table \ref{tb:timing_models}, Figure \ref{fig:timing_fits}). However, apsidal precession cannot be ruled out. We find an updated period of \NewPeriod{}, a decay rate of \Newdecayrate{}, and an orbital decay lifetime of 15.77$\pm$1.57 Myr assuming the system is undergoing orbital decay. The planetary physical parameters are also updated with greater precision than previous studies. More transit, occultation, and RV data are needed over the next few years to determine conclusively the cause of WASP-4b's changing orbit and help place the system in context with the overall hot Jupiter population.

\acknowledgments

Support for this work was provided by NASA through the NASA Hubble Fellowship grant $\#$HST-HF2-51495.001-A awarded by the Space Telescope Science Institute, which is operated by the Association of Universities for Research in Astronomy, Incorporated, under NASA contract NAS5- 26555.

We are grateful to Benjamin (BJ) Fulton for his help with \texttt{RadVel}. We thank Dong Lai and Maryame El Moutamid for useful discussions. We were inspired to pursue this project after attending a TESS hackathon hosted by the Carl Sagan Institute.

\edits{We also thank the anonymous referee for the useful comments.}

This paper includes data collected by the TESS mission, which are publicly available from the Mikulski Archive for Space Telescopes  (MAST). Funding for the TESS mission is provided by the NASA Explorer Program.

This research has made use of the Extrasolar Planet Encyclopaedia, NASA's Astrophysics Data System Bibliographic Services, and the the NASA Exoplanet Archive, which is operated by the California Institute of Technology, under contract with the National Aeronautics and Space Administration under the Exoplanet Exploration Program.  

%We also thank the anonymous referee for their comments.

%This document contains \total{citnum}\ references.   %For reference, will delete when paper is submitted


\facilities{\textit{TESS} (\citealt{Ricker2015}); \textit{Exoplanet Archive}}

\software{
          \texttt{EXOMOP} \citep{Pearson2014,Turner2016b,Turner2017}; \texttt{IDL Astronomy Users Library} \citep{Landsman1995}; \texttt{Coyote IDL} created by David Fanning and now maintained by Paulo Penteado (JPL); \texttt{RadVel} \citep{Fulton2018}
}

%We used the following {\tt Python} packages: {\tt Astropy} \citep{Astropy18}, {\tt SciPy} \citep{Jones01}, {\tt NumPy} \citep{Colbert11, Oliphant15}} %\citep{Landsman1995}
%\ \texttt{molefit} \citep{Smette2015} 


\appendix
\restartappendixnumbering

\section{Transit fits to individual TESS transit events} \label{app:individual_transits}
\begin{figure*}
\centering
\begin{tabular}{cc}
 \includegraphics[width=0.50\textwidth]{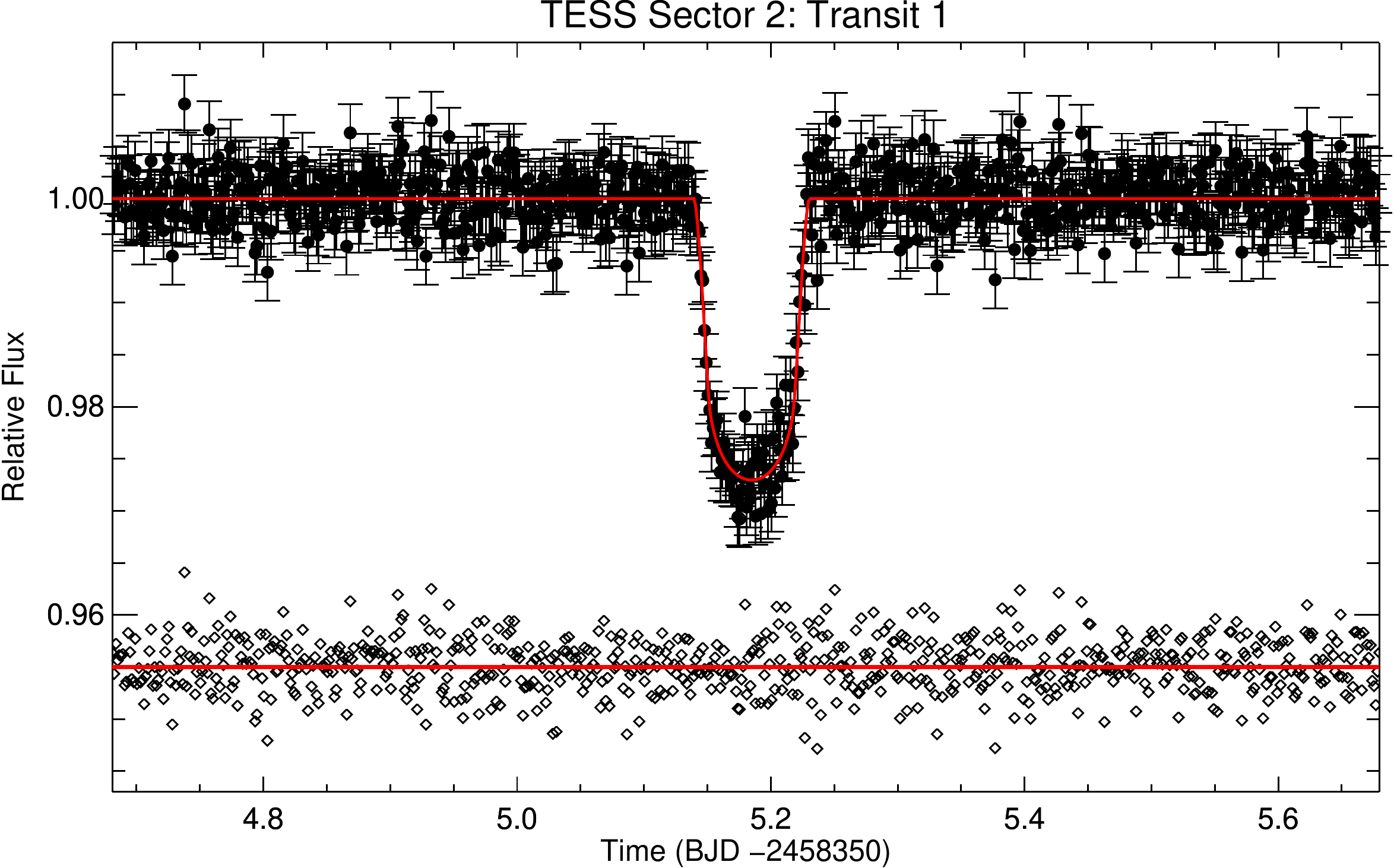} & \includegraphics[width=0.50\textwidth]{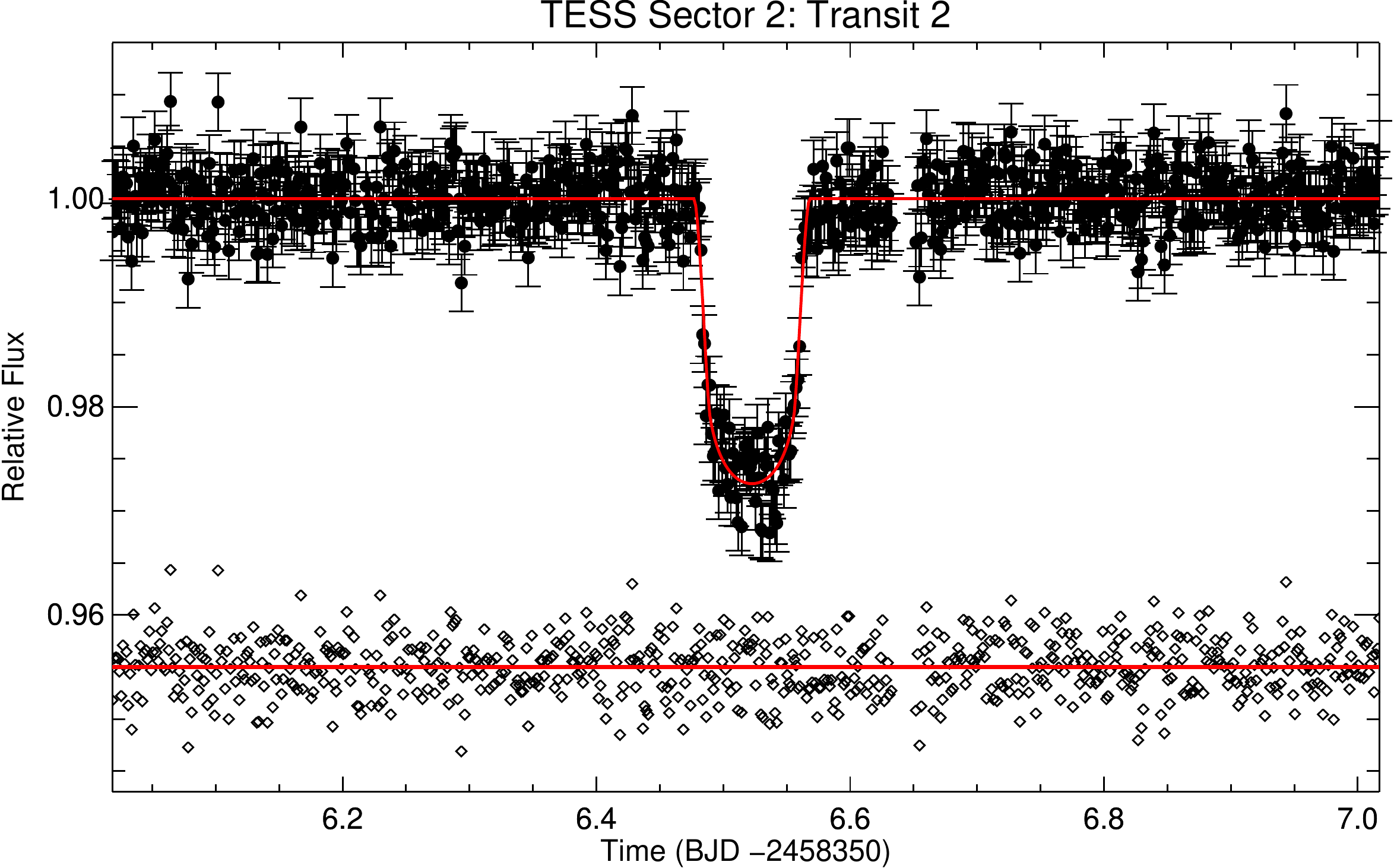}\\
 \includegraphics[width=0.50\textwidth]{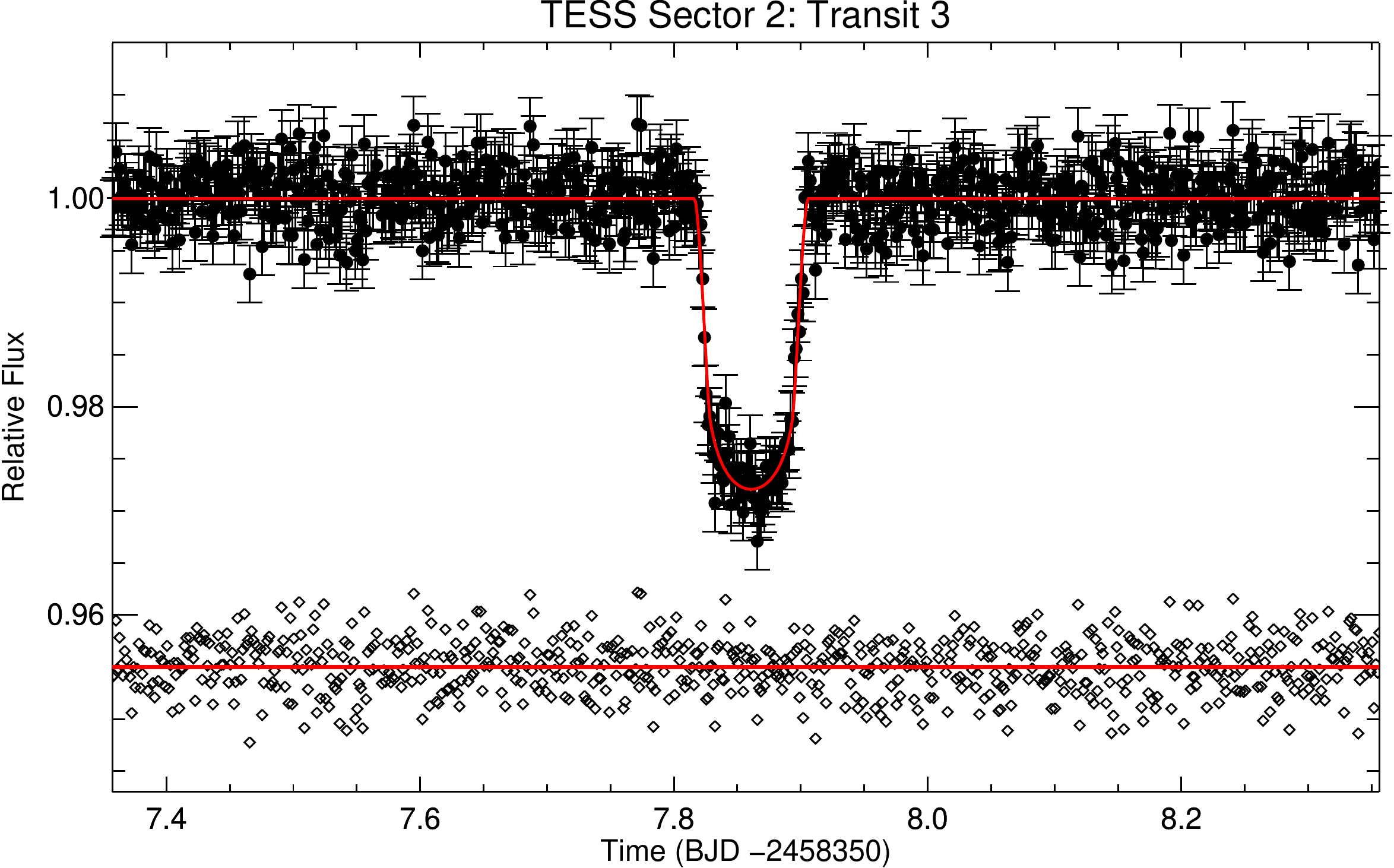} & \includegraphics[width=0.50\textwidth]{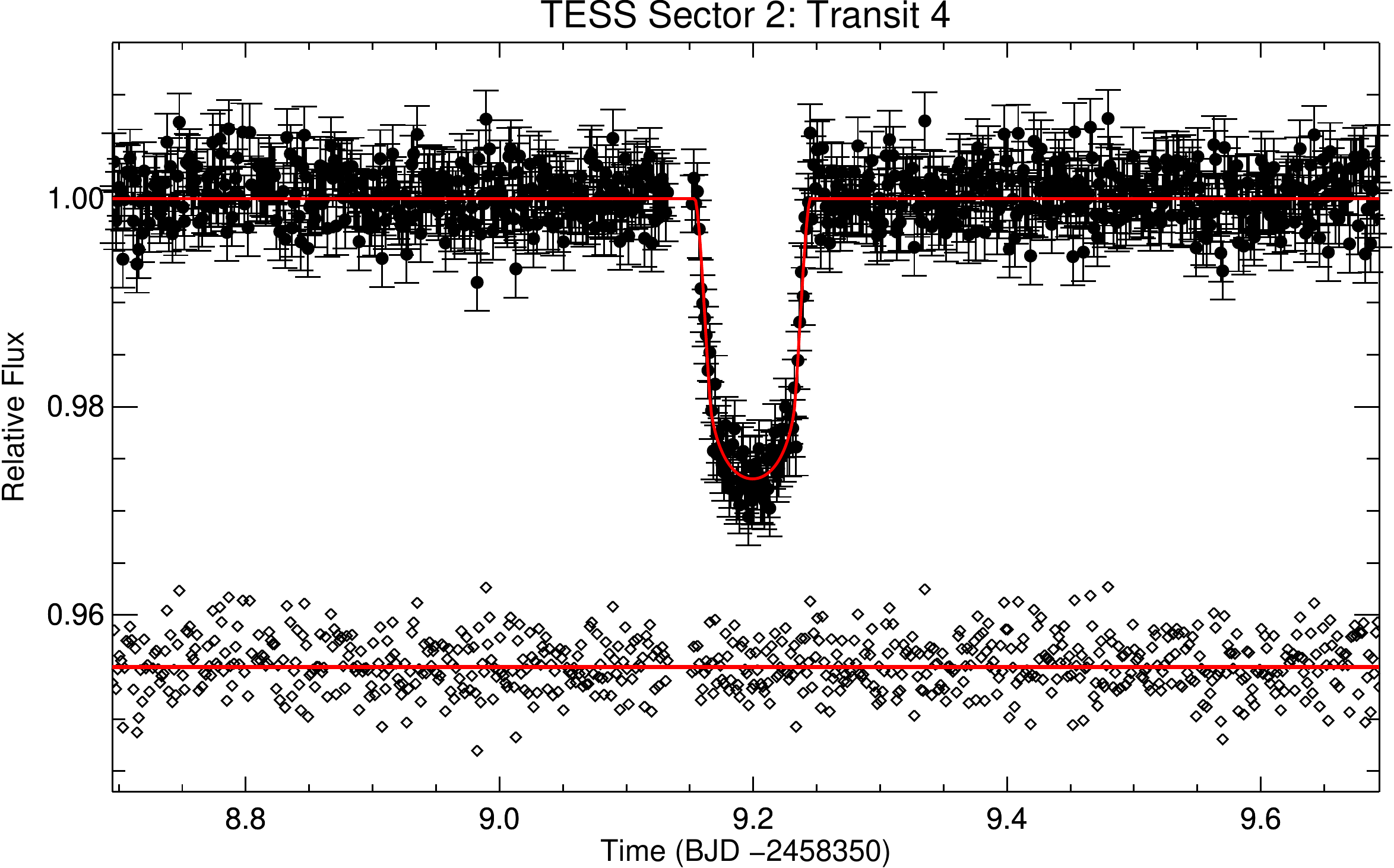}\\
  \includegraphics[width=0.50\textwidth]{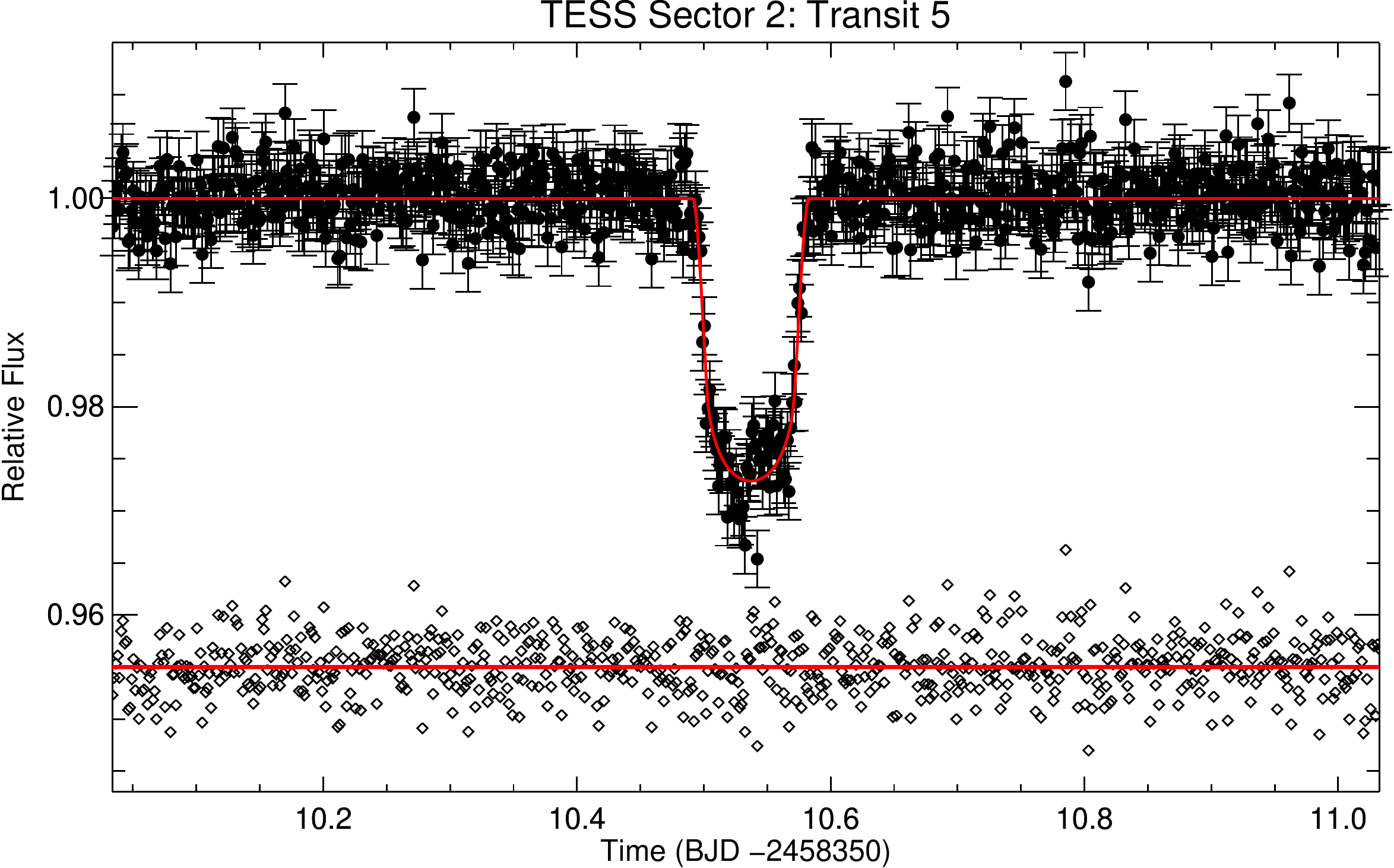} & \includegraphics[width=0.50\textwidth]{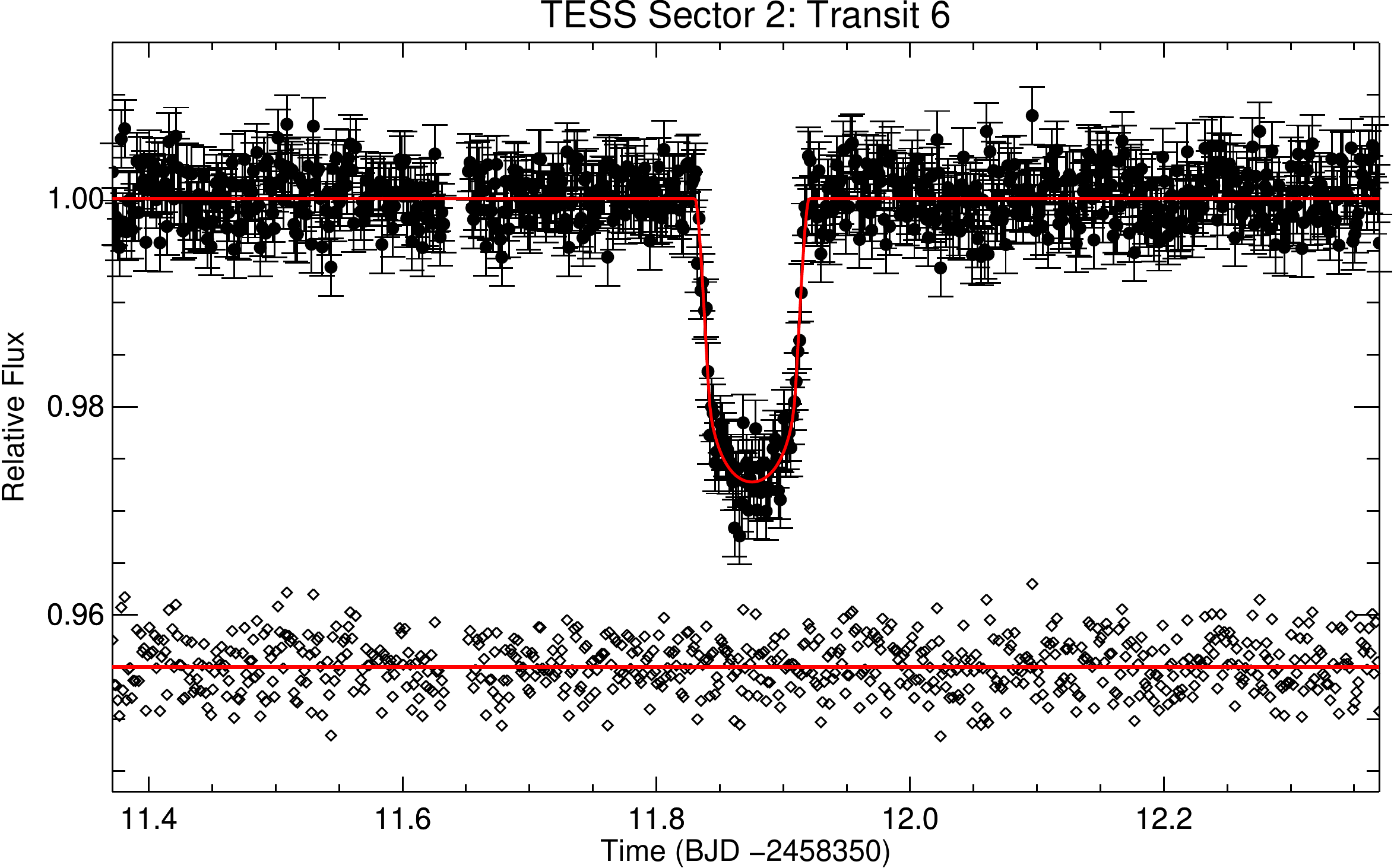}\\
  \includegraphics[width=0.50\textwidth]{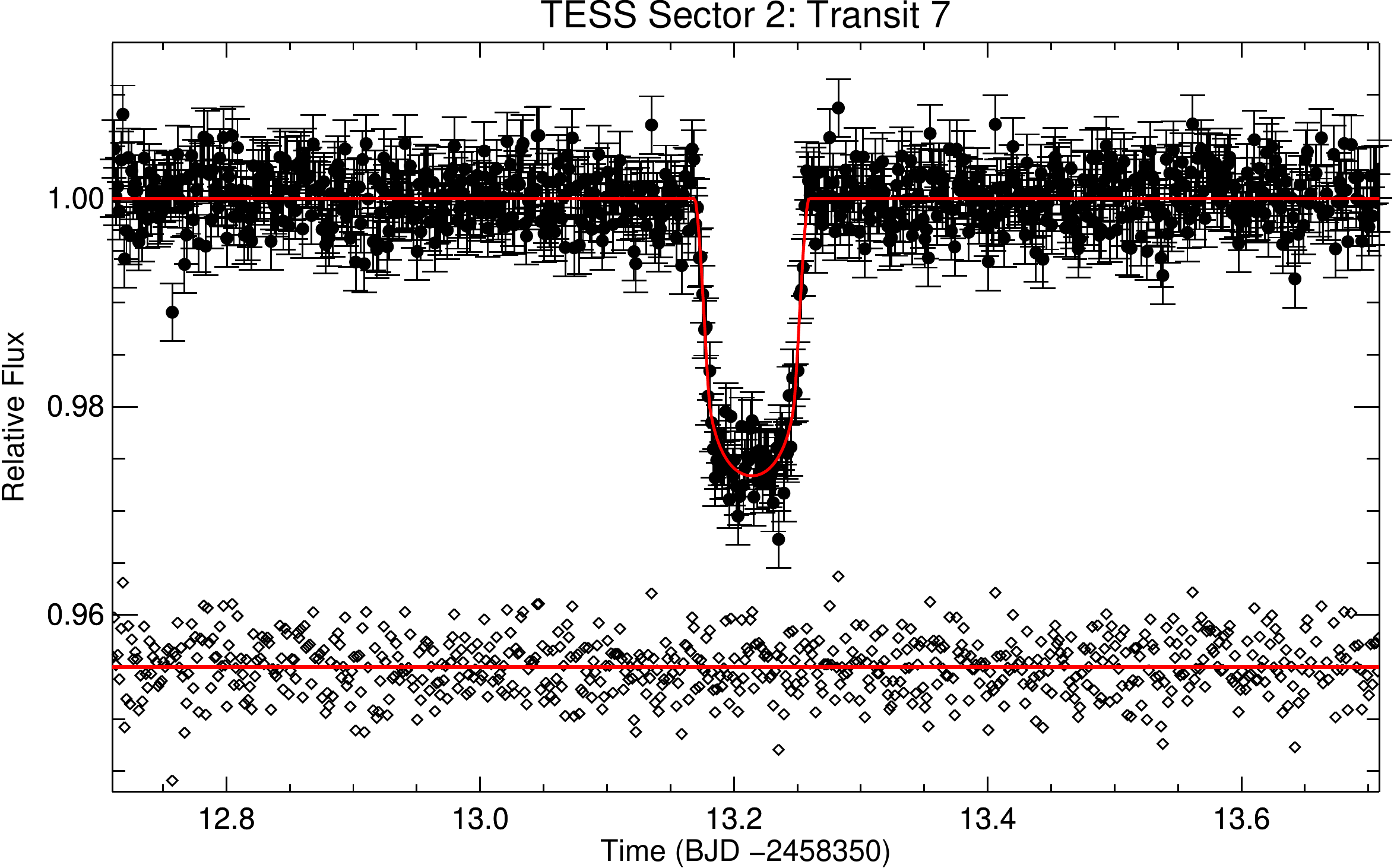} & \includegraphics[width=0.50\textwidth]{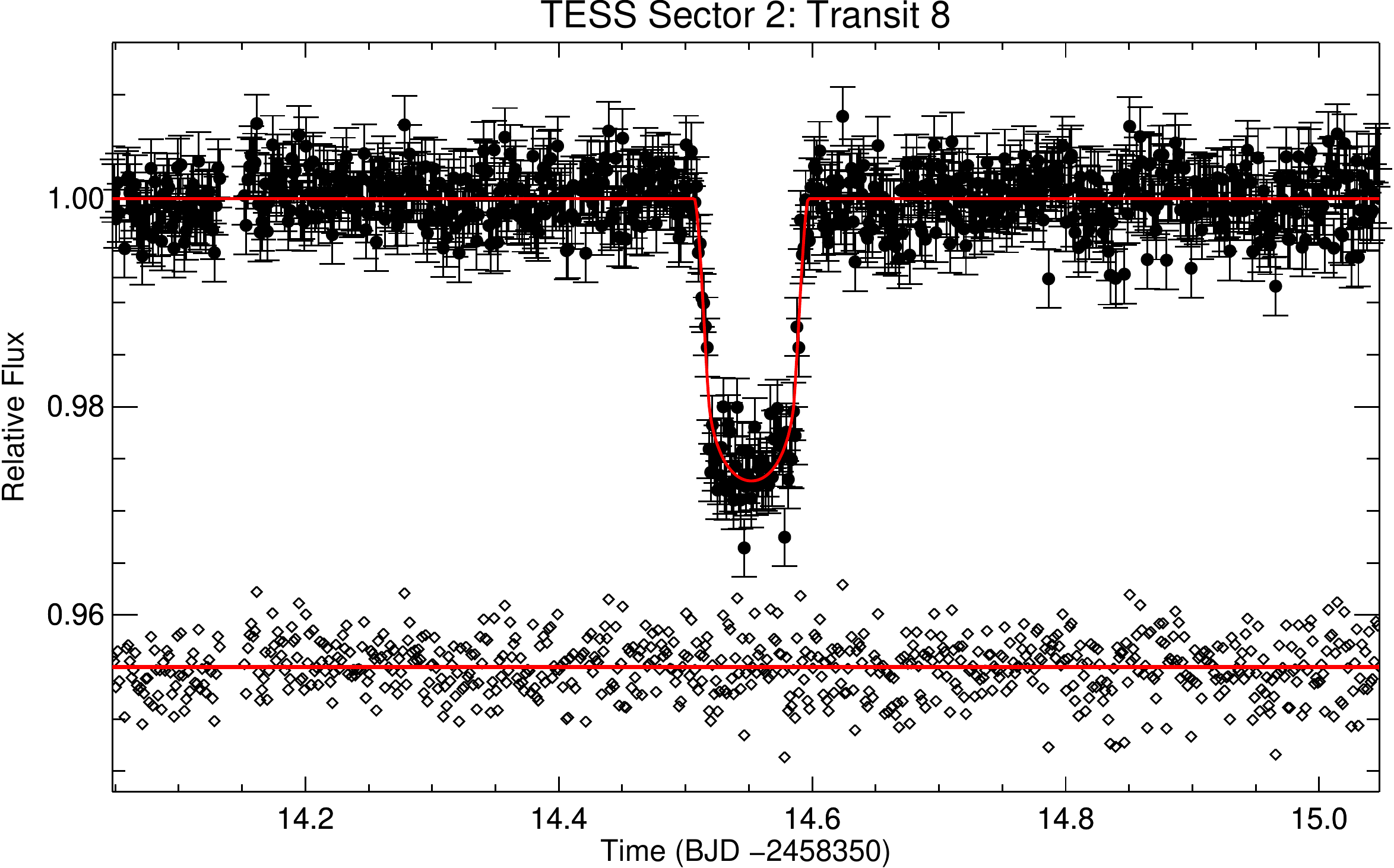}\\
  \end{tabular}
\caption{Individual TESS transit events (1-8) from Sector 2 of WASP-4b. The best-fitting model obtained from the EXOplanet MOdeling Package (\texttt{EXOMOP}) is shown as a solid red line. The residuals (light curve - model) are shown below the light curve.}
\label{fig:ind_transits_sec2_1}
\end{figure*}

\begin{figure*}
\centering
\begin{tabular}{cc}
 \includegraphics[width=0.50\textwidth]{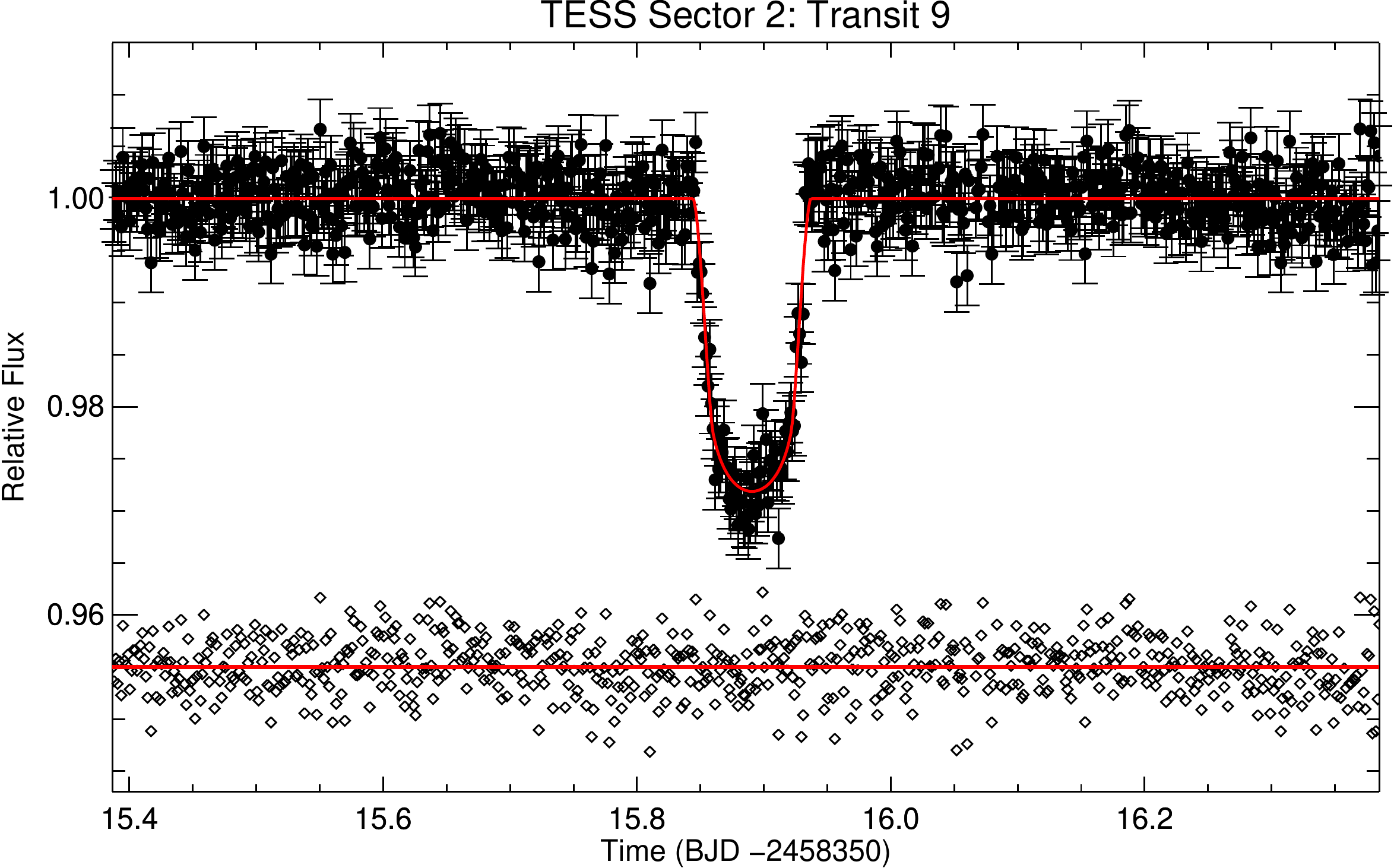} & \includegraphics[width=0.50\textwidth]{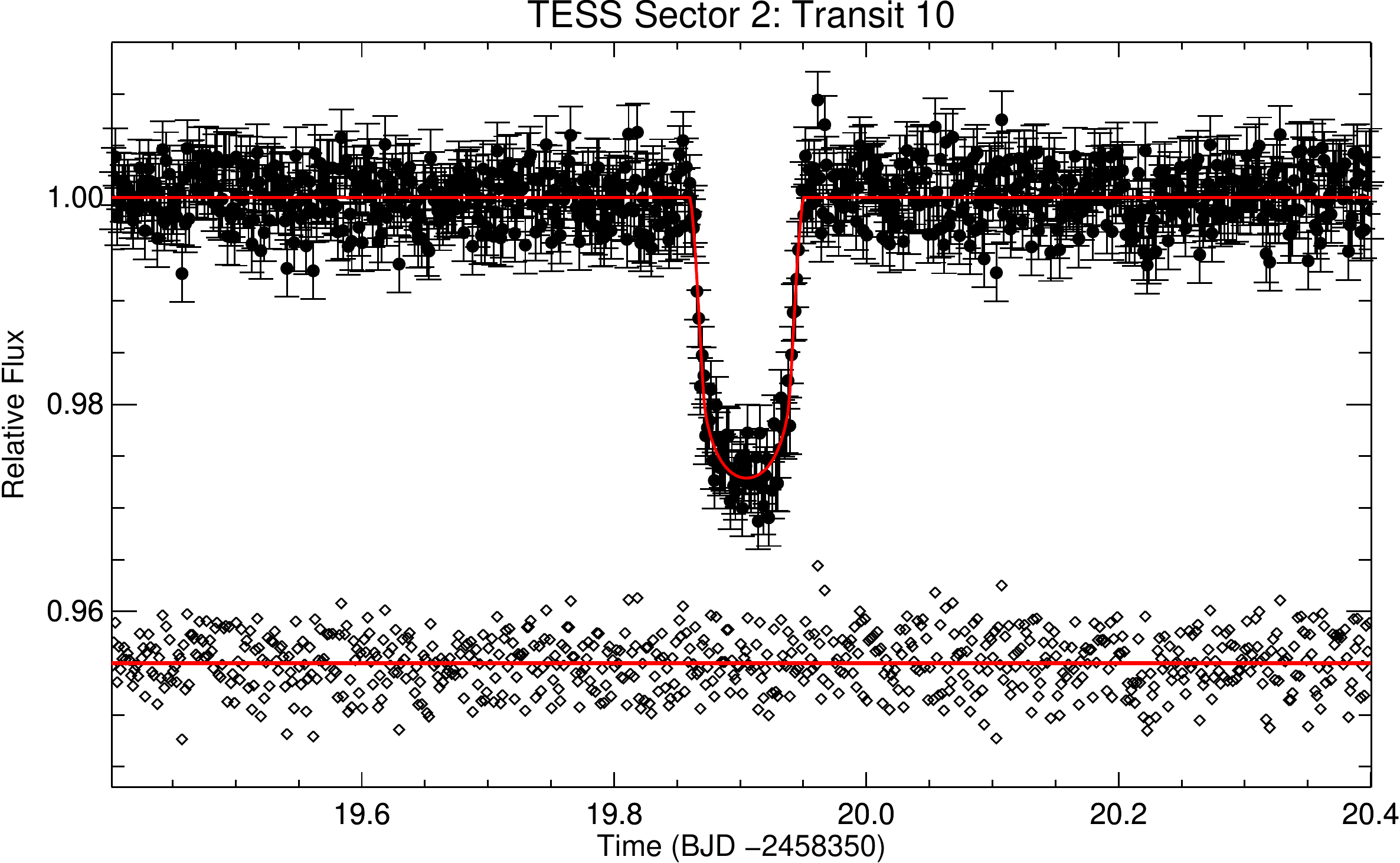}\\
 \includegraphics[width=0.50\textwidth]{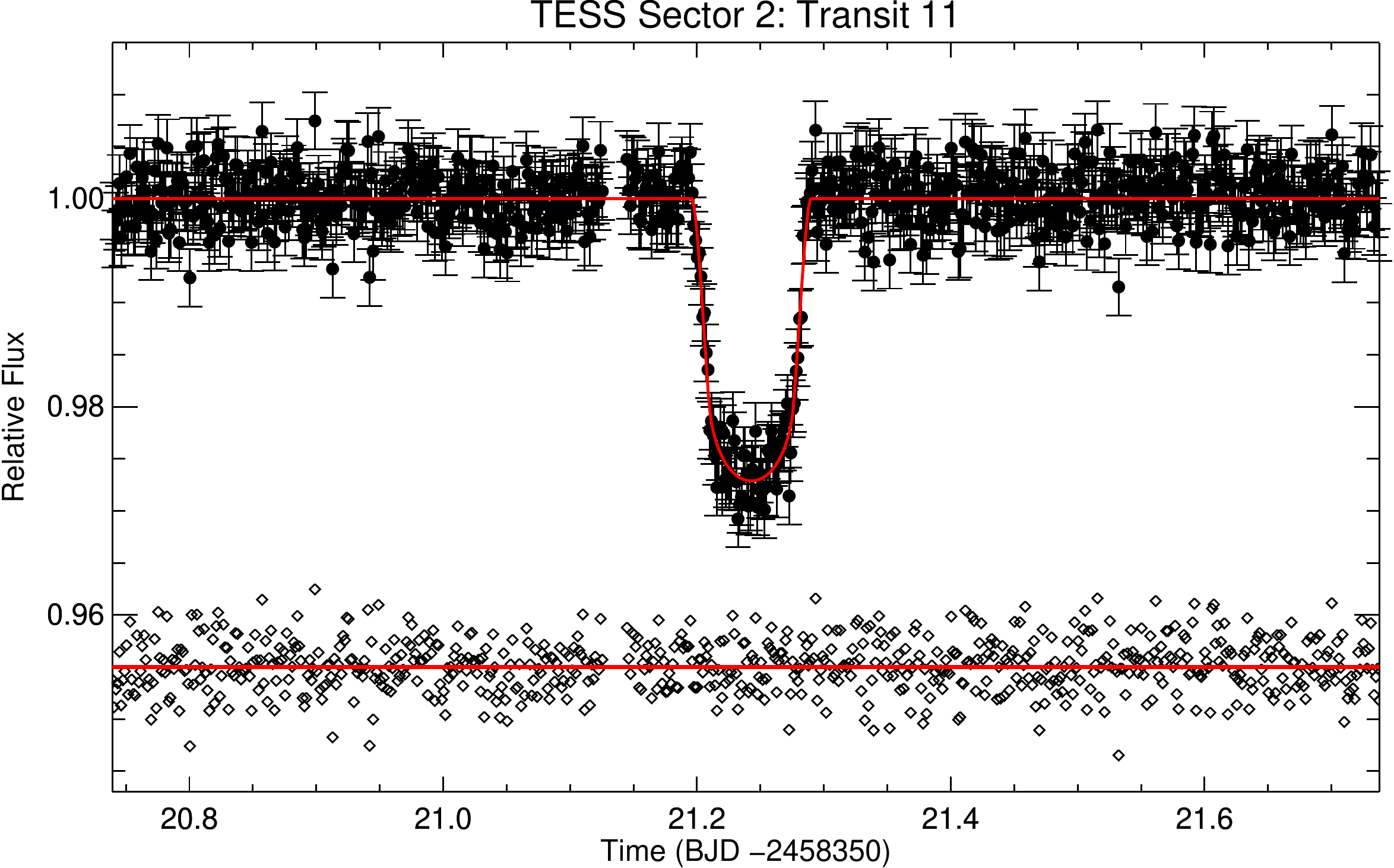} & \includegraphics[width=0.50\textwidth]{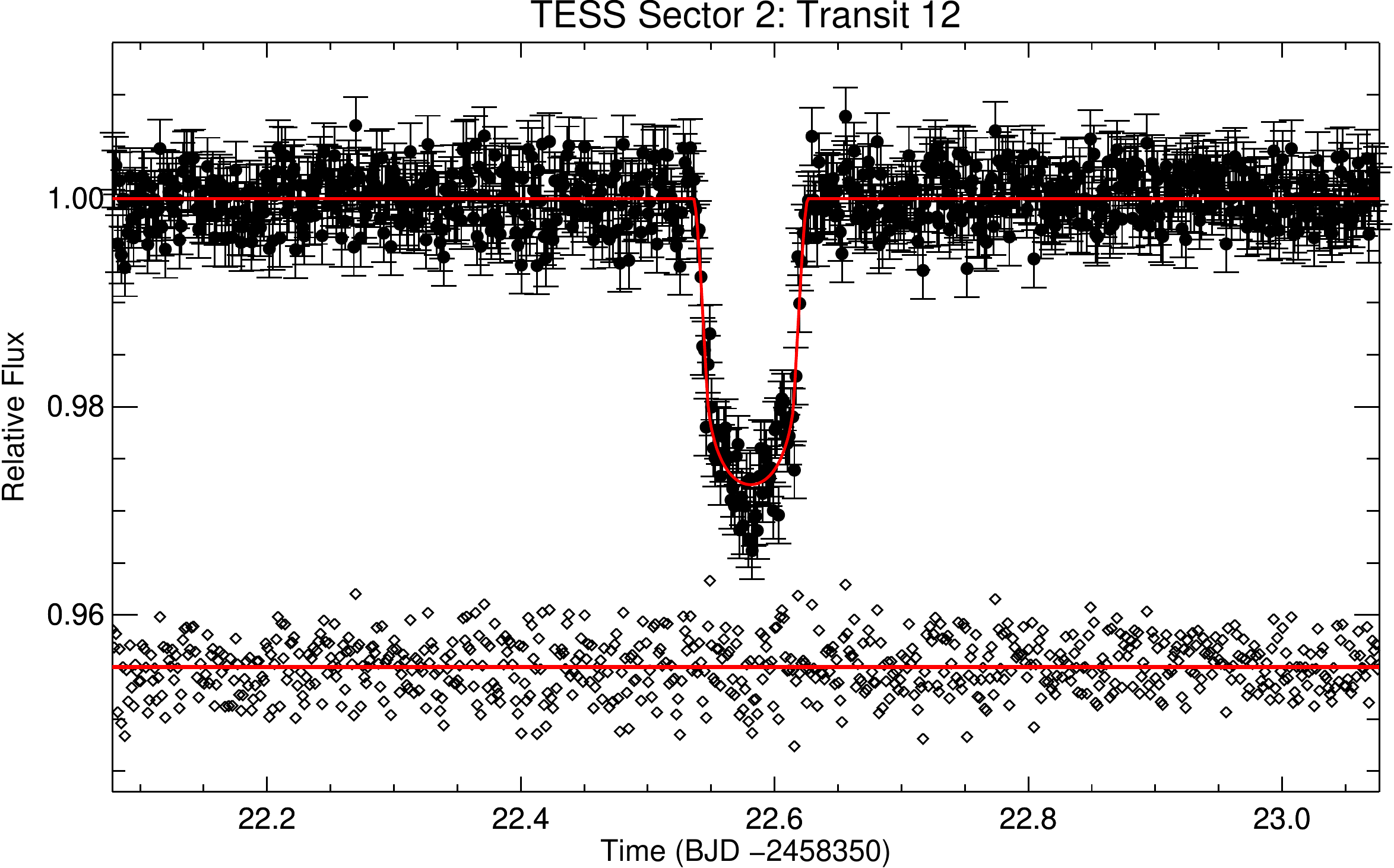}\\
  \includegraphics[width=0.50\textwidth]{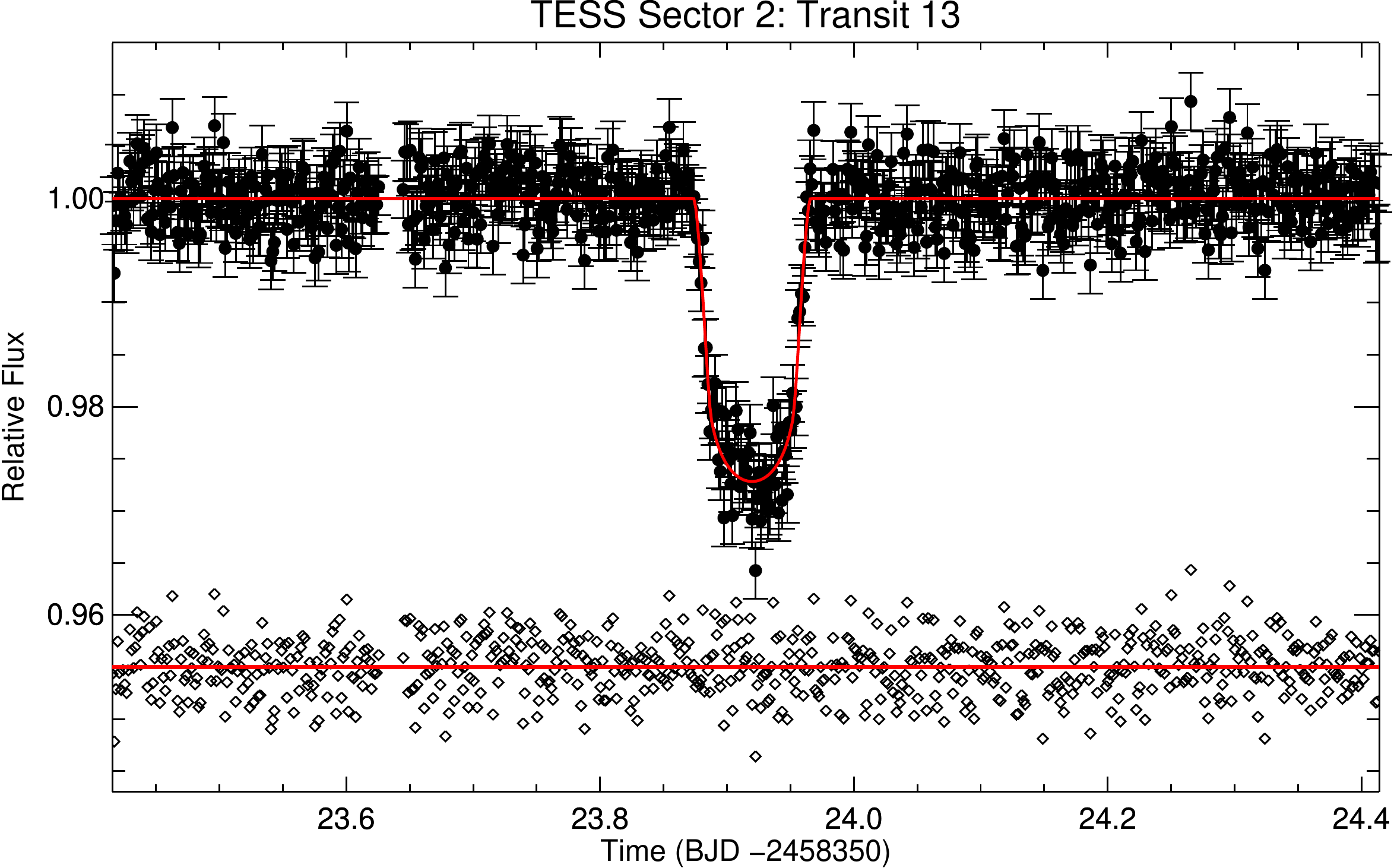} & \includegraphics[width=0.50\textwidth]{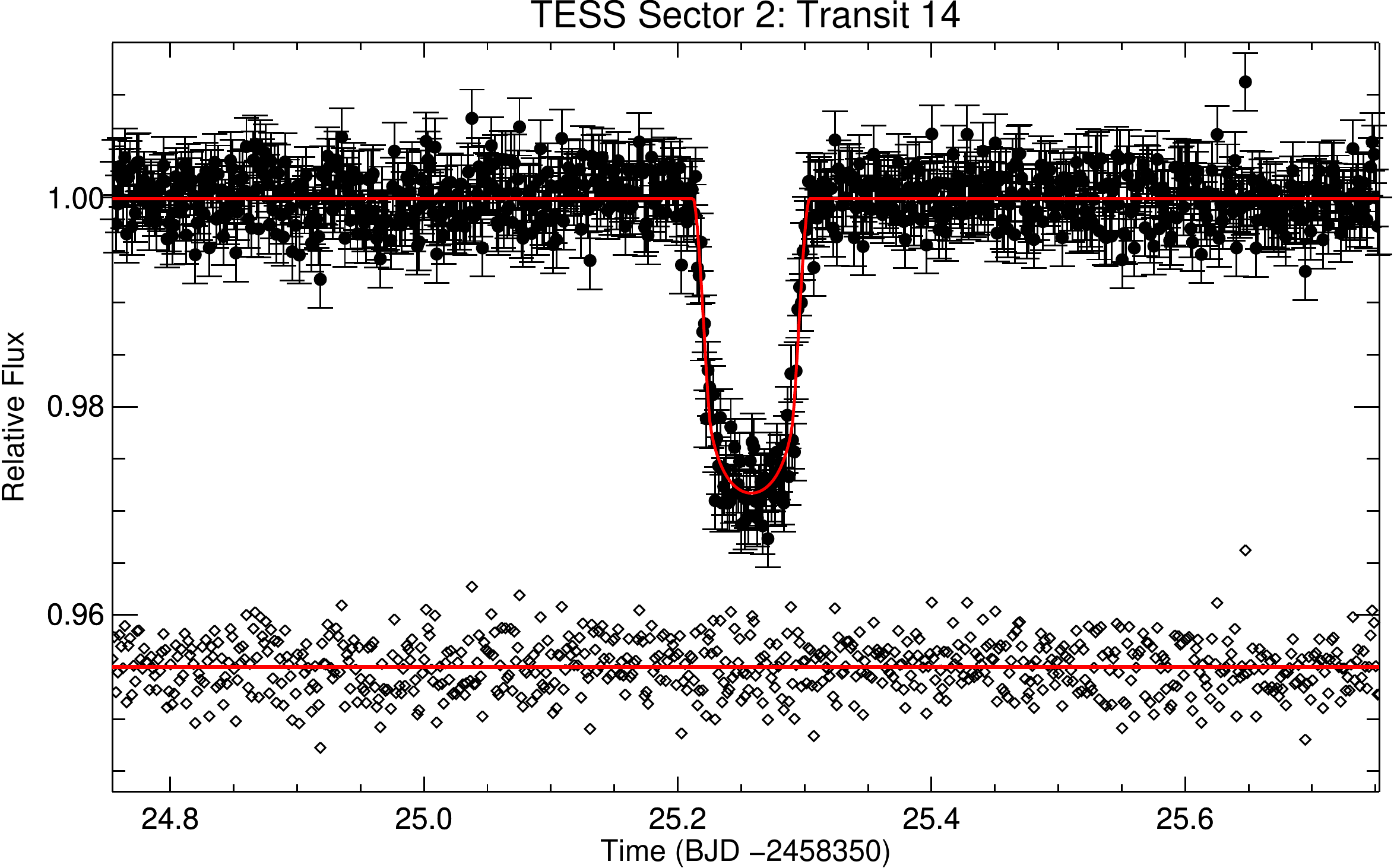}\\
  \includegraphics[width=0.50\textwidth]{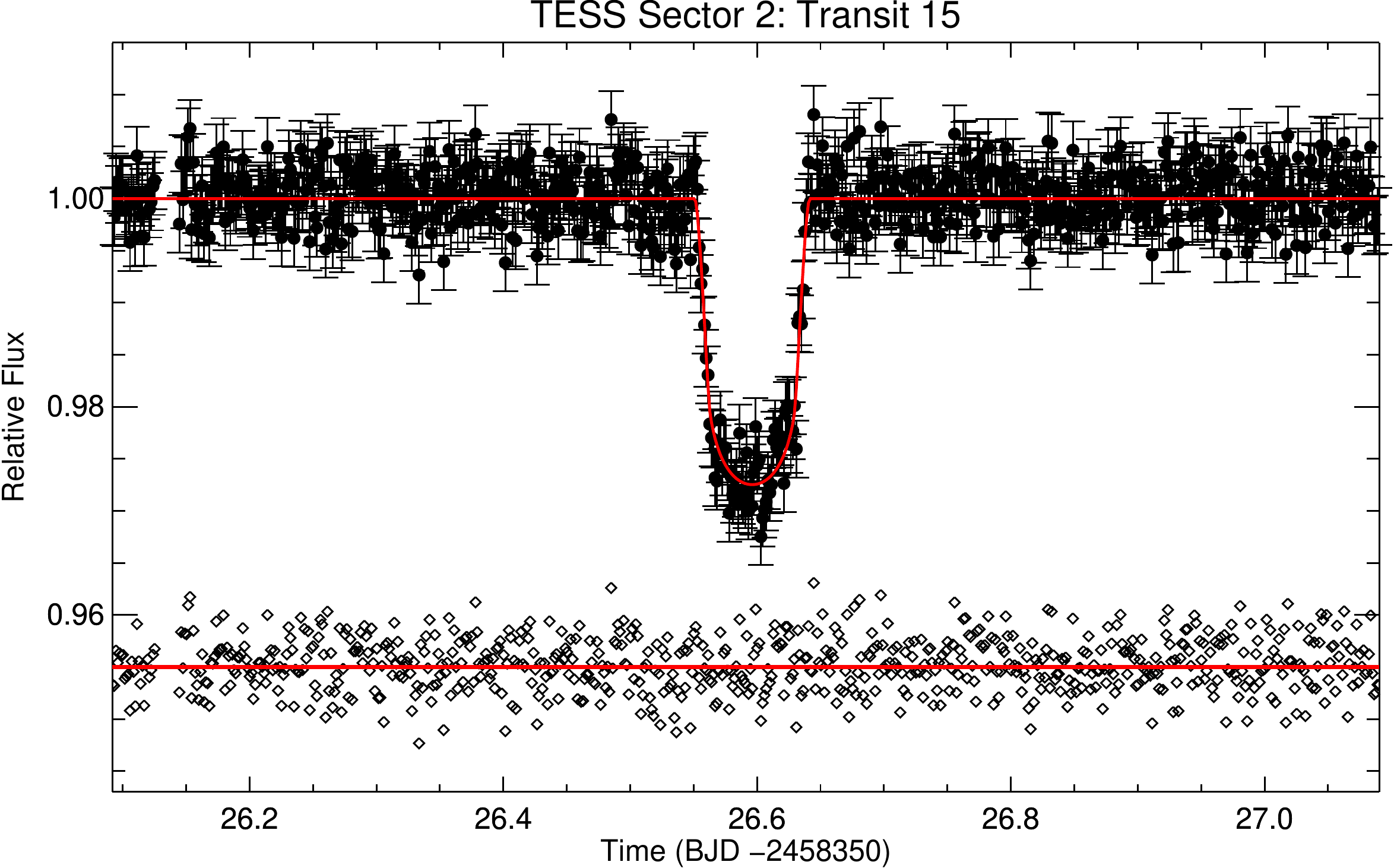} & \includegraphics[width=0.50\textwidth]{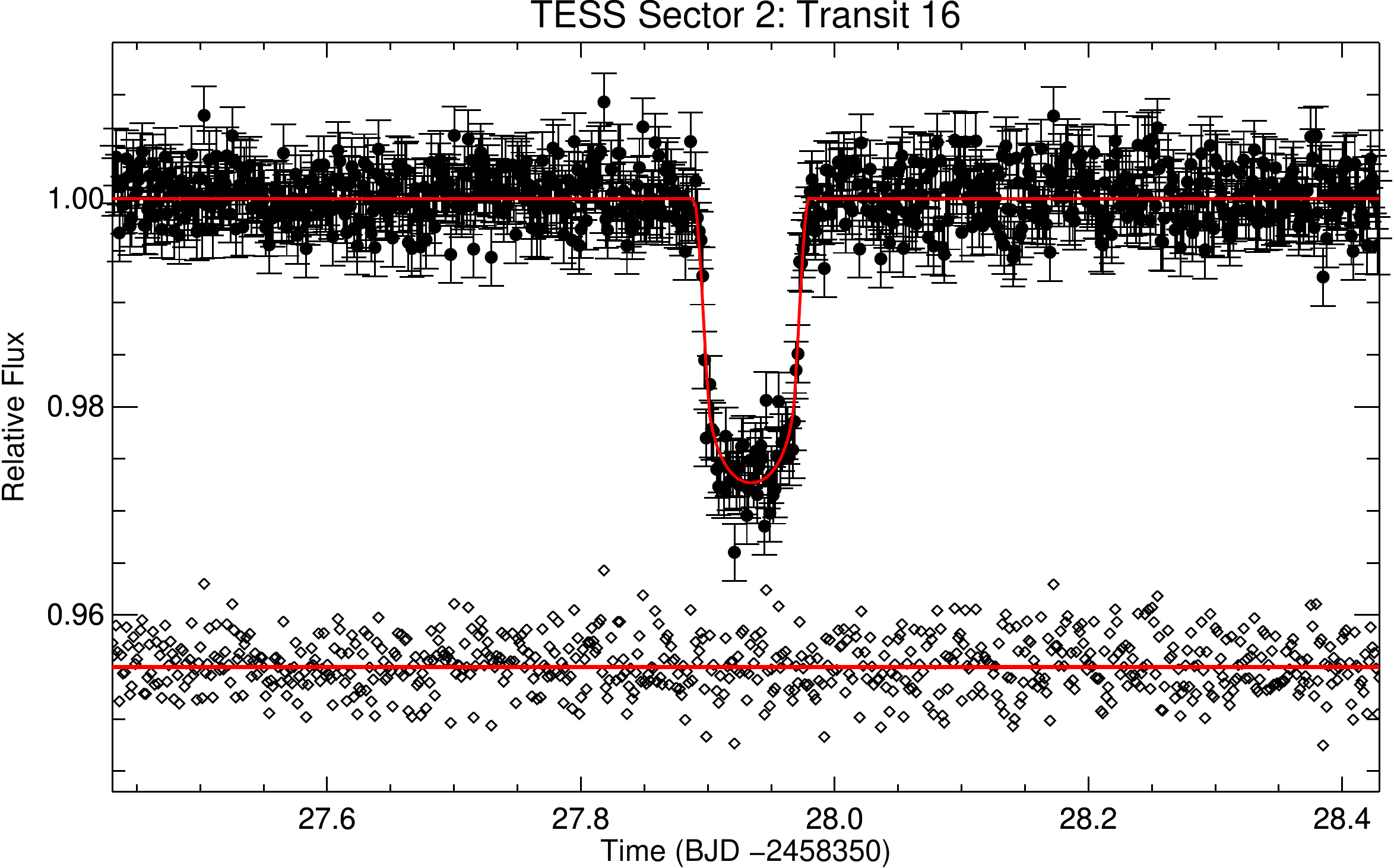}\\
  \end{tabular}
\caption{Individual TESS transit events (8-16) from Sector 20 of WASP-4b. Other comments are the same as Figure \ref{fig:ind_transits_sec2_1}.
}
\label{fig:ind_transits_sec2_2}
\end{figure*}

\begin{figure*}
\centering
\begin{tabular}{cc}
 \includegraphics[width=0.50\textwidth]{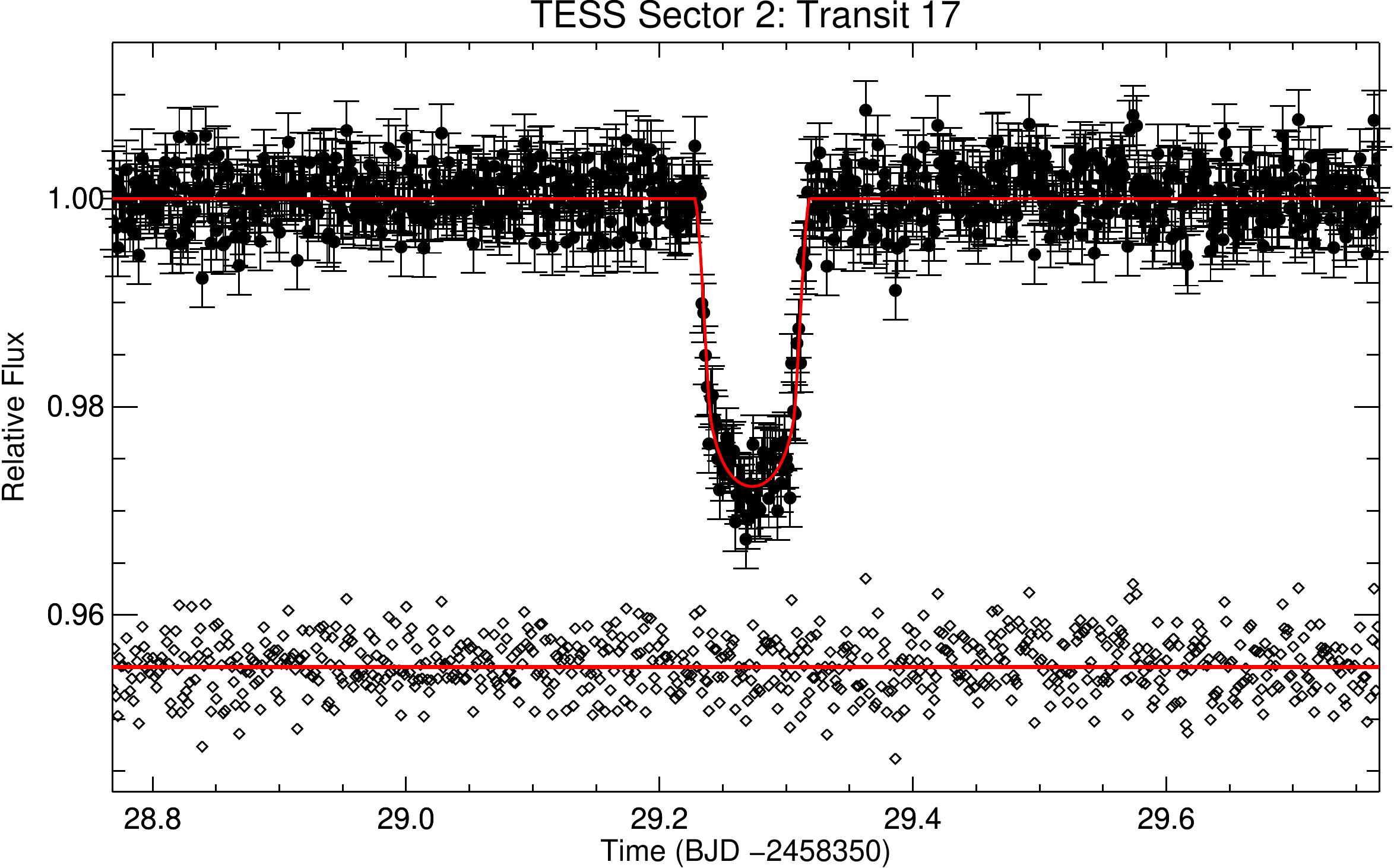} & \includegraphics[width=0.50\textwidth]{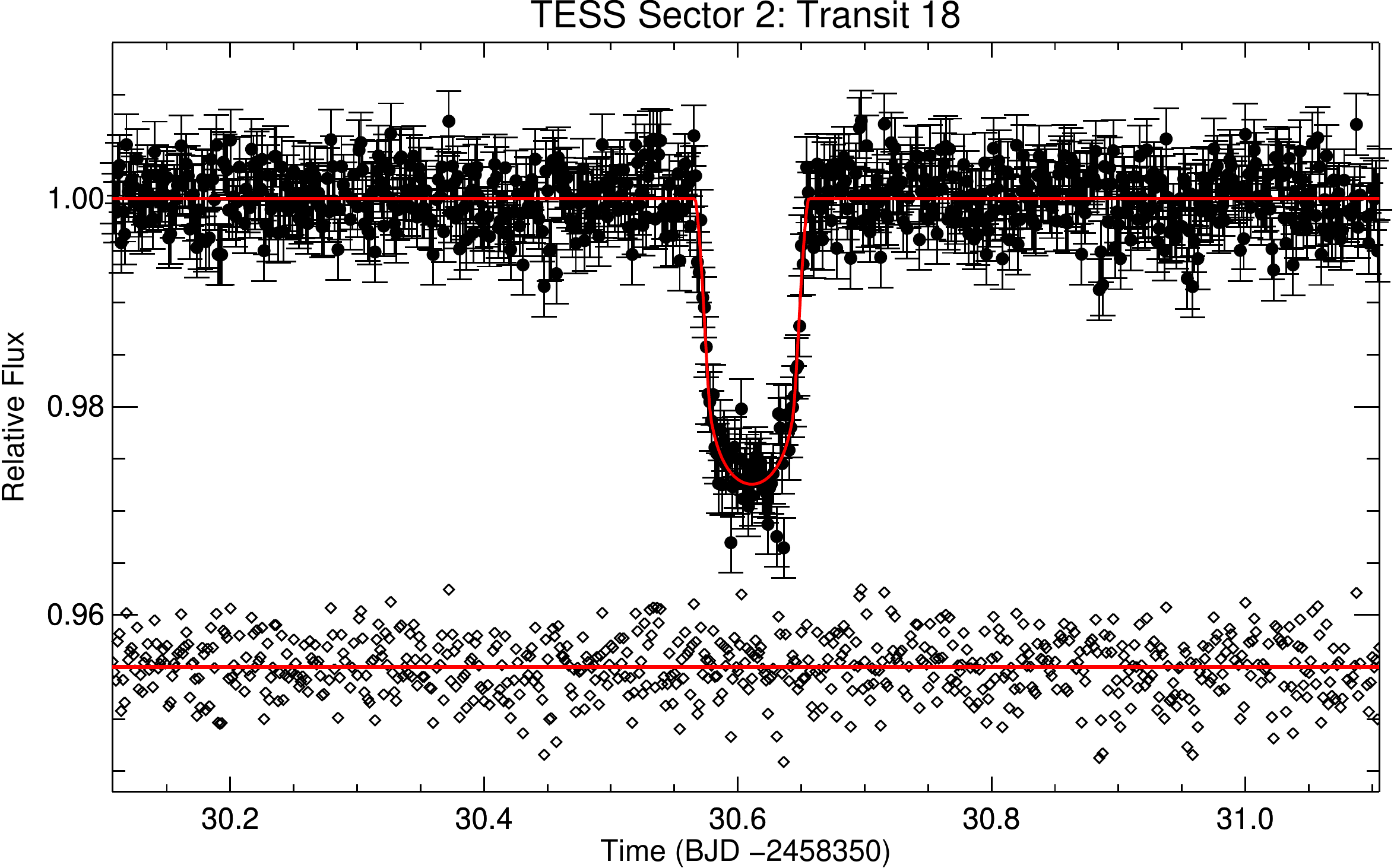}\\
  \end{tabular}
\caption{Individual TESS transit events (17-18) from Sector 2 of WASP-4b. Other comments are the same as Figure \ref{fig:ind_transits_sec2_1}.
}
\label{fig:ind_transits_sec2_3}
\end{figure*}

\begin{figure*}
\centering
\begin{tabular}{cc}
  \includegraphics[width=0.50\textwidth]{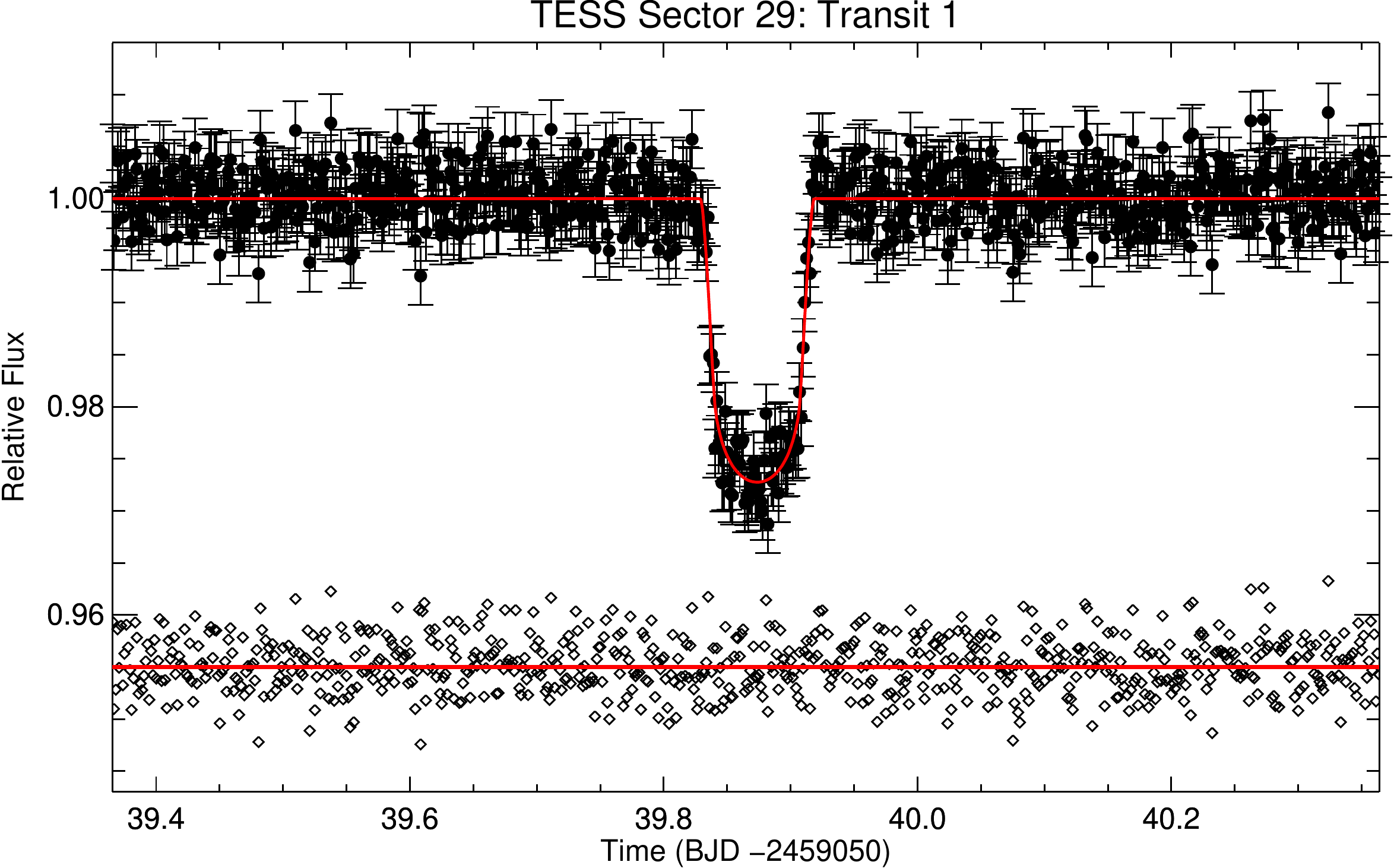} & \includegraphics[width=0.50\textwidth]{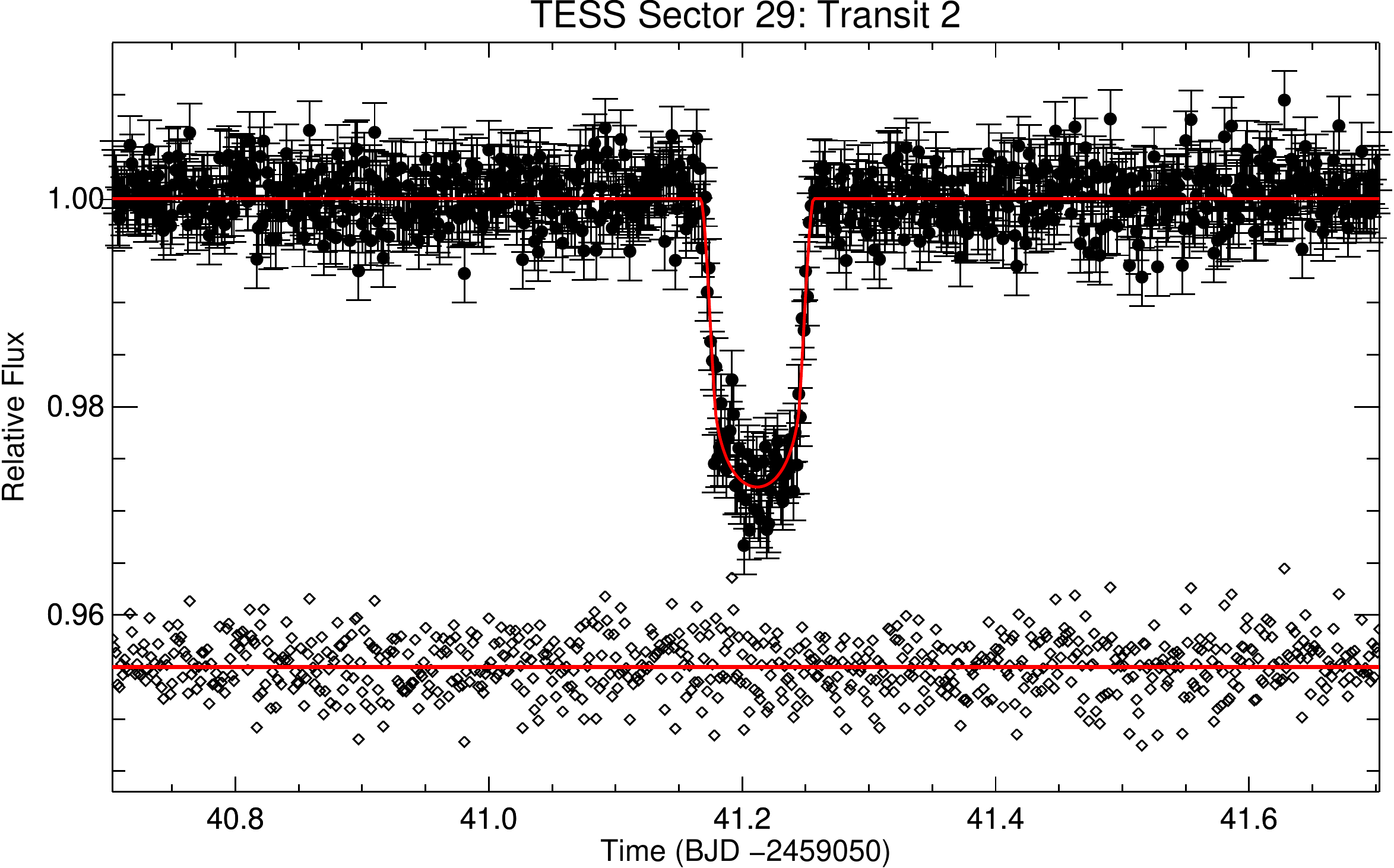} \\ \includegraphics[width=0.50\textwidth]{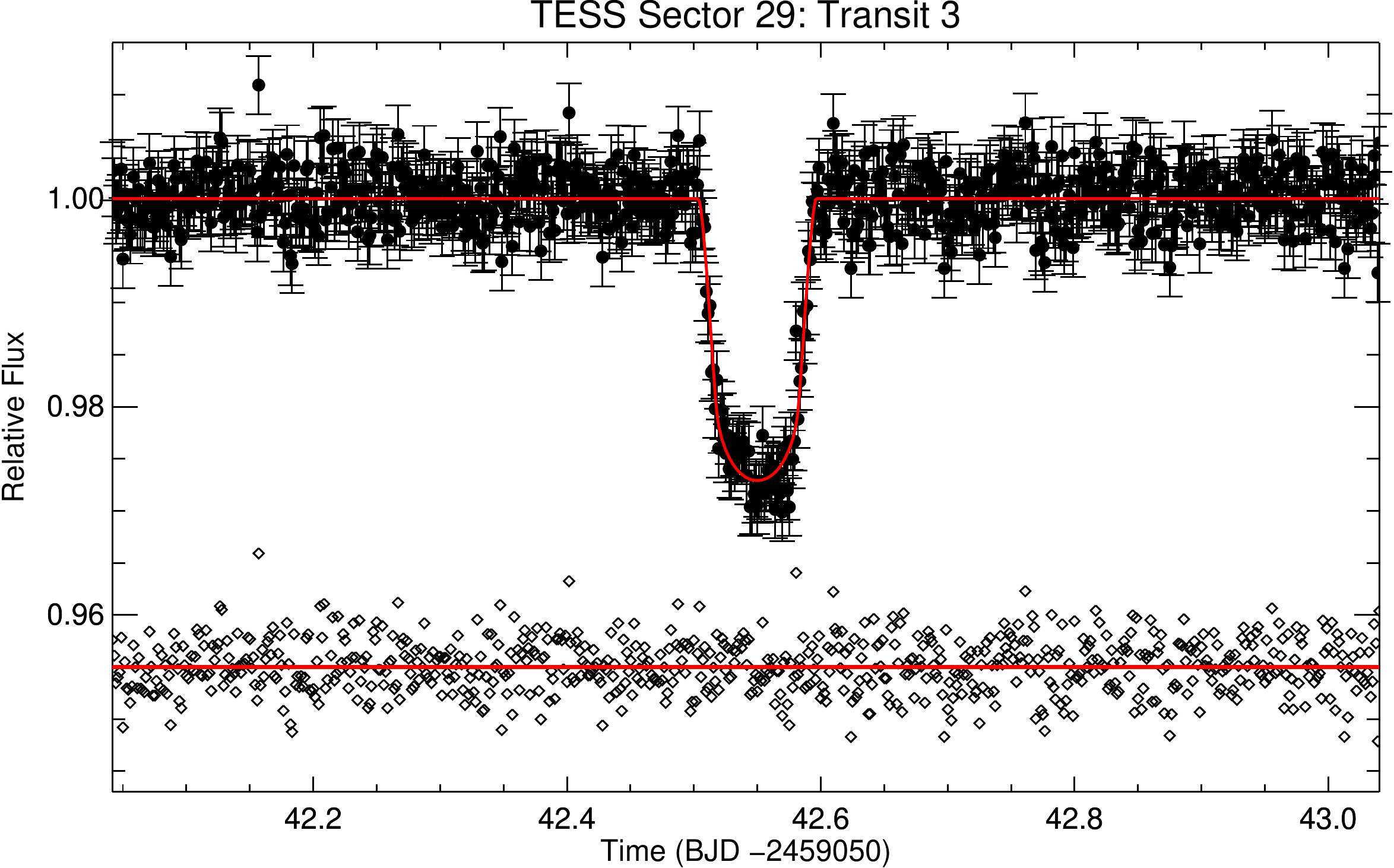} &
  \includegraphics[width=0.50\textwidth]{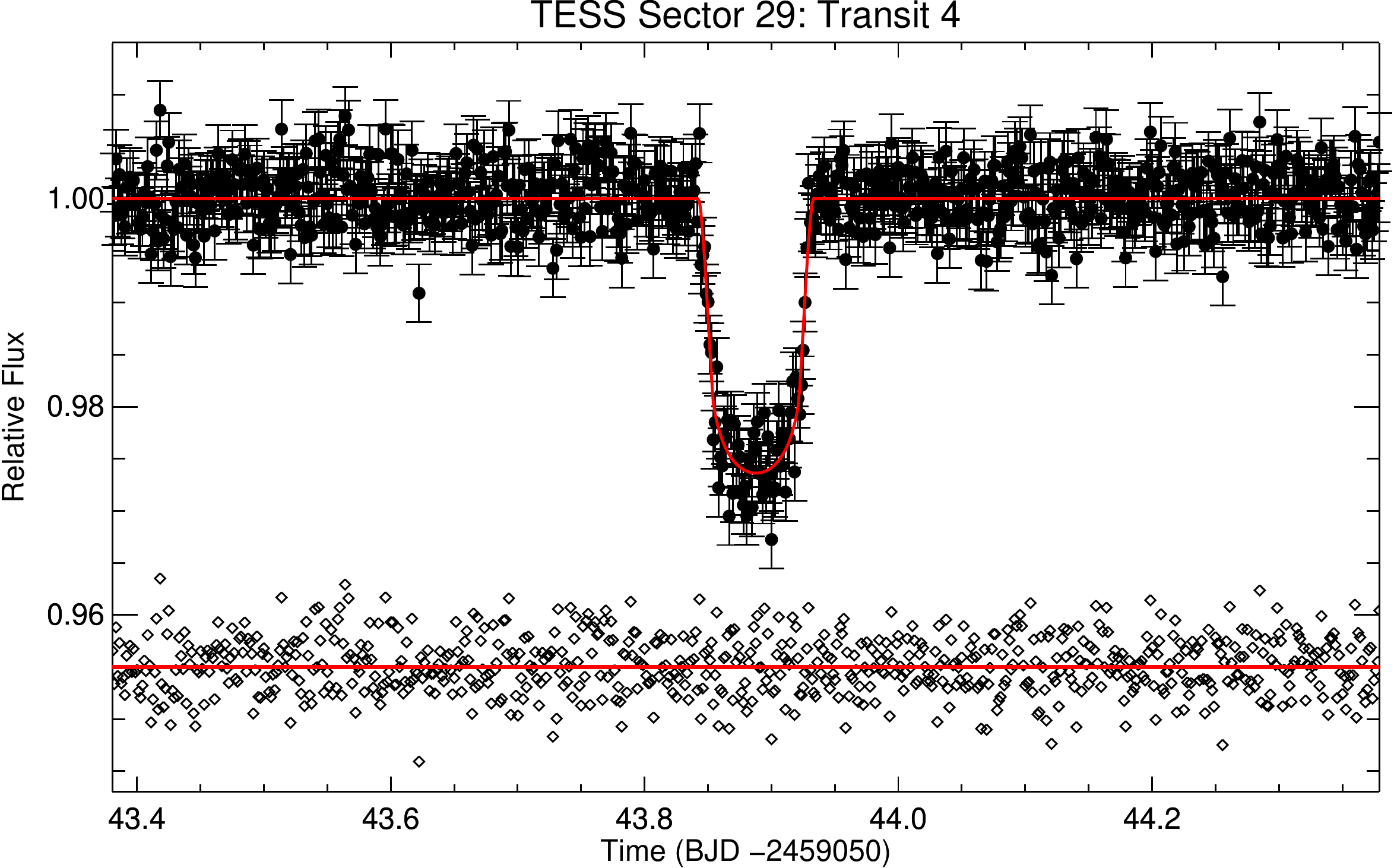} \\
    \includegraphics[width=0.50\textwidth]{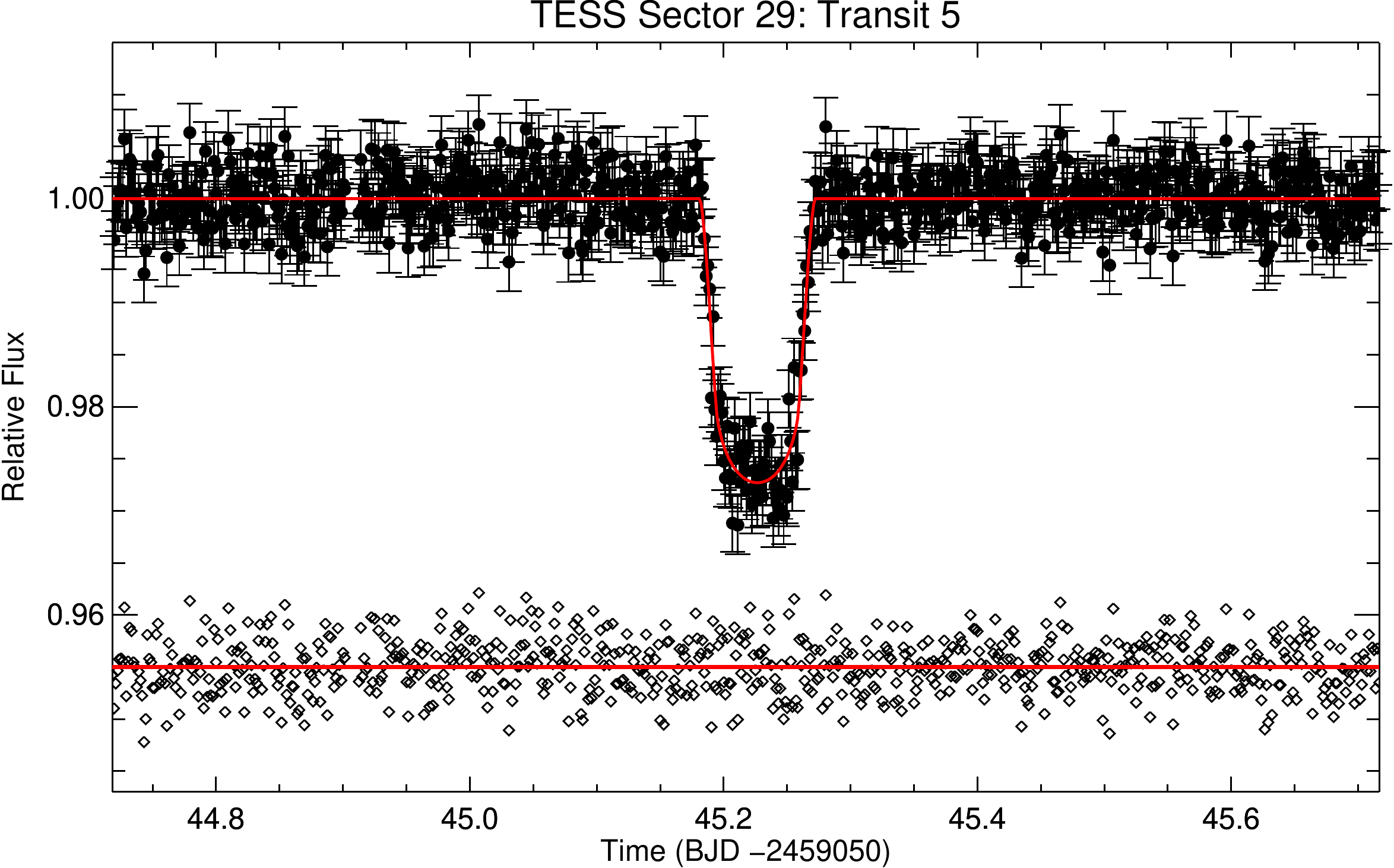} & \includegraphics[width=0.50\textwidth]{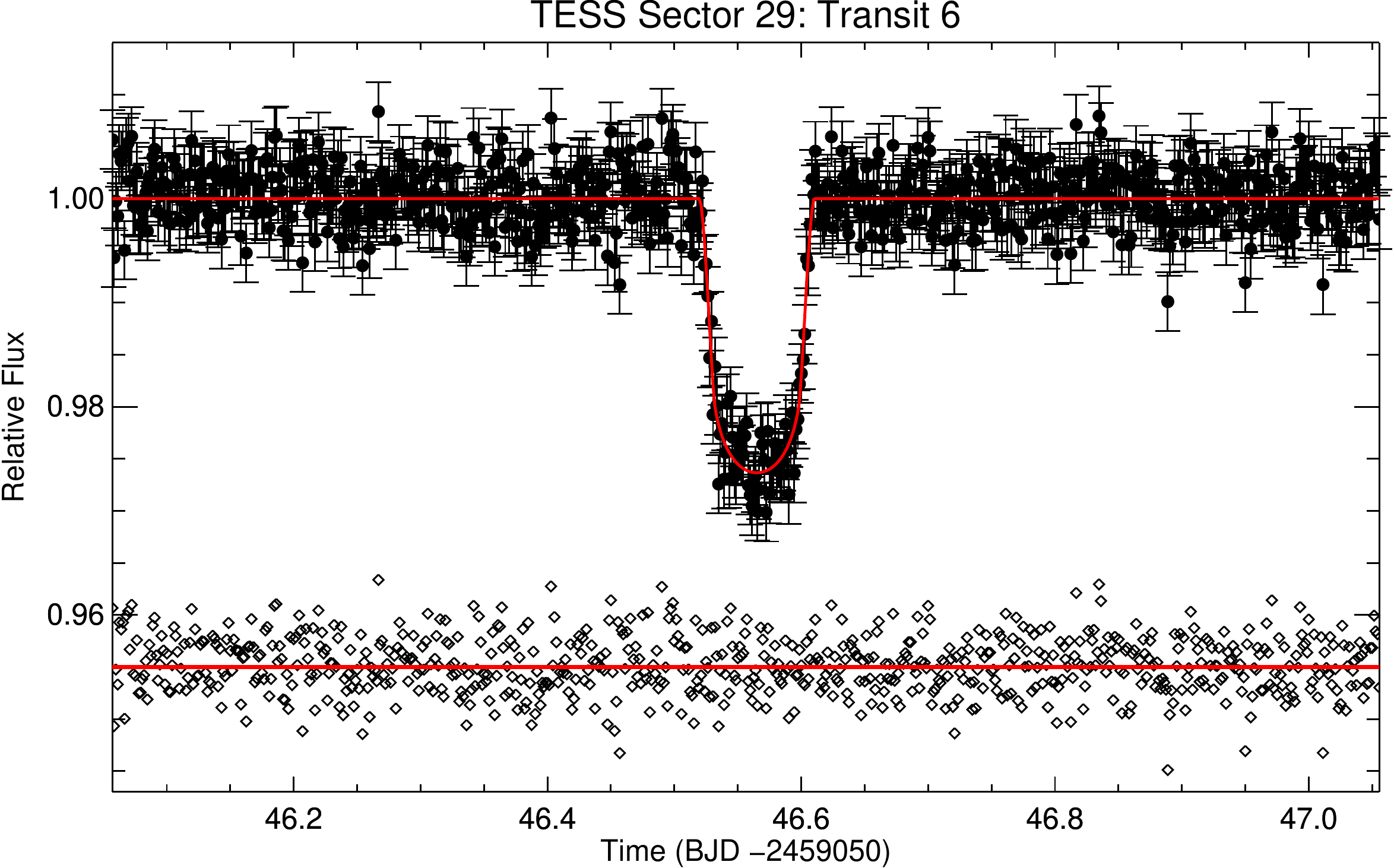} \\ \includegraphics[width=0.50\textwidth]{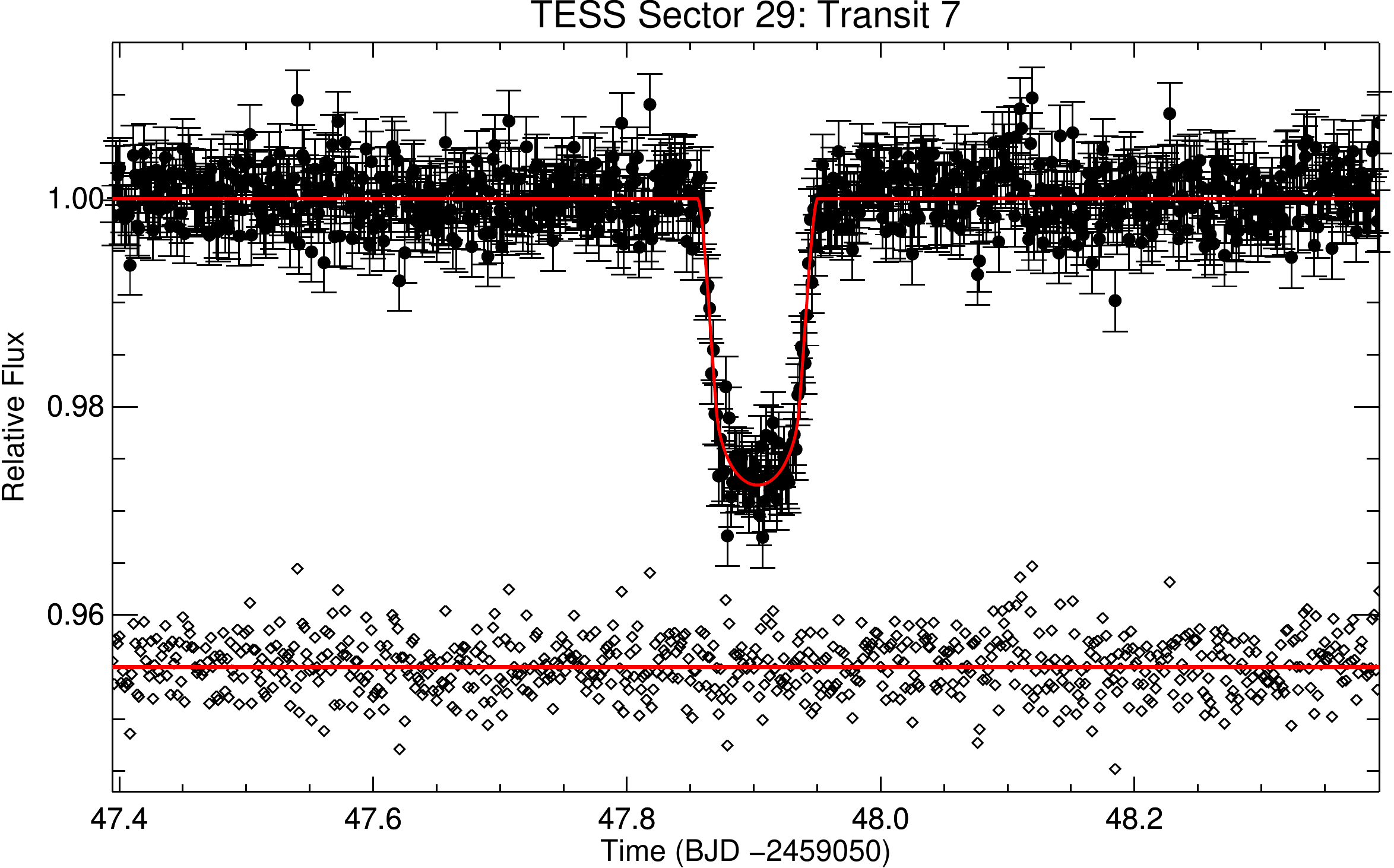} &
  \includegraphics[width=0.50\textwidth]{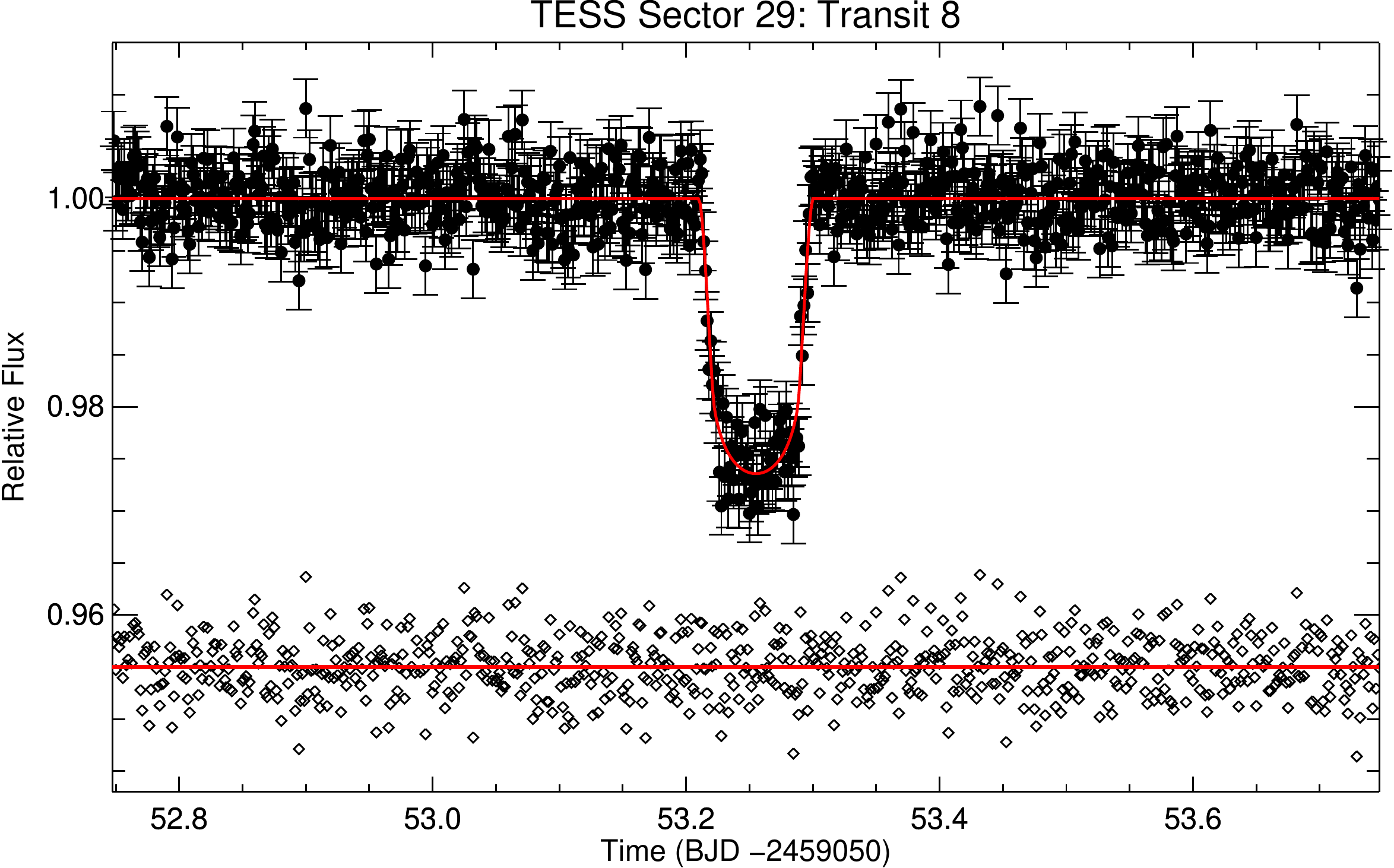} \\
  \end{tabular}
\caption{Individual TESS transit events (1-8) from Sector 29 of WASP-4b. Other comments are the same as Figure \ref{fig:ind_transits_sec2_1}.}
\label{fig:ind_transits_sec29_1}
\end{figure*}

\begin{figure*}
\centering
\begin{tabular}{cc}
  \includegraphics[width=0.50\textwidth]{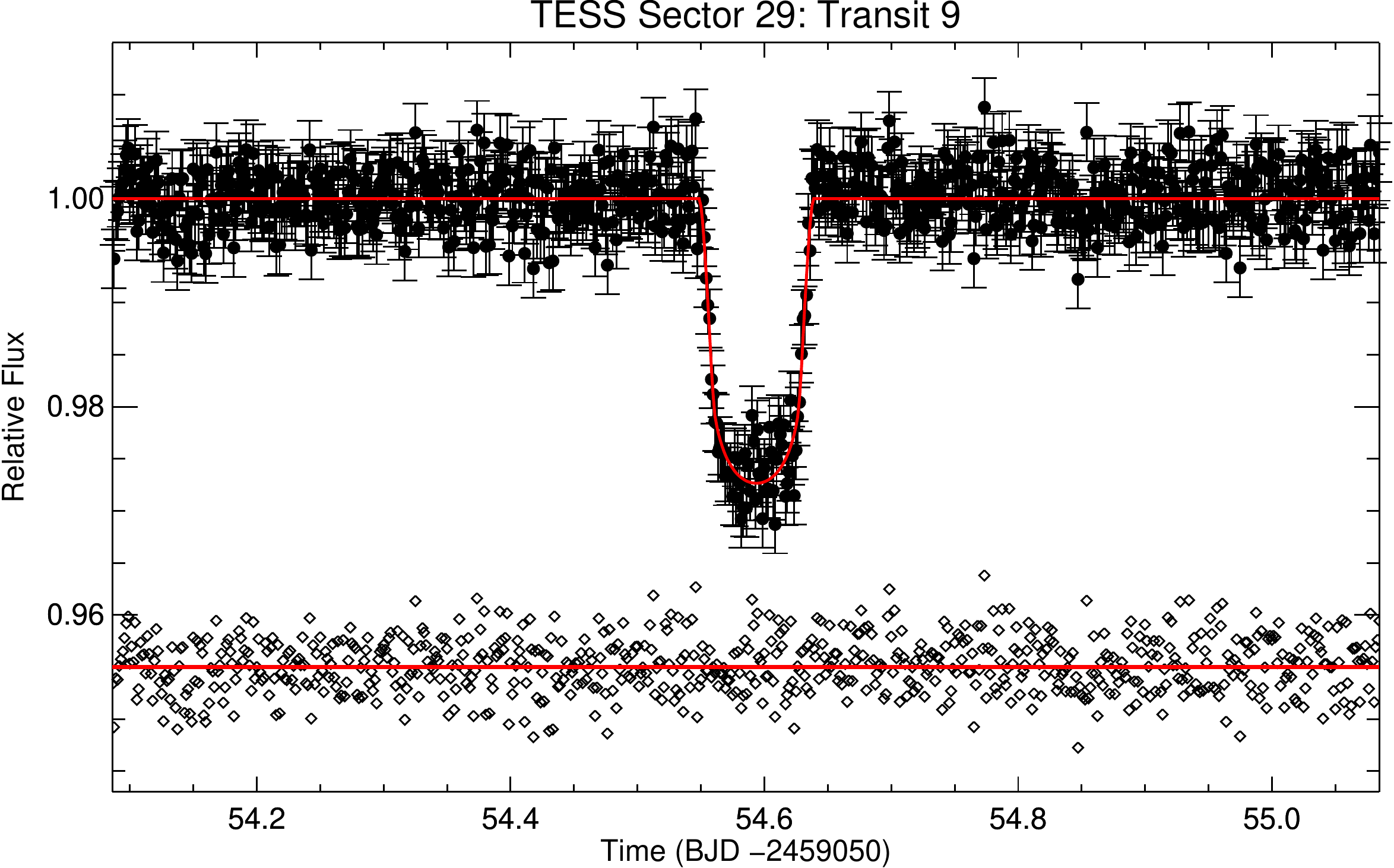} & \includegraphics[width=0.50\textwidth]{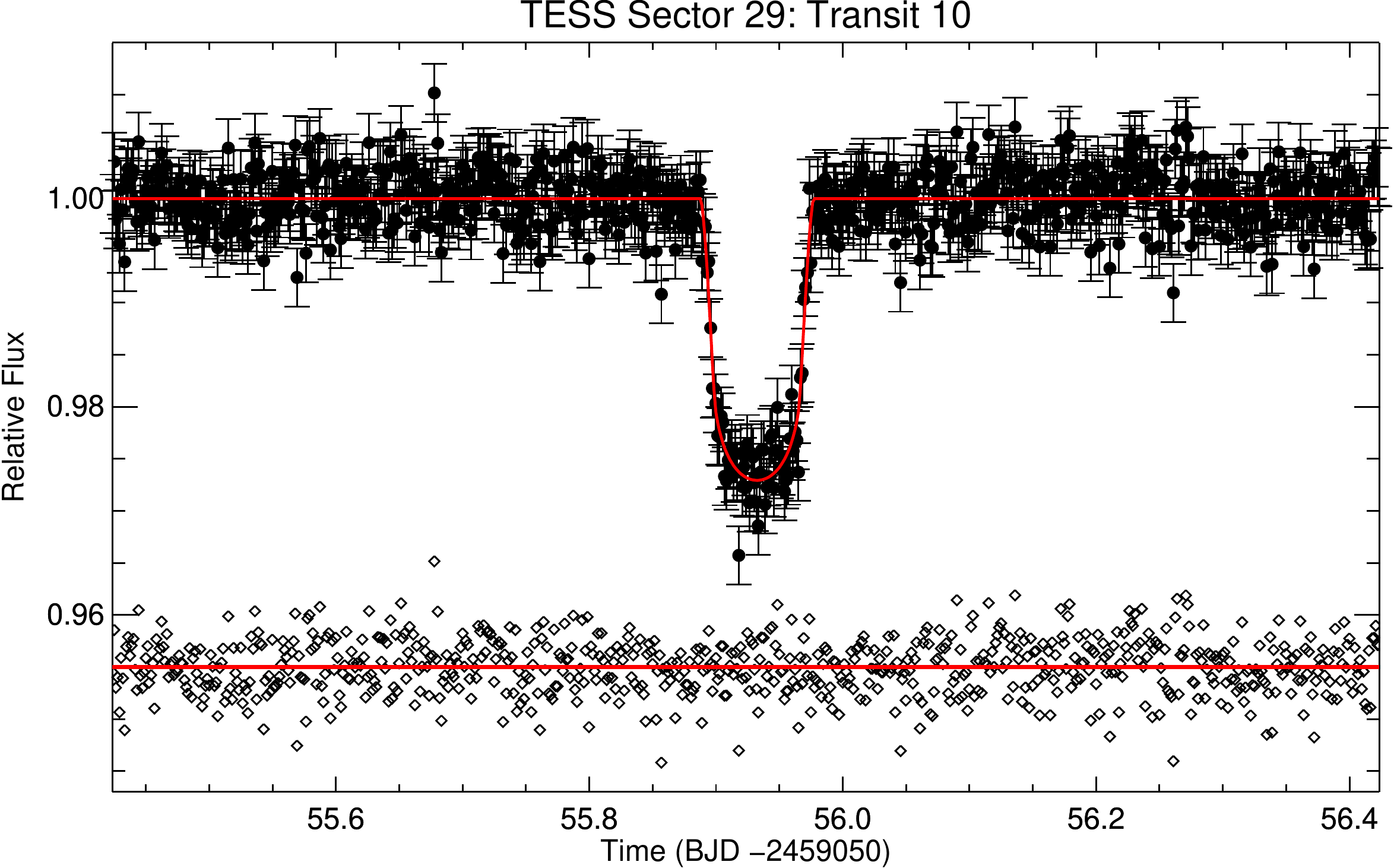} \\ \includegraphics[width=0.50\textwidth]{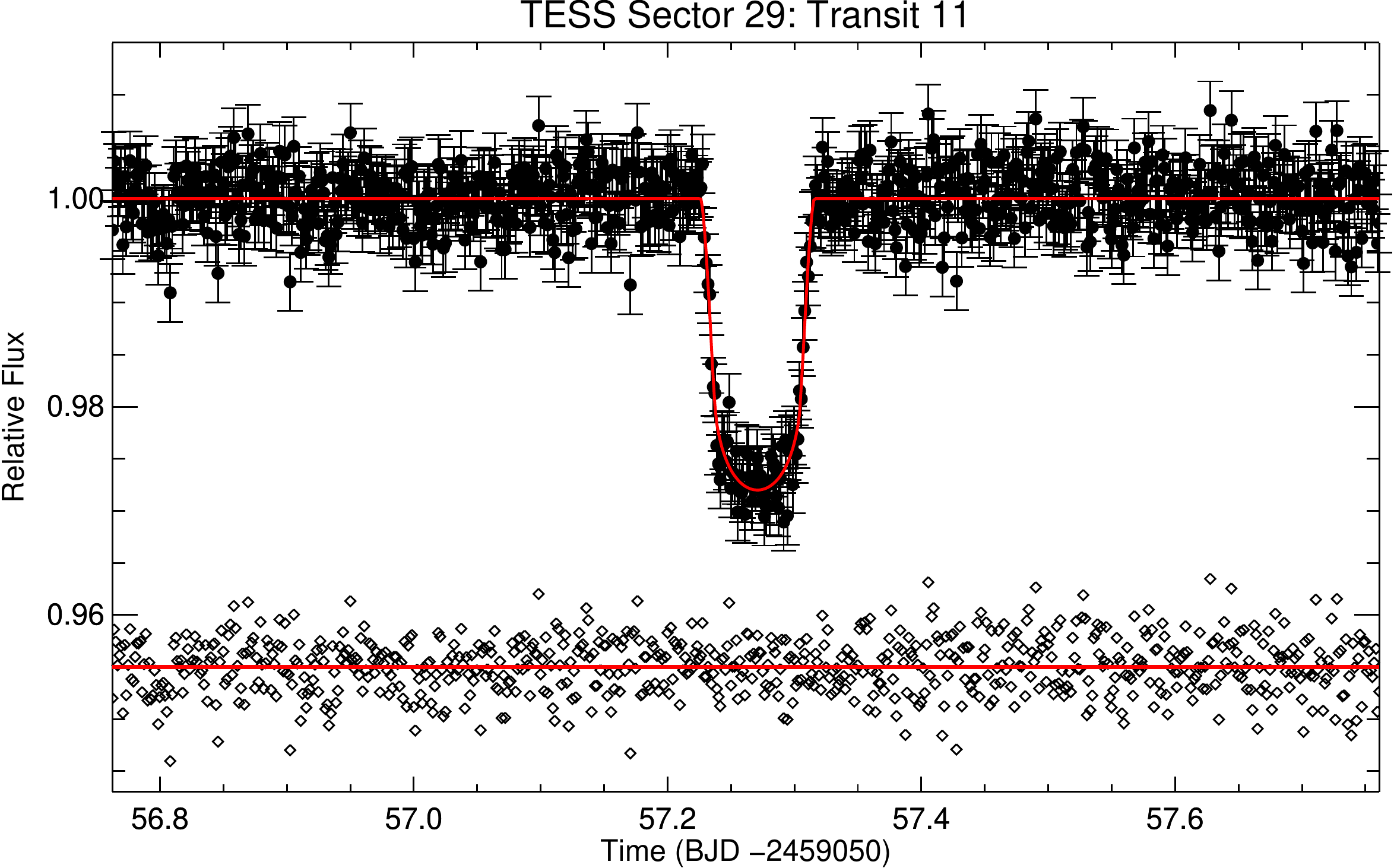} &
  \includegraphics[width=0.50\textwidth]{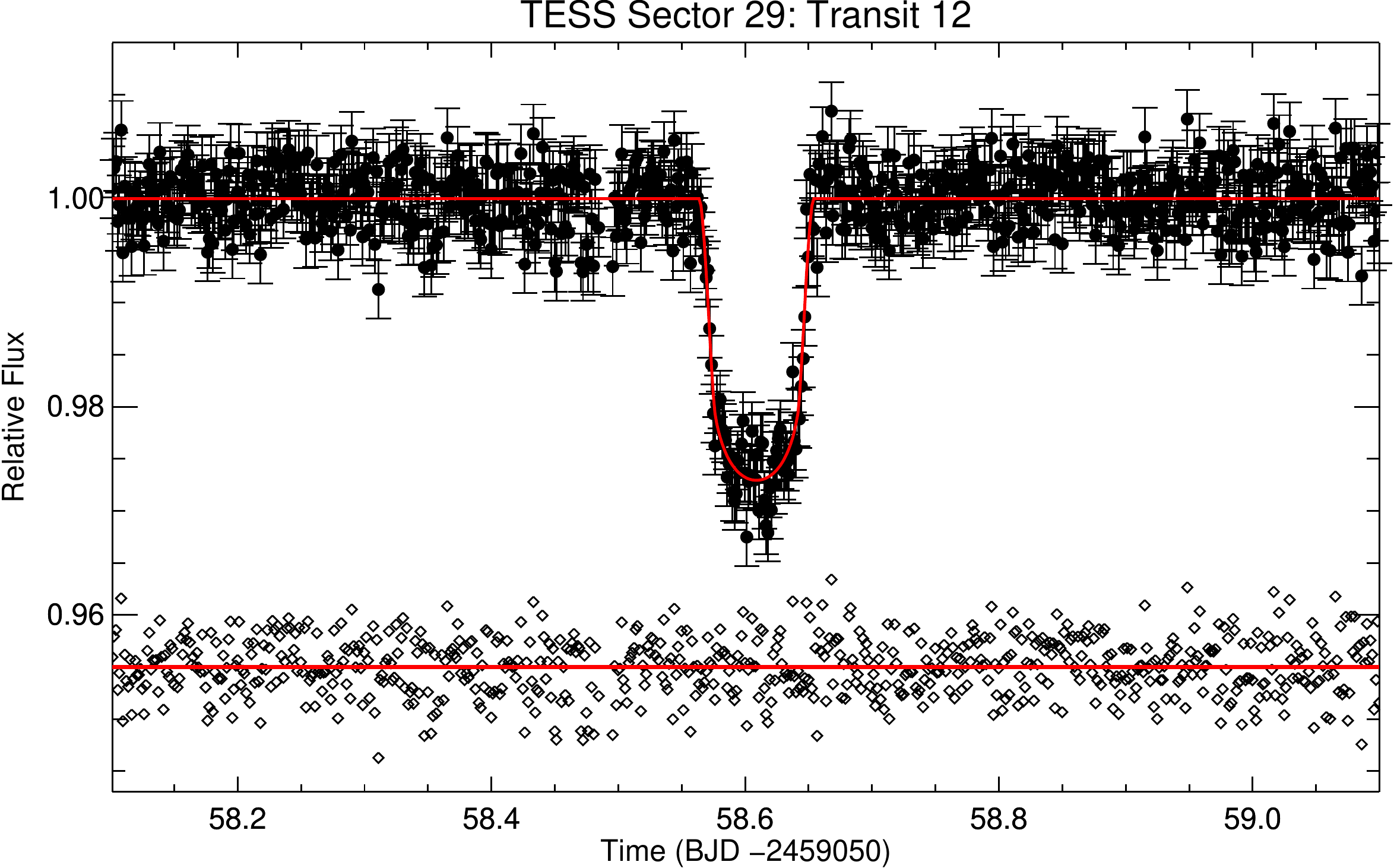} \\
    \includegraphics[width=0.50\textwidth]{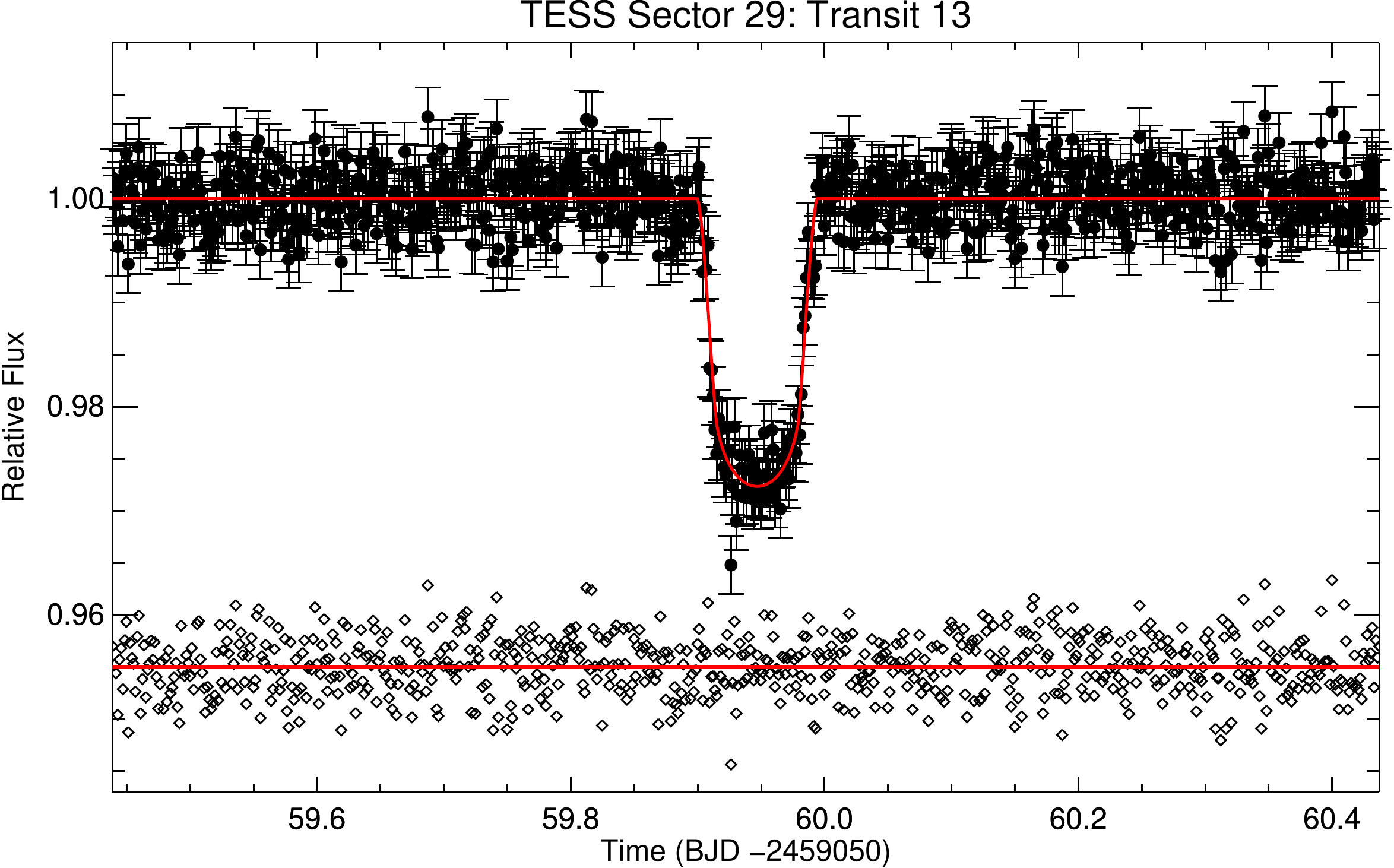} & \includegraphics[width=0.50\textwidth]{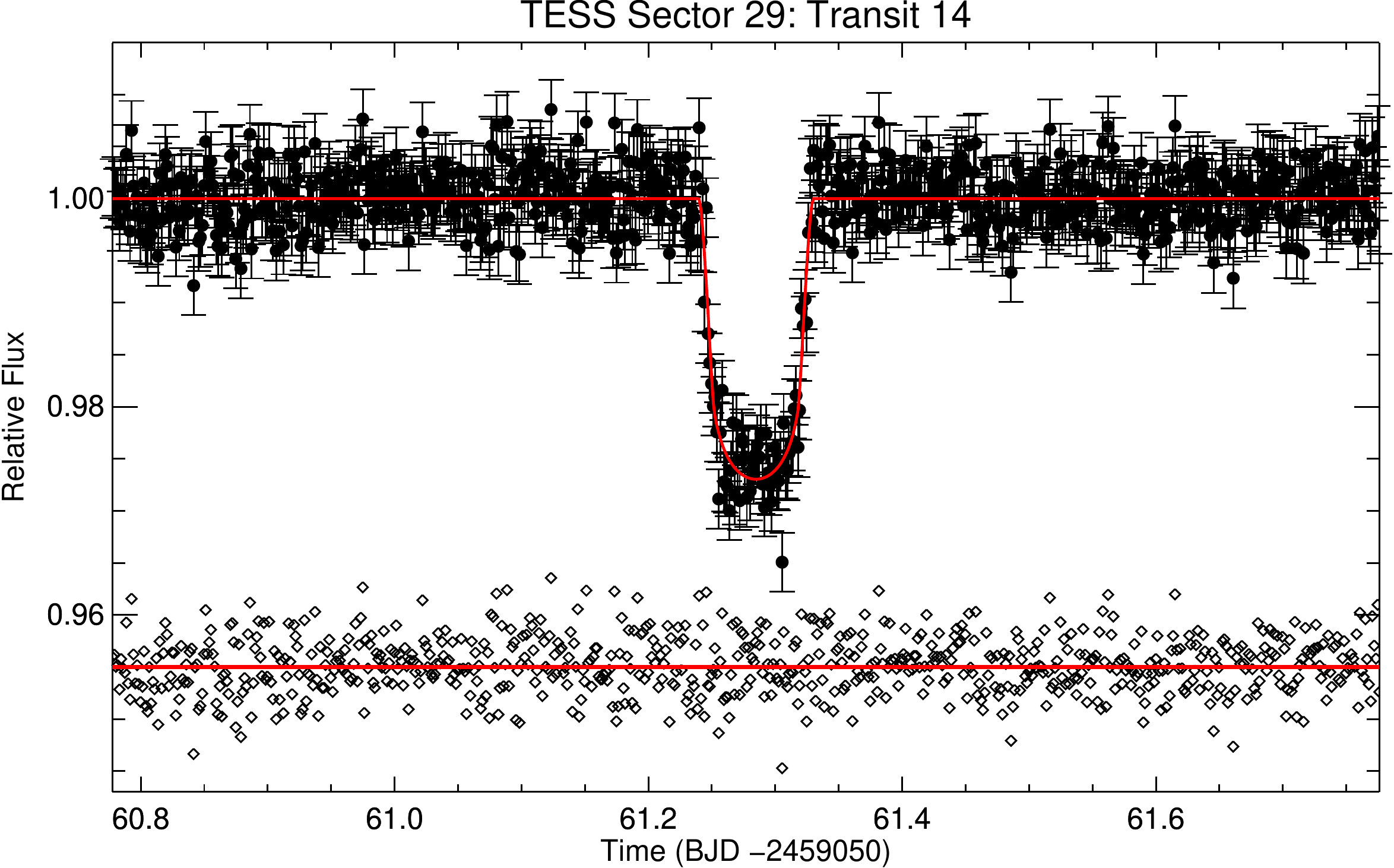} %\includegraphics[width=0.50\textwidth]{WASP-4b_Best-Fit_Sector2829_29.pdf} 
  \end{tabular}
\caption{Individual TESS transit events (9-14) from Sector 29 of WASP-4b. Other comments are the same as Figure \ref{fig:ind_transits_sec2_1}.}
\label{fig:ind_transits_sec29_2}
\end{figure*}

The parameters for each transit fit can be found in Tables \ref{tb:lighcurve_model_TESS}-\ref{tb:lighcurve_model_TESS3}. The light curves and \texttt{EXOMOP} model fits can be found in Figures \ref{fig:ind_transits_sec2_1}--\ref{fig:ind_transits_sec29_2}.

%******************* Individual TESS Sector 2%*******************
\begin{table*}[phtb]
\caption{Individual TESS Sector 2 transit parameters for WASP-4b derived using \texttt{EXOMOP}  }
    \centering
    \begin{tabular}{lccc}
    \hline
    \hline 
Transit	&	1	&	2	&	3		\\
T$_{c}$ (BJD$_{TDB}$-2458350)	&	5.1848698$\pm$0.0003176	&	6.5225253$\pm$0.00034	&	7.8609667$\pm$0.00032		\\
R$_p$/R$_\ast$	&	0.1516$\pm$0.0016	&	0.1523$\pm$0.0015	&	0.1538$\pm$0.0016		\\
a/R$_\ast$	&	5.32$\pm$0.27	&	5.36$\pm$0.16	&	5.46$\pm$0.18		\\
Inclination ($\degree$) 	&	87.35$\pm$2.64	&	89.93$\pm$2.36	&	89.96$\pm$2.56		\\
Duration (mins)	&	127.97$\pm$2.84	&	127.97$\pm$2.93	&	126.04$\pm$2.85		\\
$\chi_{reduced}^{2}$  & 1.02 & 1.05 & 0.98 \\
\hline								
Transit	&	4	&	5	&	6		\\
T$_{c}$ (BJD$_{TDB}$-2458350)	&	9.1994963$\pm$0.00033	&	10.5370341$\pm$0.00037	&	11.8754331$\pm$0.00037		\\
R$_p$/R$_\ast$	&	0.1513$\pm$0.0019	&	0.1505$\pm$0.0017	&	0.1513$\pm$0.0014		\\
a/R$_\ast$	&	5.31$\pm$0.31	&	5.40$\pm$0.21	&	5.014$\pm$0.063		\\
Inclination ($\degree$) 	&	87.25$\pm$2.27	&	90.05$\pm$2.88	&	89.0$\pm$4.23		\\
Duration (mins)	&	129.90$\pm$2.93	&	128.00$\pm$2.89	&	127.97$\pm$2.90		\\
$\chi_{reduced}^{2}$ & 0.97 & 1.00 & 0.97\\

\hline								
Transit	&	7	&	8	&	9		\\
T$_{c}$ (BJD$_{TDB}$-2458350)	&	13.2140842$\pm$0.00036	&	14.5520029$\pm$0.00036	&	15.8906537$\pm$0.00035		\\
R$_p$/R$_\ast$	&	0.1505$\pm$0.0017	&	0.1515$\pm$0.0015	&	0.1568$\pm$0.0028		\\
a/R$_\ast$	&	5.40$\pm$0.21	&	5.42$\pm$0.17	&	5.02$\pm$0.30		\\
Inclination ($\degree$) 	&	90.05$\pm$2.87	&	90.00$\pm$2.41	&	85.17$\pm$2.66		\\
Duration (mins)	&	127.97$\pm$2.99	&	128.00$\pm$2.83	&	132.01$\pm$2.87		\\
$\chi_{reduced}^{2}$ & 0.99 & 1.04 & 1.03 \\
\hline								
Transit	&	10	&	11	&	12		\\
T$_{c}$ (BJD$_{TDB}$-2458350)	&	19.9050877$\pm$0.00036	&	21.2429237$\pm$0.00039	&	22.5812197$\pm$0.00038		\\
R$_p$/R$_\ast$	&	0.1519$\pm$0.0019	&	0.1525$\pm$0.0025	&	0.1528$\pm$0.0021		\\
a/R$_\ast$	&	5.35$\pm$0.25	&	5.14$\pm$0.32	&	5.40$\pm$0.26		\\
Inclination ($\degree$) 	&	89.99$\pm$3.11	&	89.88$\pm$4.60	&	89.81$\pm$3.06		\\
Duration (mins)	&	127.97$\pm$2.85	&	132.01$\pm$2.88	&	127.97$\pm$2.84		\\
$\chi_{reduced}^{2}$  & 1.01  & 0.97  & 1.09 \\
\hline								
Transit	&	13	&	14	&	15		\\
T$_{c}$ (BJD$_{TDB}$-2458350)	&	23.9197516$\pm$0.00036	&	25.2580051$\pm$0.00037	&	26.5961751$\pm$0.00029		\\
R$_p$/R$_\ast$	&	0.1525$\pm$0.0021	&	0.1554$\pm$0.0024	&	0.1519$\pm$0.0014		\\
a/R$_\ast$	&	5.18$\pm$0.30	&	5.30$\pm$0.32	&	5.43$\pm$0.13		\\
Inclination ($\degree$) 	&	86.50$\pm$2.85	&	89.97$\pm$3.77	&	90.03$\pm$2.10		\\
Duration (mins)	&	132.01$\pm$2.90	&	130.08$\pm$2.84	&	126.04$\pm$2.89		\\
$\chi_{reduced}^{2}$  & 1.01  & 1.01  & 1.07 \\
\hline								
Transit	&	16	&	17	&	18		\\
T$_{c}$ (BJD$_{TDB}$-2458350)	&	27.9342669$\pm$0.00033	&	29.2731926$\pm$0.00033	&	30.6110336$\pm$0.00035		\\
R$_p$/R$_\ast$	&	0.1519$\pm$0.0014	&	0.1530$\pm$0.0015	&	0.1521$\pm$0.0018		\\
a/R$_\ast$	&	5.43$\pm$0.13	&	5.40$\pm$0.16	&	5.46$\pm$0.30		\\
Inclination ($\degree$) 	&	90.02$\pm$2.11	&	89.99$\pm$2.38	&	88.36$\pm$1.91		\\
Duration (mins)	&	126.04$\pm$2.89	&	127.97$\pm$2.84	&	127.98$\pm$2.95		\\
$\chi_{reduced}^{2}$  & 1.02 & 1.01 & 1.03 \\
    \hline
    \end{tabular}
    \tablecomments{The linear and quadratic limb darkening coefficient used in the analysis are 0.382 and 0.210 \citep{Claret2017}}
    \label{tb:lighcurve_model_TESS}
\end{table*}

\newpage 
%******************* Individual TESS Sector 28%*******************
\begin{table*}[tb]
\caption{Individual TESS Sector 28 transit parameters for WASP-4b derived using \texttt{EXOMOP}  }
    \centering
    \begin{tabular}{lccc}
    \hline
    \hline 
Transit	&	1	&	2	&	3		\\
T$_{c}$ (BJD$_{TDB}$-2459050)	&	13.1085457$\pm$0.00043	&	14.44734$\pm$0.00041	&	15.7854034$\pm$0.00043		\\
R$_p$/R$_\ast$	&	0.1534$\pm$0.0027	&	0.1541$\pm$0.0022	&	0.1524$\pm$0.0028		\\
a/R$_\ast$	&	4.92$\pm$0.49	&	5.41$\pm$0.27	&	5.25$\pm$0.33		\\
Inclination ($\degree$) 	&	85.32$\pm$3.66	&	90.12$\pm$3.35	&	89.80$\pm$4.81		\\
Duration (mins)	&	133.95$\pm$2.82	&	126.21$\pm$2.83	&	127.97$\pm$2.82		\\
$\chi_{reduced}^{2}$   & 0.88  & 0.98  & 0.85 \\

\hline								
Transit	&	4	&	5	&	6		\\
T$_{c}$ (BJD$_{TDB}$-2459050)	&	17.1240632$\pm$0.00036	&	18.4619638$\pm$0.00035	&	19.8001969$\pm$0.00044		\\
R$_p$/R$_\ast$	&	0.1532$\pm$0.0018	&	0.1511$\pm$0.0028	&	0.1507$\pm$0.0029		\\
a/R$_\ast$	&	5.46$\pm$0.20	&	5.36$\pm$0.38	&	5.12$\pm$0.34		\\
Inclination ($\degree$) 	&	90.06$\pm$2.65	&	90.12$\pm$4.37	&	89.97$\pm$5.19		\\
Duration (mins)	&	126.21$\pm$2.82	&	127.97$\pm$2.86	&	131.84$\pm$2.83		\\
$\chi_{reduced}^{2}$  & 0.85  & 0.98  & 0.97   \\
\hline								
Transit	&	7	&	8	&	9		\\
T$_{c}$ (BJD$_{TDB}$-2459050)	&	21.1379261$\pm$0.00049	&	26.4905036$\pm$0.00043	&	27.8293474$\pm$0.00040		\\
R$_p$/R$_\ast$	&	0.1527$\pm$0.0029	&	0.1517$\pm$0.0035	&	0.1513$\pm$0.0025		\\
a/R$_\ast$	&	5.30$\pm$0.40	&	5.05$\pm$0.37	&	5.33$\pm$0.31		\\
Inclination ($\degree$) 	&	89.96$\pm$4.27	&	89.74$\pm$5.80	&	89.93$\pm$3.78		\\
Duration (mins)	&	126.21$\pm$2.82	&	131.84$\pm$2.83	&	127.97$\pm$2.83		\\
$\chi_{reduced}^{2}$  & 0.95   & 0.92  & 0.92  \\
\hline								
Transit	&	10	&	11	&	12		\\
T$_{c}$ (BJD$_{TDB}$-2459050)	&	29.1681276$\pm$0.00042	&	30.5064451$\pm$0.00044	&	31.8447896$\pm$0.00040		\\
R$_p$/R$_\ast$	&	0.1521$\pm$0.0025	&	0.1534$\pm$0.0036	&	0.1548$\pm$0.0021		\\
a/R$_\ast$	&	5.24$\pm$0.32	&	5.23$\pm$0.44	&	5.62$\pm$0.25		\\
Inclination ($\degree$) 	&	90.13$\pm$3.79	&	90.00$\pm$4.75	&	90.06$\pm$2.89		\\
Duration (mins)	&	131.84$\pm$2.83	&	130.08$\pm$2.83	&	121.99$\pm$2.83		\\
$\chi_{reduced}^{2}$  &  0.92  & 0.87  & 0.99  \\
\hline								
Transit	&	13	&	14	&	15		\\
T$_{c}$ (BJD$_{TDB}$-2459050)	&	33.1816361$\pm$0.00041	&	34.5203618$\pm$0.00044	&	38.5347781$\pm$0.00034		\\
R$_p$/R$_\ast$	&	0.1536$\pm$0.0021	&	0.1494$\pm$0.0028	&	0.1503$\pm$0.0015		\\
a/R$_\ast$	&	5.47$\pm$0.27	&	5.18$\pm$0.35	&	5.43$\pm$0.16		\\
Inclination ($\degree$) 	&	90.04$\pm$3.14	&	90.09$\pm$4.41	&	90.00$\pm$2.37		\\
Duration (mins)	&	125.86$\pm$2.83	&	130.08$\pm$2.83	&	127.97$\pm$2.83		\\
$\chi_{reduced}^{2}$  & 0.95  & 0.87  & 0.89\\
    \hline
    \end{tabular}
    \tablecomments{The linear and quadratic limb darkening coefficient used in the analysis are 0.382 and 0.210 \citep{Claret2017}}
    \label{tb:lighcurve_model_TESS2}
\end{table*}

\newpage 
%******************* Individual TESS Sector 29%*******************
\begin{table*}[tb]
\caption{Individual TESS Sector 29 transit parameters for WASP-4b derived using \texttt{EXOMOP}  }
    \centering
    \begin{tabular}{lccc}
    \hline
    \hline 
Transit	&	1	&	2	&	3		\\
T$_{c}$ (BJD$_{TDB}$-2459050)	&	39.873712$\pm$0.00032	&	41.2118022$\pm$0.00034	&	42.5500092$\pm$0.00037		\\
R$_p$/R$_\ast$	&	0.1519$\pm$0.0015	&	0.1532$\pm$0.0017	&	0.1531$\pm$0.0034		\\
a/R$_\ast$	&	5.44$\pm$0.18	&	5.47$\pm$0.19	&	5.07$\pm$0.37		\\
Inclination ($\degree$) 	&	90.00$\pm$2.53	&	89.84$\pm$2.71	&	89.61$\pm$5.78		\\
Duration (mins)	&	127.97$\pm$2.84	&	125.86$\pm$2.83	&	131.84$\pm$2.83		\\
$\chi_{reduced}^{2}$   & 0.91  & 0.92  & 0.86   \\
\hline								
Transit	&	4	&	5	&	6		\\
T$_{c}$ (BJD$_{TDB}$-2459050)	&	43.8878265$\pm$0.00033	&	45.2266667$\pm$0.00034	&	46.5650437$\pm$0.00036		\\
R$_p$/R$_\ast$	&	0.1495$\pm$0.0016	&	0.1529$\pm$0.0023	&	0.1495$\pm$0.0018		\\
a/R$_\ast$	&	5.37$\pm$0.19	&	5.30$\pm$0.30	&	5.37$\pm$0.21		\\
Inclination ($\degree$) 	&	89.96$\pm$2.58	&	89.83$\pm$4.23	&	90.19$\pm$2.87		\\
Duration (mins)	&	127.97$\pm$2.83	&	127.97$\pm$2.86	&	127.97$\pm$2.83		\\
$\chi_{reduced}^{2}$   & 0.98  & 0.84  & 0.96 \\
\hline								
Transit	&	7	&	8	&	9		\\
T$_{c}$ (BJD$_{TDB}$-2459050)	&	47.9030583$\pm$0.00037	&	53.2551368$\pm$0.00033	&	54.5937406$\pm$0.00032		\\
R$_p$/R$_\ast$	&	0.1540$\pm$0.0029	&	0.1495$\pm$0.0015	&	0.1523$\pm$0.0016		\\
a/R$_\ast$	&	5.09$\pm$0.34	&	5.42$\pm$0.15	&	5.40$\pm$0.20		\\
Inclination ($\degree$) 	&	89.55$\pm$5.49	&	90.09$\pm$2.28	&	90.03$\pm$2.71		\\
Duration (mins)	&	133.95$\pm$2.83	&	127.97$\pm$2.84	&	125.86$\pm$2.83		\\
$\chi_{reduced}^{2}$   & 0.86  & 1.05 & 0.92  \\ 
\hline								
Transit	&	10	&	11	&	12		\\
T$_{c}$ (BJD$_{TDB}$-2459050)	&	55.9324389$\pm$0.00033	&	57.2708858$\pm$0.00033	&	58.6088955$\pm$0.00035		\\
R$_p$/R$_\ast$	&	0.1516$\pm$0.0018	&	0.1542$\pm$0.0016	&	0.1514$\pm$0.0016		\\
a/R$_\ast$	&	5.42$\pm$0.25	&	5.43$\pm$0.20	&	5.42$\pm$019		\\
Inclination ($\degree$) 	&	90.00$\pm$3.07	&	90.04$\pm$2.77	&	90.03$\pm$2.45		\\
Duration (mins)	&	125.86$\pm$2.83	&	127.97$\pm$2.83	&	125.90$\pm$2.88		\\
$\chi_{reduced}^{2}$  & 0.91  & 0.94 & 0.99 \\
\hline								
Transit	&	13	&	14				\\
T$_{c}$ (BJD$_{TDB}$-2459050)	&	59.9469868$\pm$0.00040	&	61.2848162$\pm$0.00040				\\
R$_p$/R$_\ast$	&	0.1544$\pm$0.0025	&	0.1511$\pm$0.0016				\\
a/R$_\ast$	&	5.07$\pm$0.31	&	5.41$\pm$0.18				\\
Inclination ($\degree$) 	&	89.40$\pm$5.43	&	89.94$\pm$2.49				\\
Duration (mins)	&	135.70$\pm$2.84	&	127.97$\pm$2.83				\\
$\chi_{reduced}^{2}$ & 0.96  & 0.93 \\
    \hline
    \end{tabular}
    \tablecomments{The linear and quadratic limb darkening coefficient used in the analysis are 0.382 and 0.210 \citep{Claret2017}}
    \label{tb:lighcurve_model_TESS3}
\end{table*}

%average(0.88 , 0.98 , 0.85, 0.85  , 0.98  , 0.97, 0.95   , 0.92  , 0.92 , 0.92  , 0.87  , 0.99,0.95  , 0.87  , 0.89, 0.91  , 0.92  , 0.86,0.98  , 0.84  , 0.96, 0.86  ,0.96  , 0.93,  1.05 , 0.92, 0.91  ,0.94 , 0.99 )

%%%%%%%%%%%%%%%%%%%%%%%%%%%

\begin{figure*}
\centering
\begin{tabular}{cc}
 \includegraphics[width=0.50\textwidth]{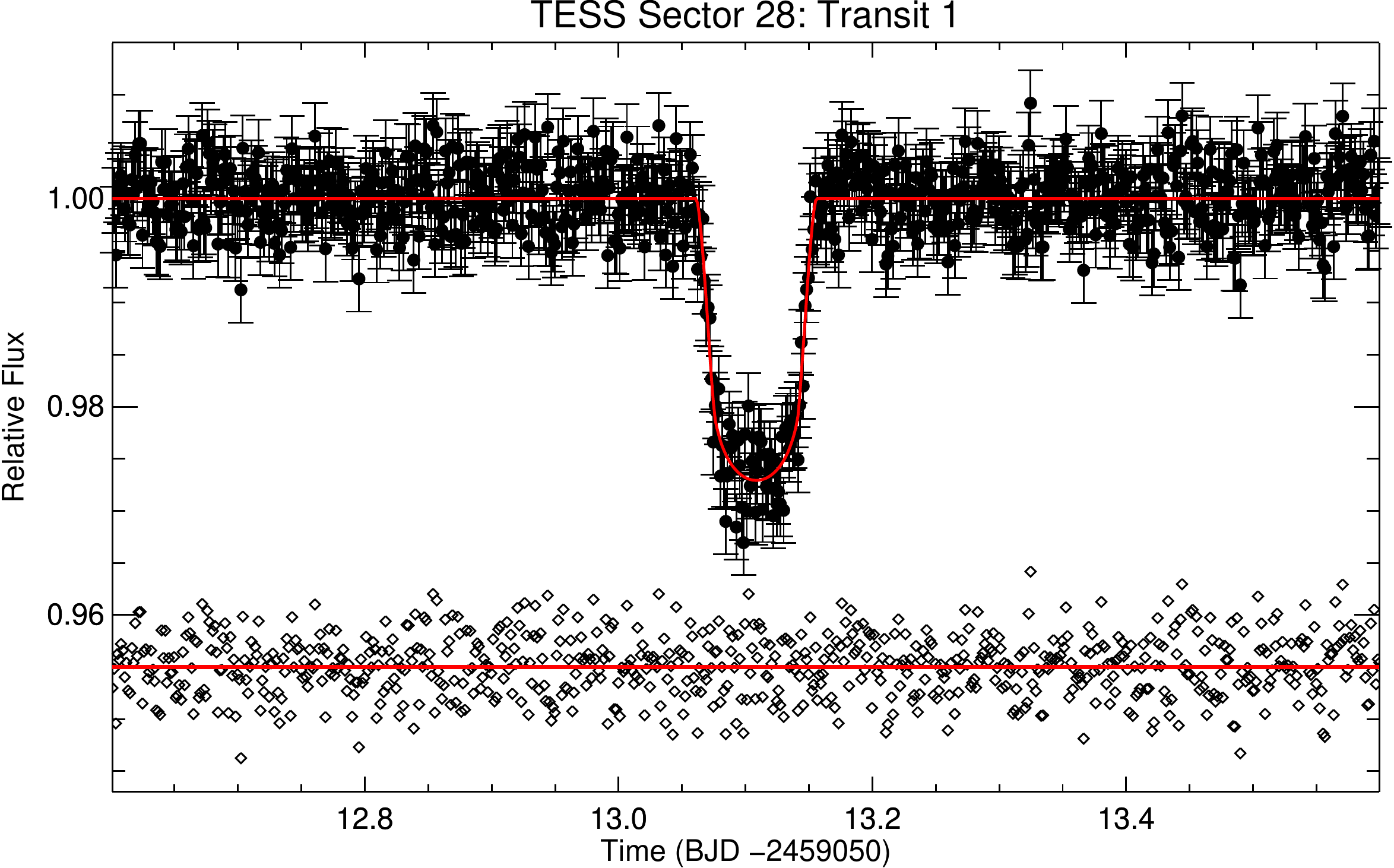} & \includegraphics[width=0.50\textwidth]{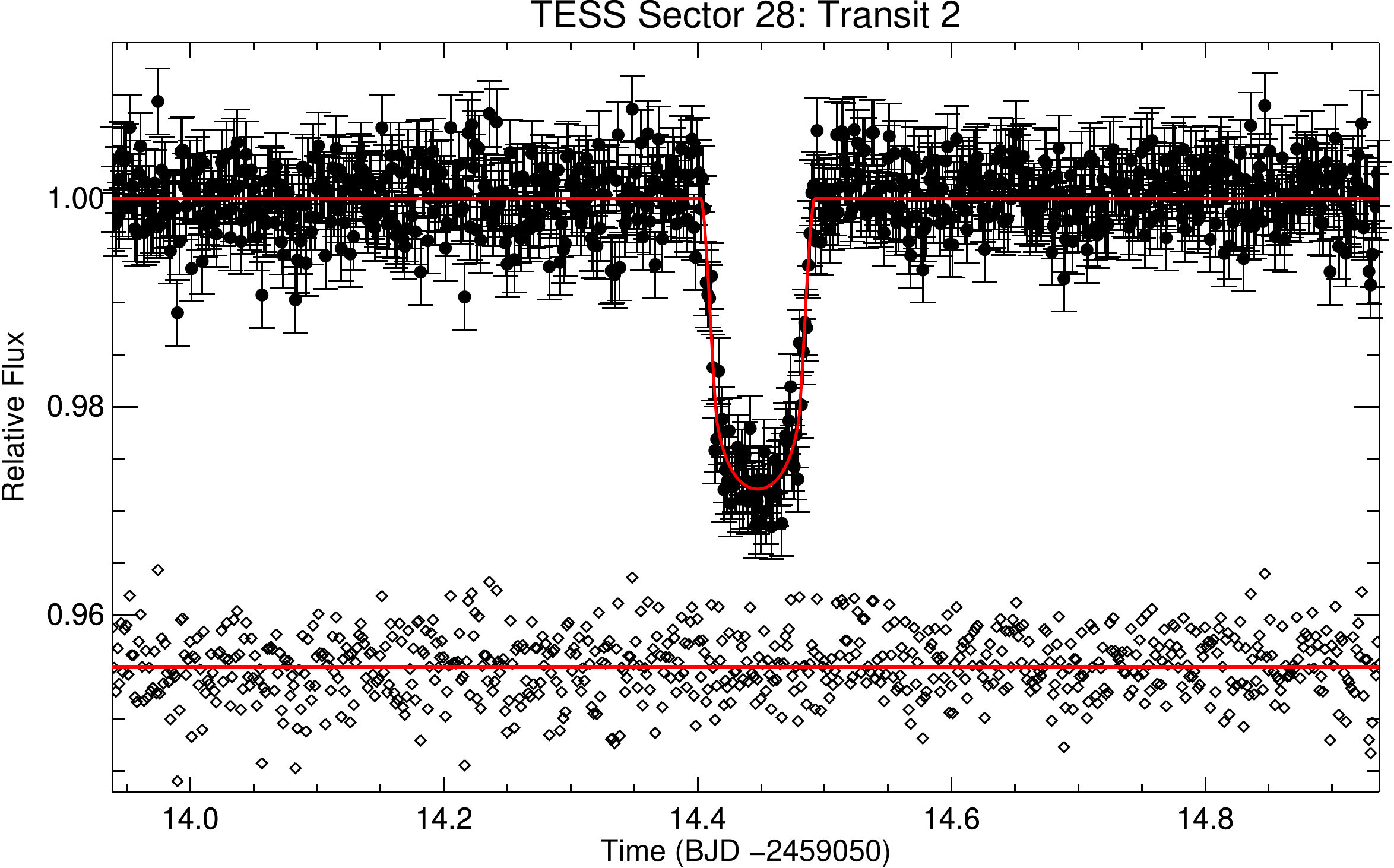}\\
 \includegraphics[width=0.50\textwidth]{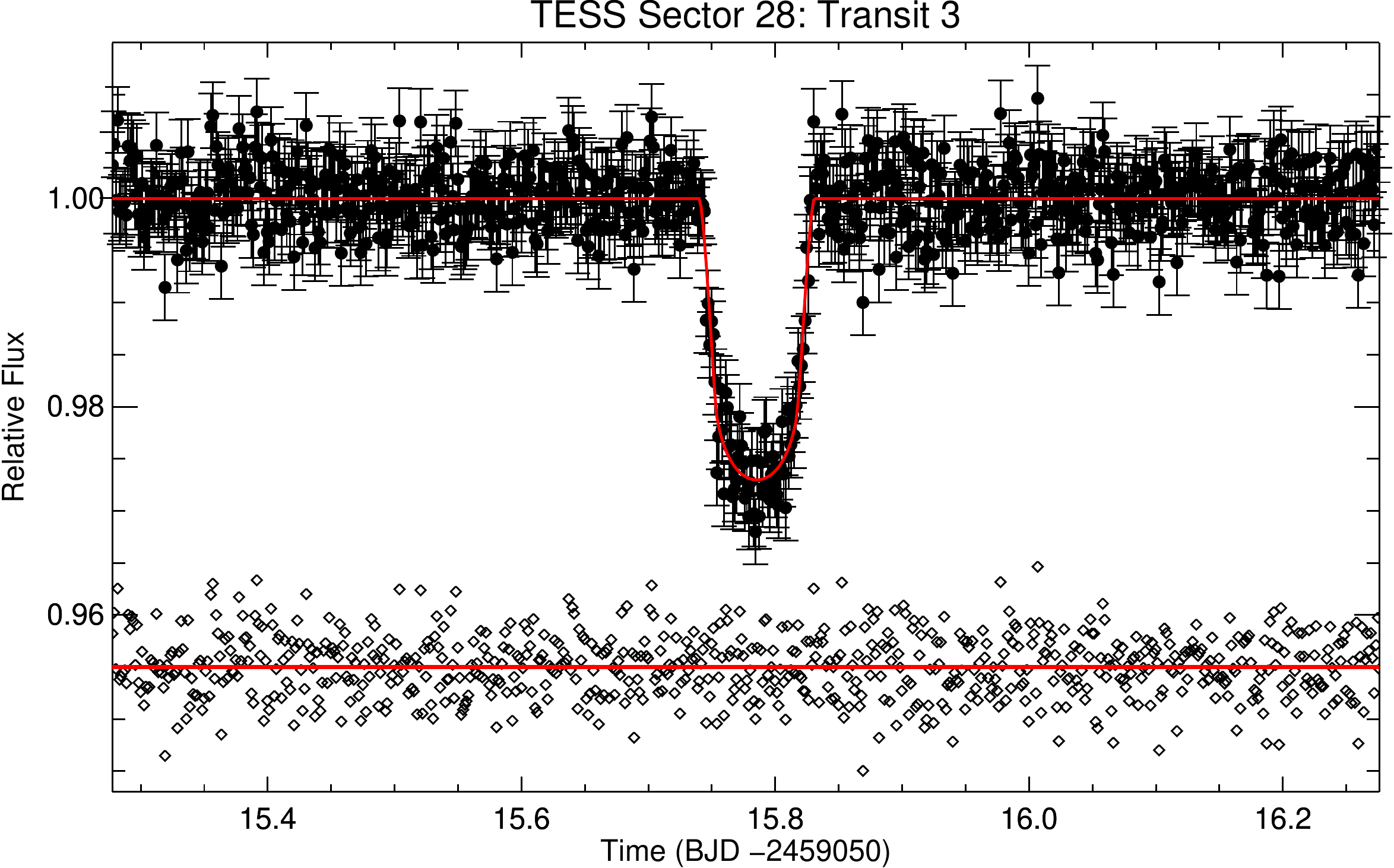} & \includegraphics[width=0.50\textwidth]{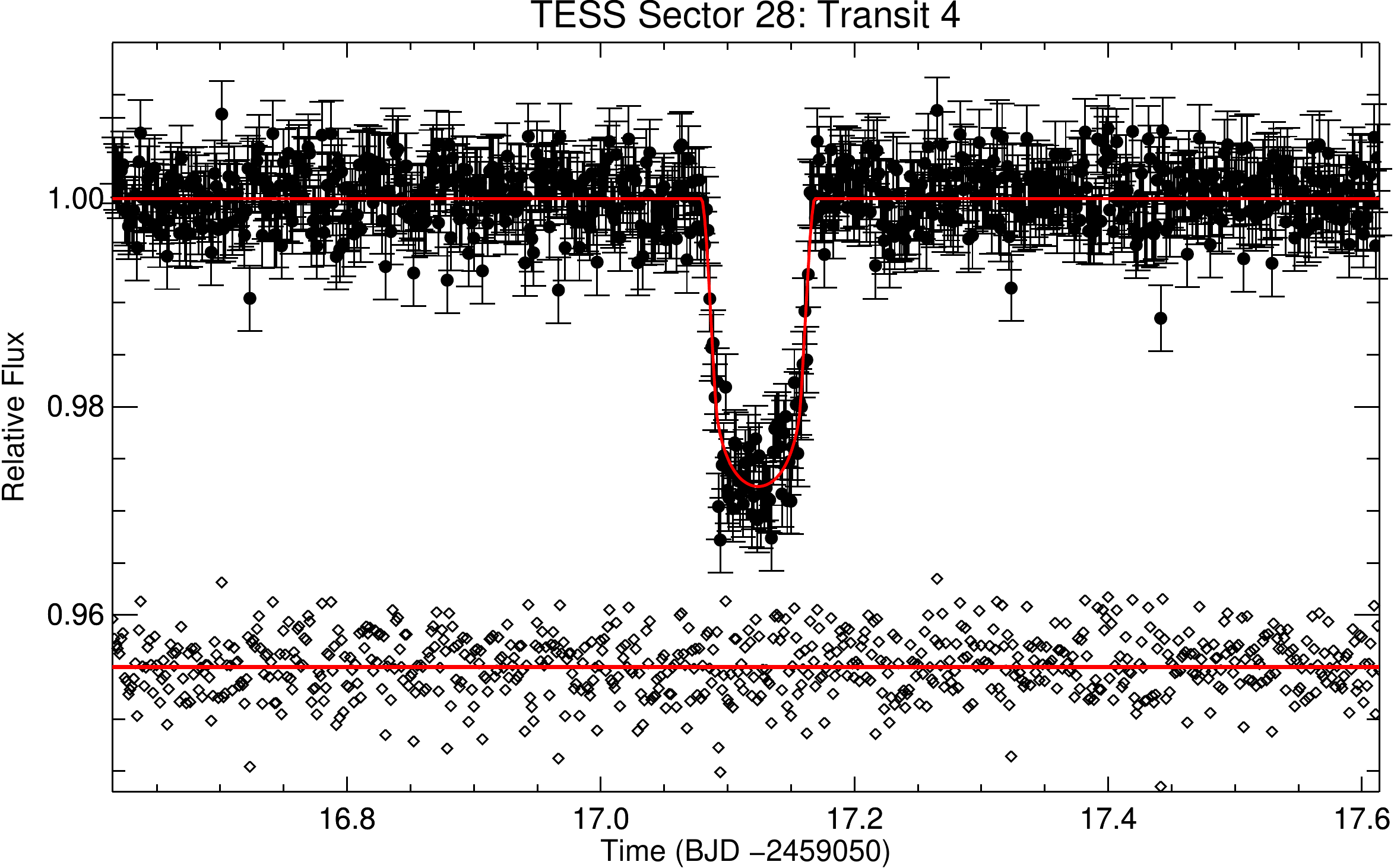}\\
  \includegraphics[width=0.50\textwidth]{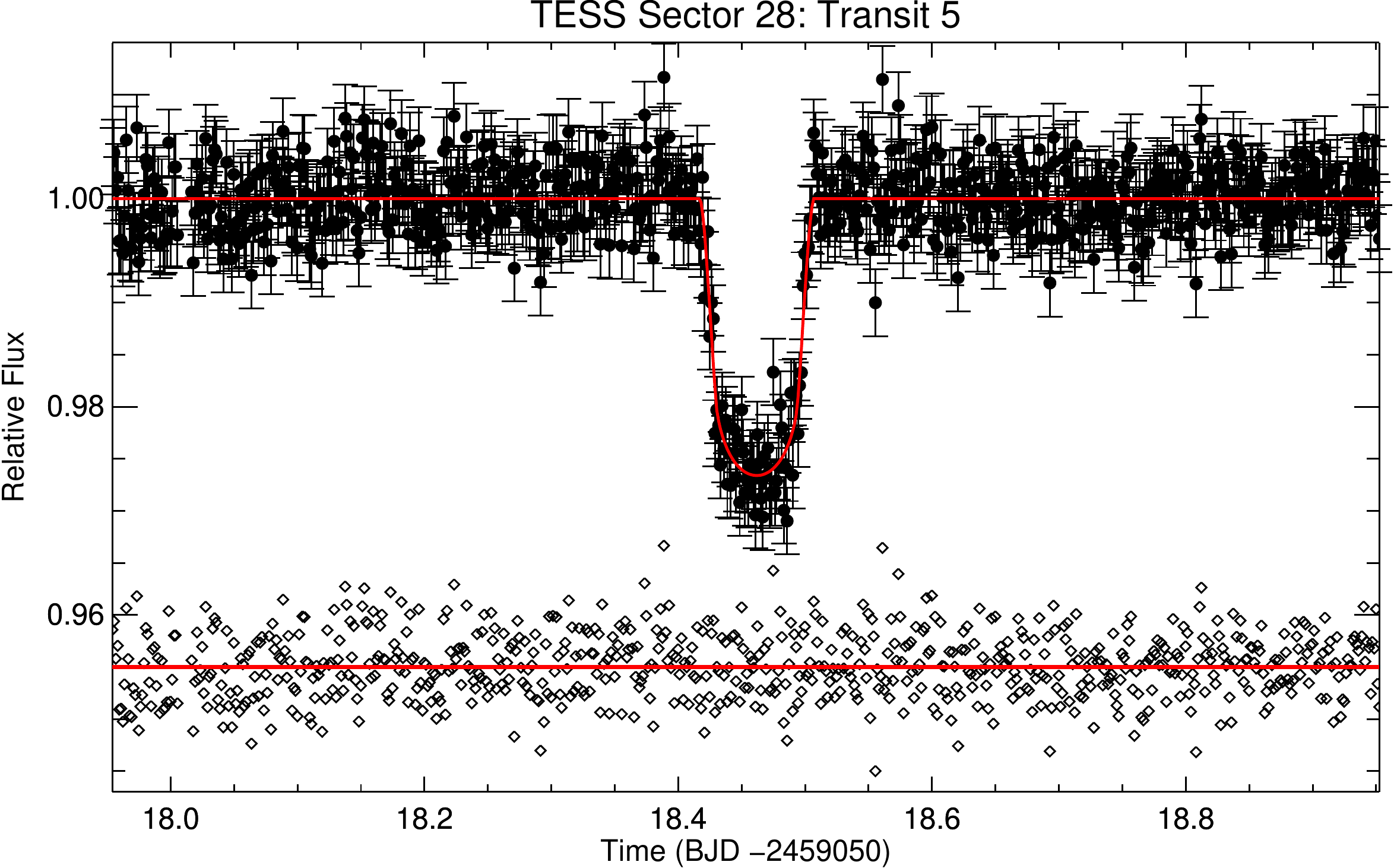} & \includegraphics[width=0.50\textwidth]{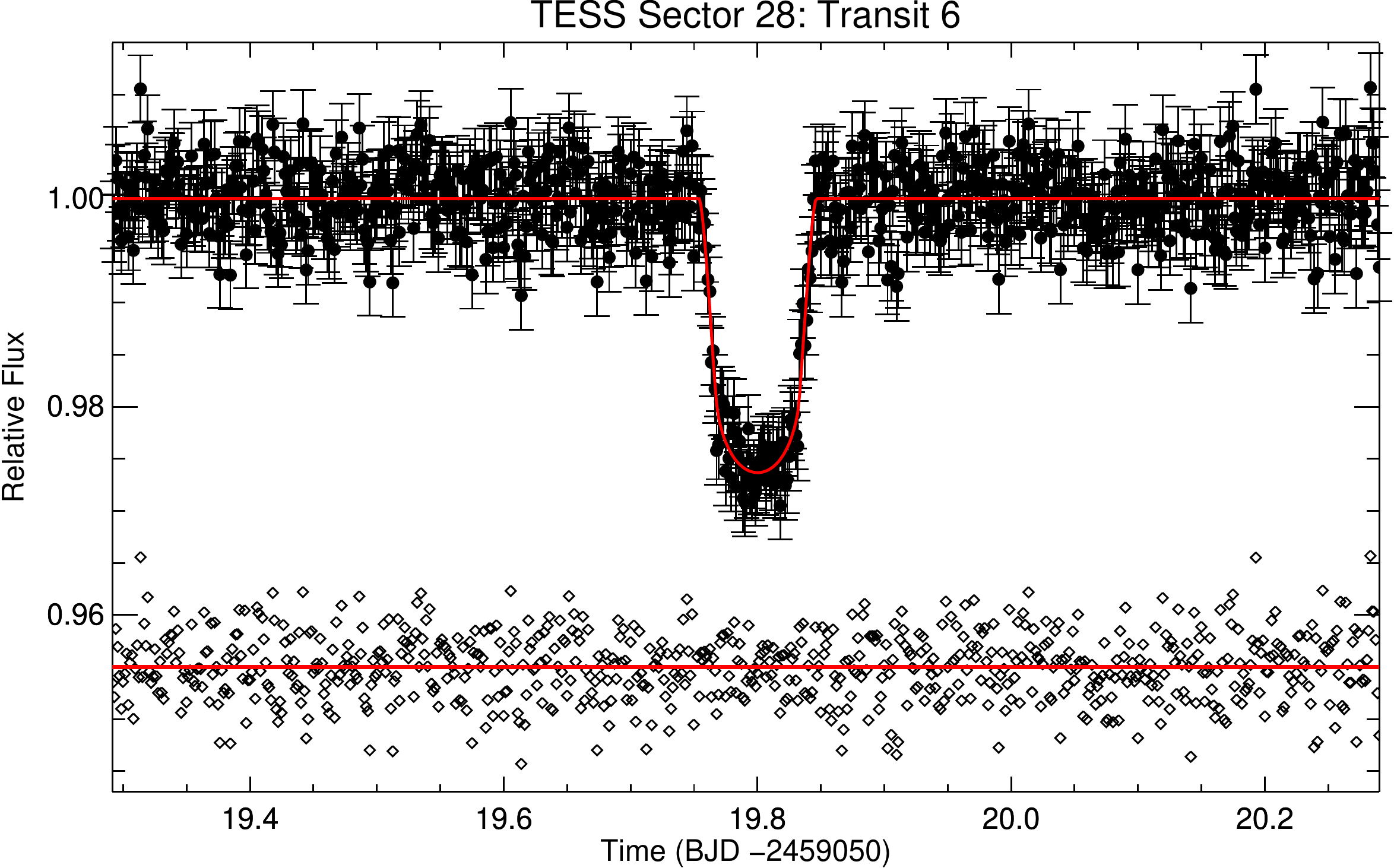}\\
  \includegraphics[width=0.50\textwidth]{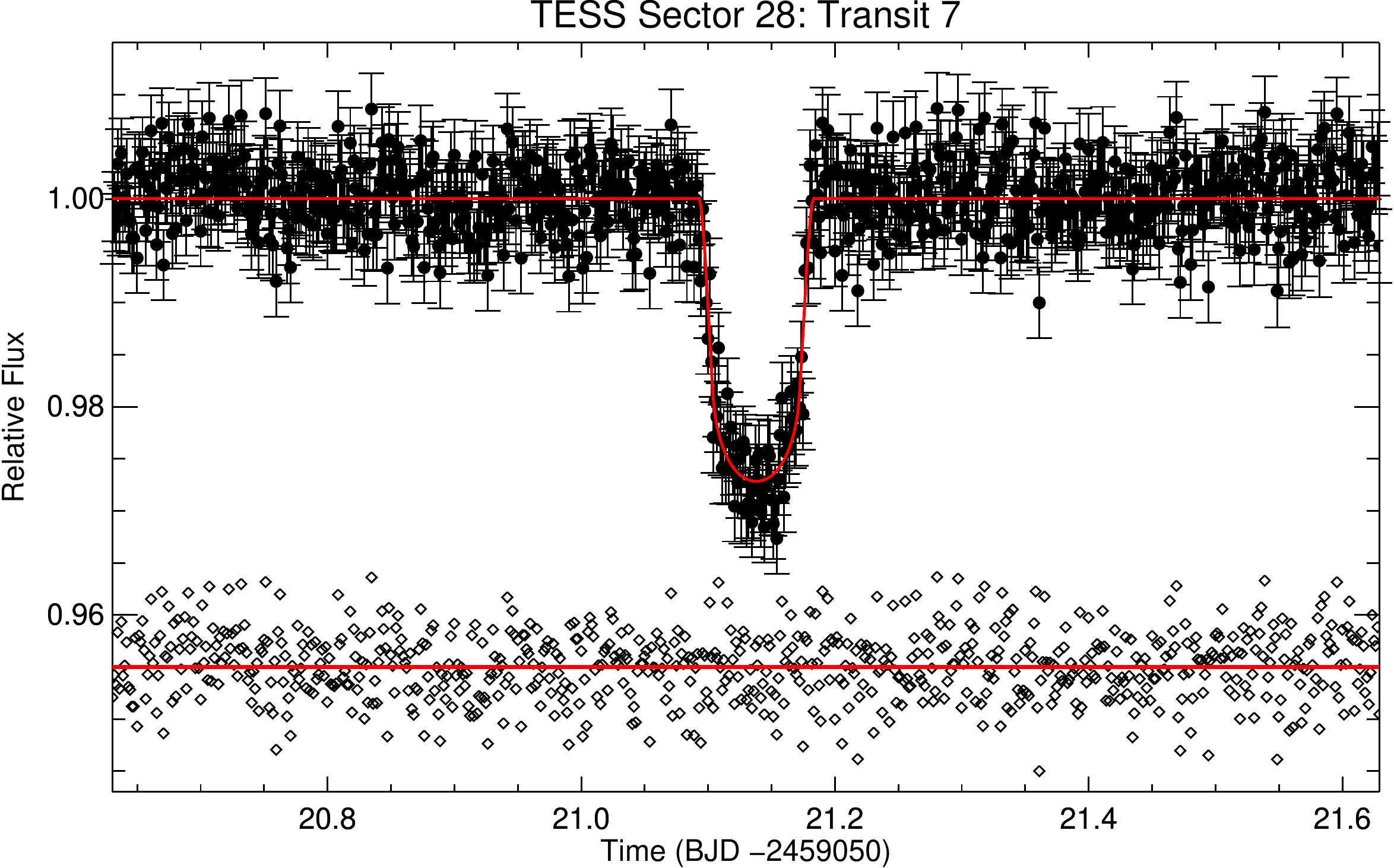} & \includegraphics[width=0.50\textwidth]{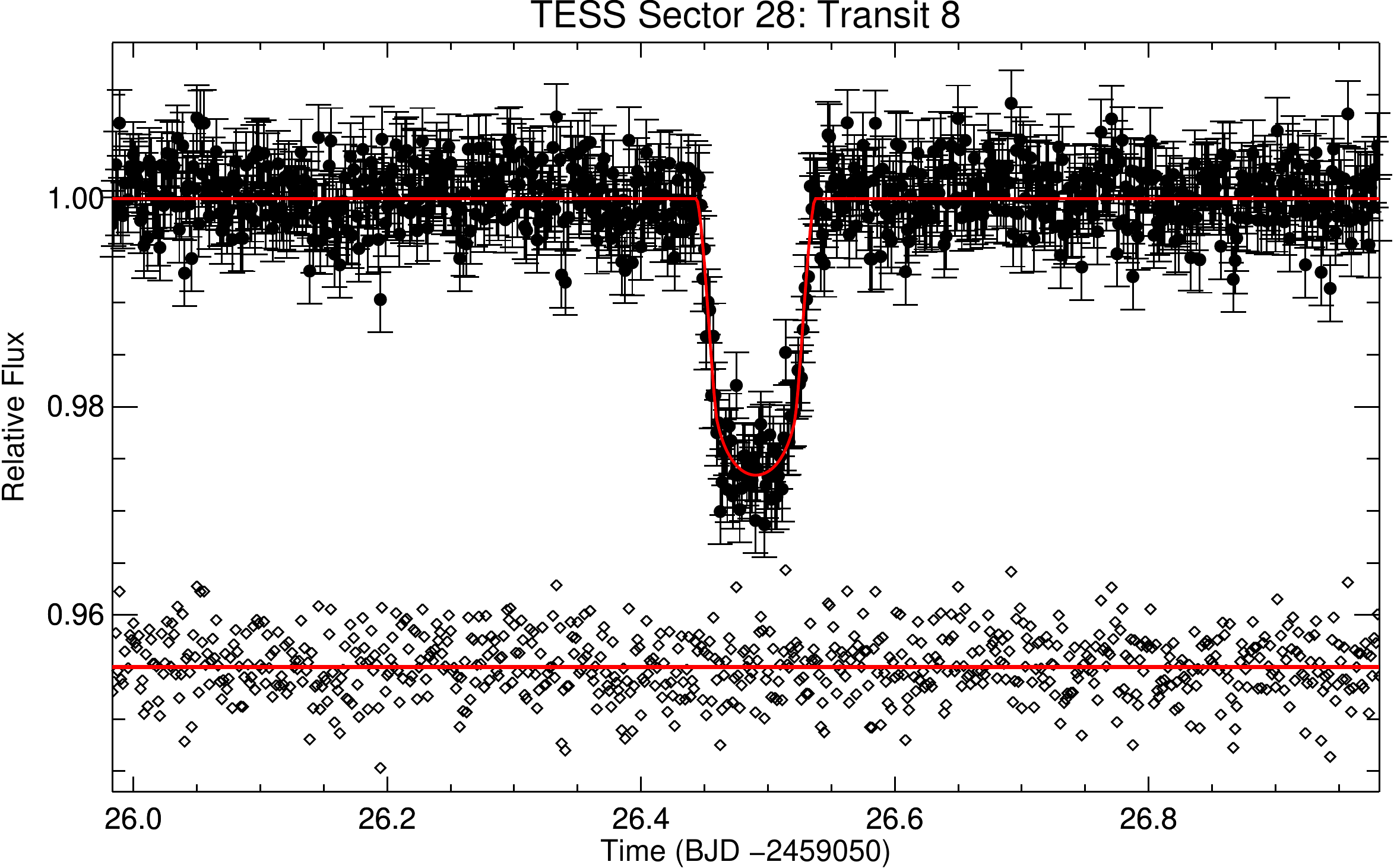}\\
  \end{tabular}
\caption{Individual TESS transit events (1-8) from Sector 28 of WASP-4b. Other comments are the same as Figure \ref{fig:ind_transits_sec2_1}.}
\label{fig:ind_transits_sec28_1}
\end{figure*}

\begin{figure*}
\centering
\begin{tabular}{cc}
 \includegraphics[width=0.50\textwidth]{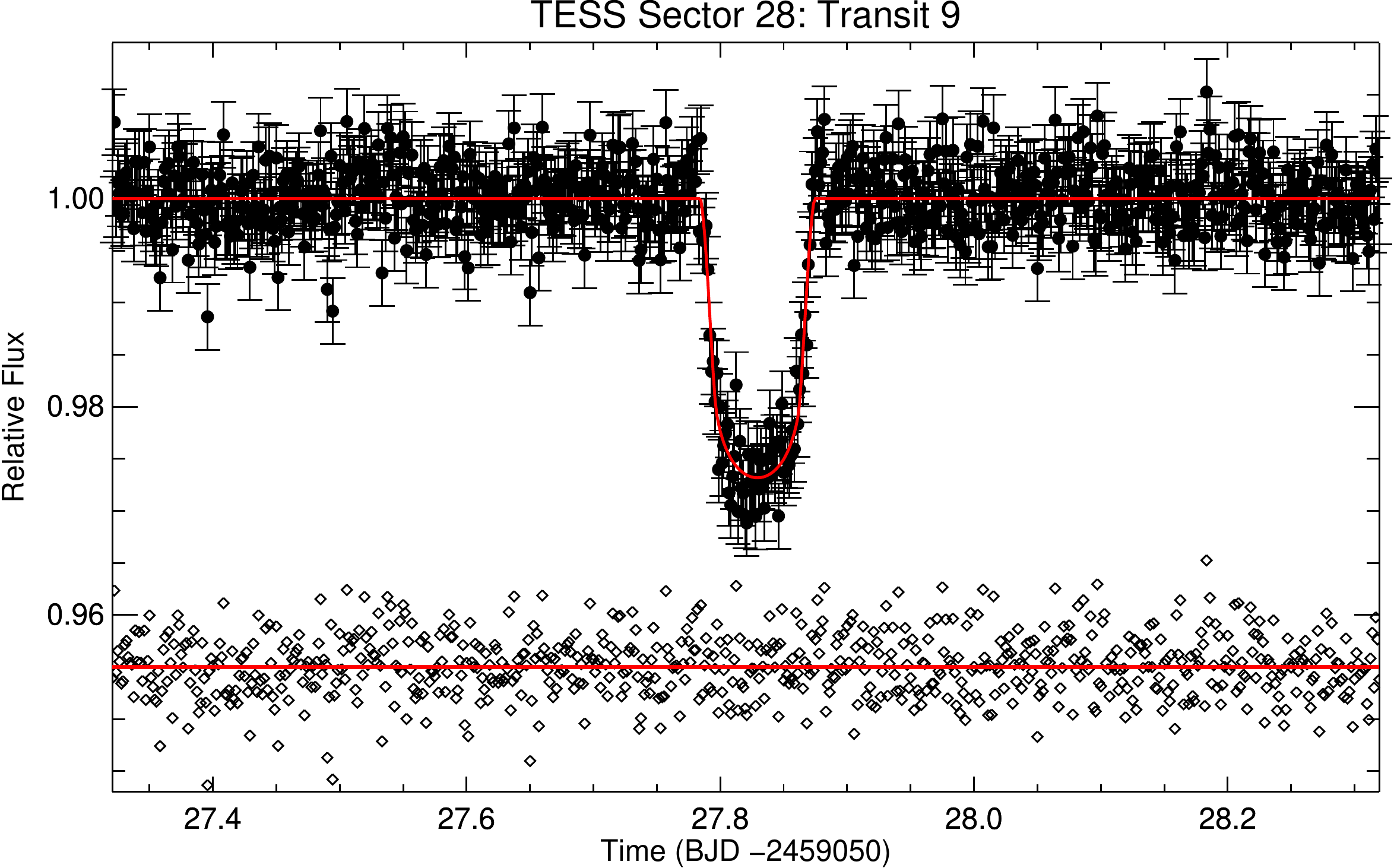} & \includegraphics[width=0.50\textwidth]{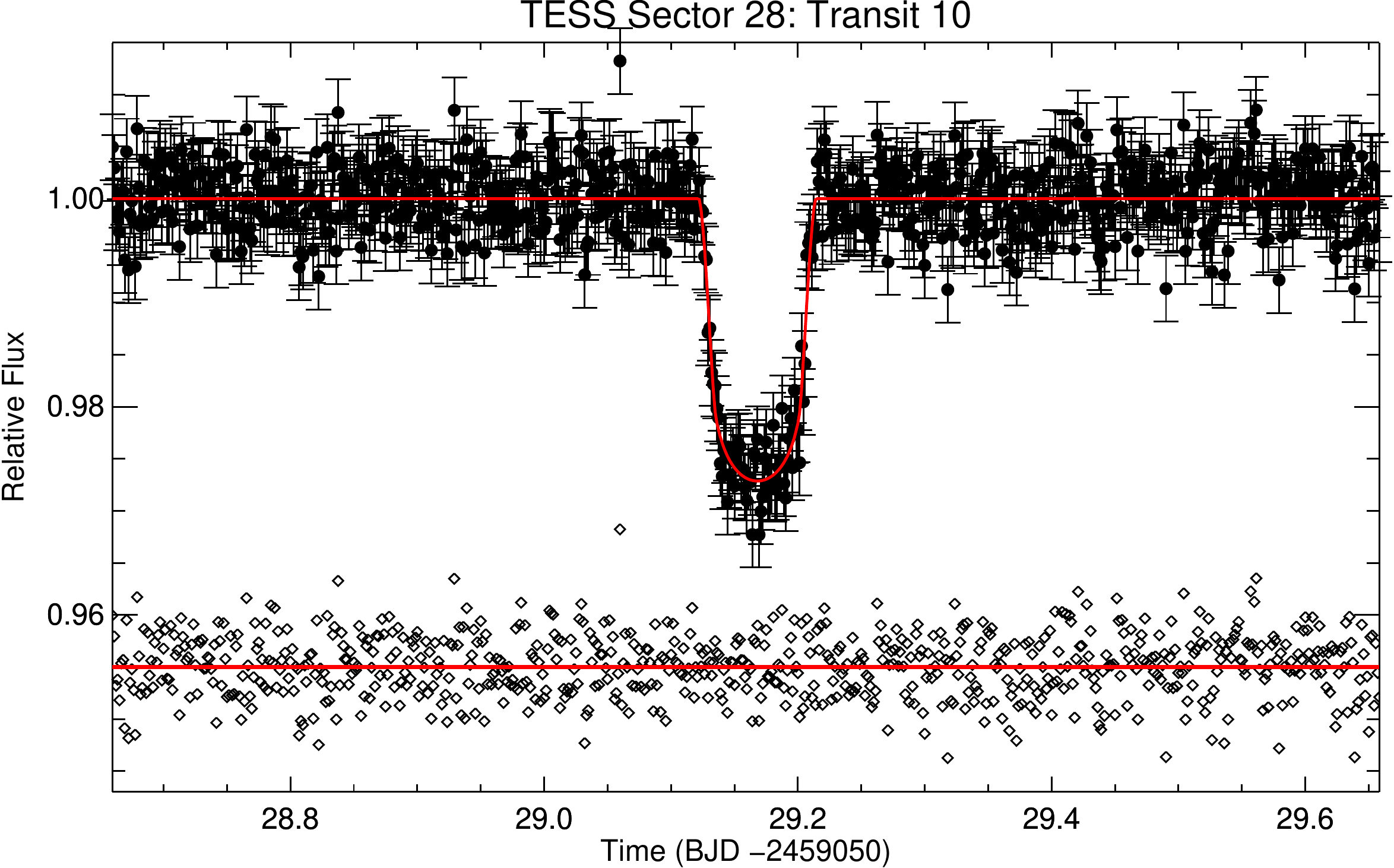}\\
 \includegraphics[width=0.50\textwidth]{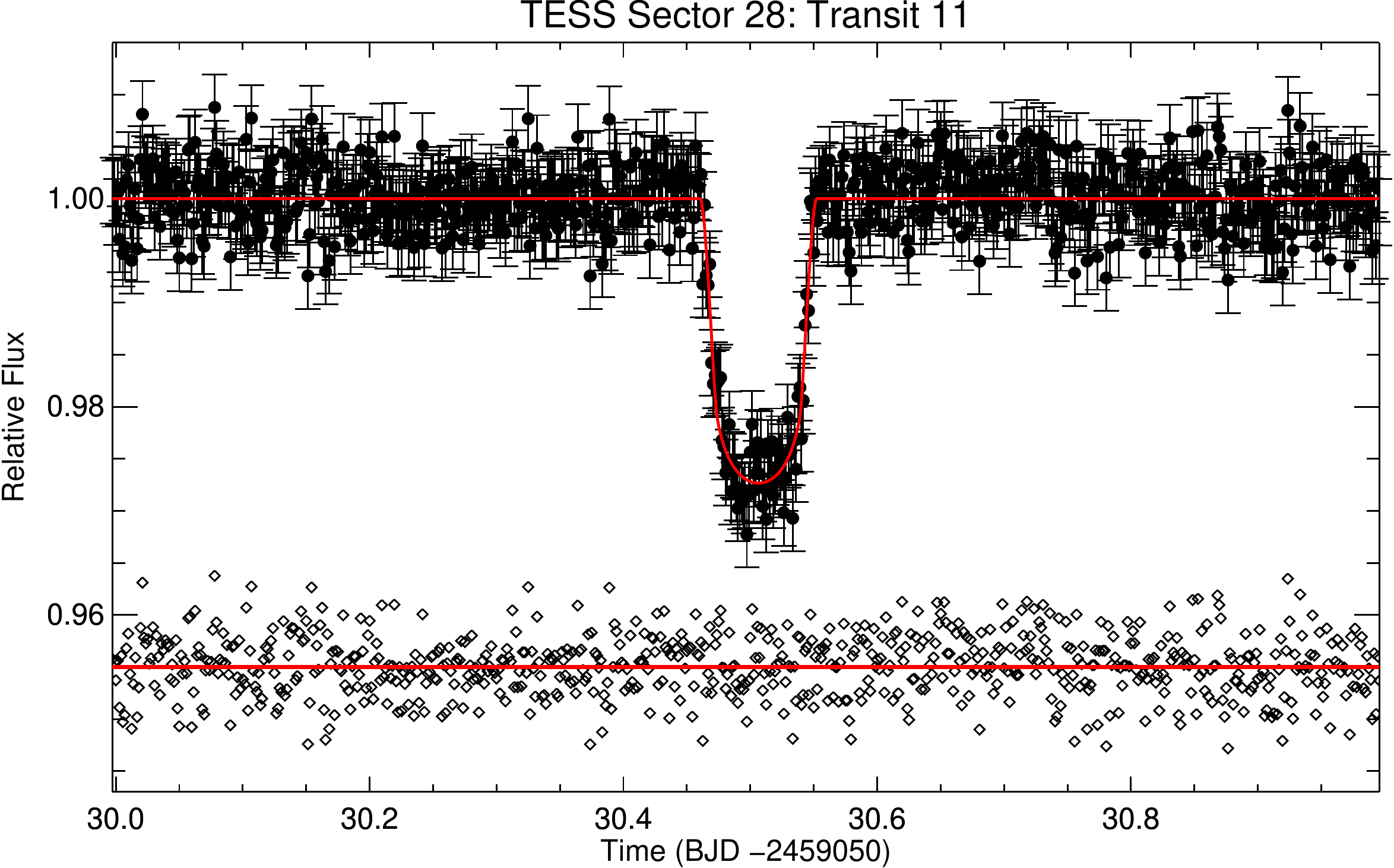} & \includegraphics[width=0.50\textwidth]{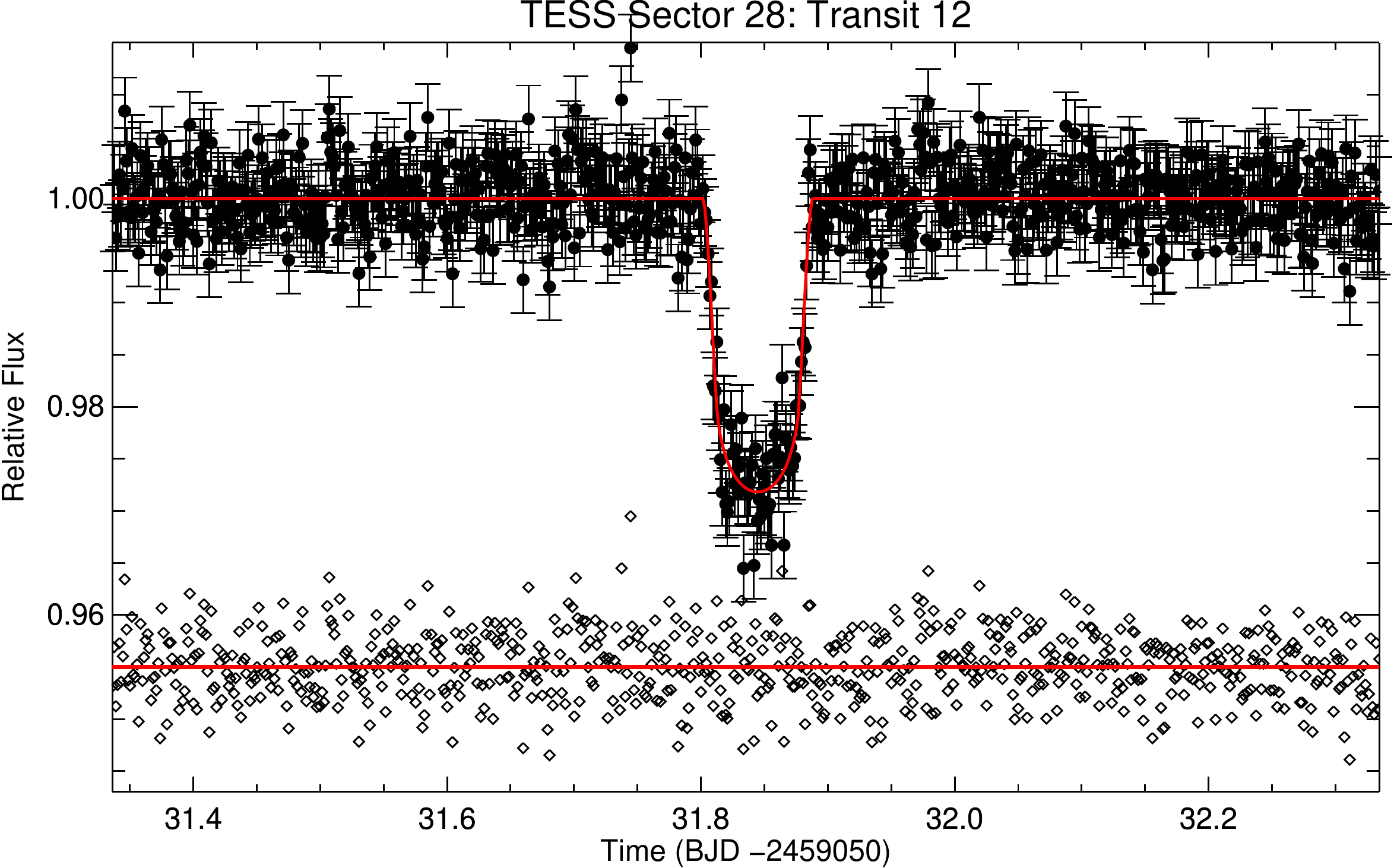}\\
  \includegraphics[width=0.50\textwidth]{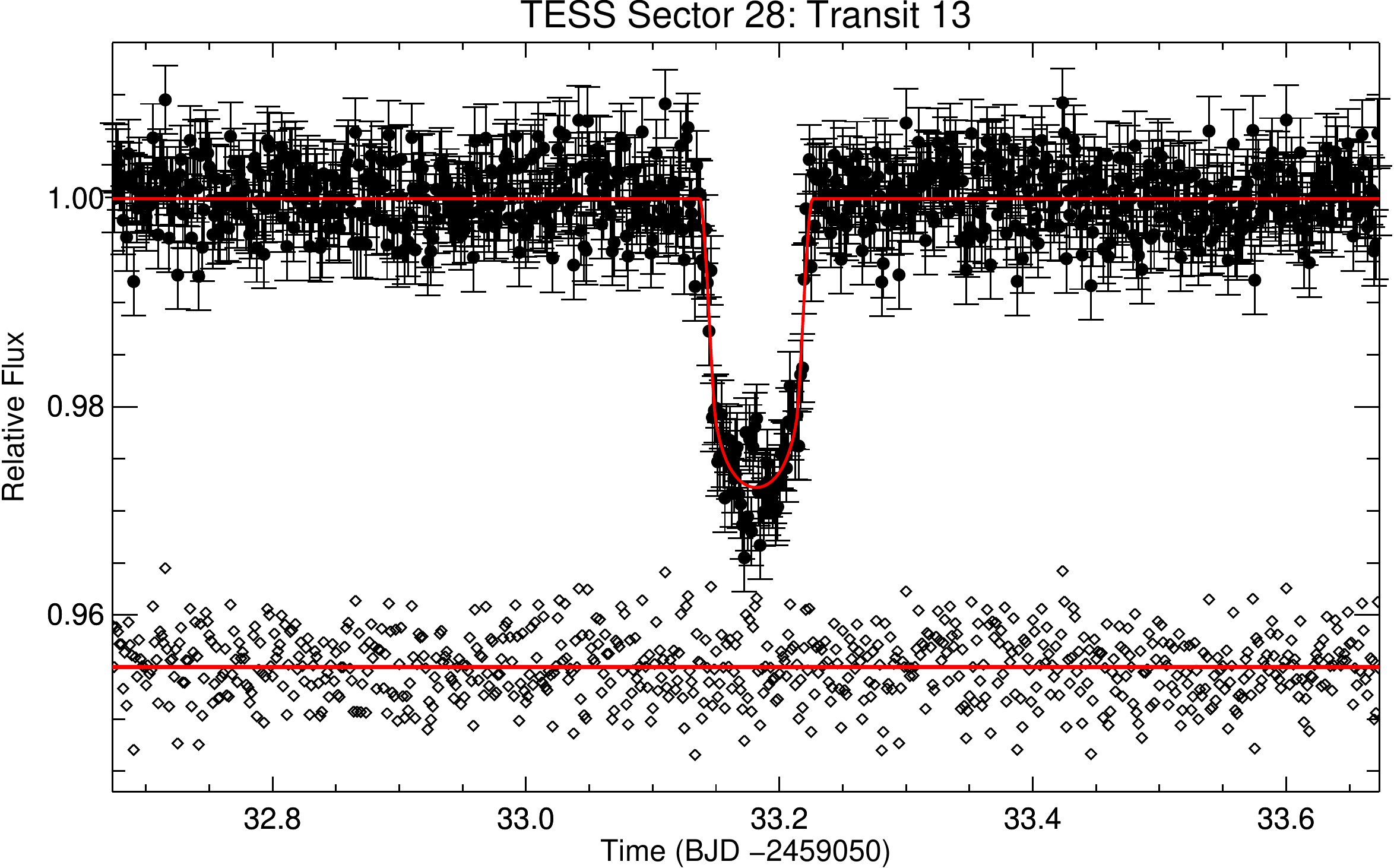} & \includegraphics[width=0.50\textwidth]{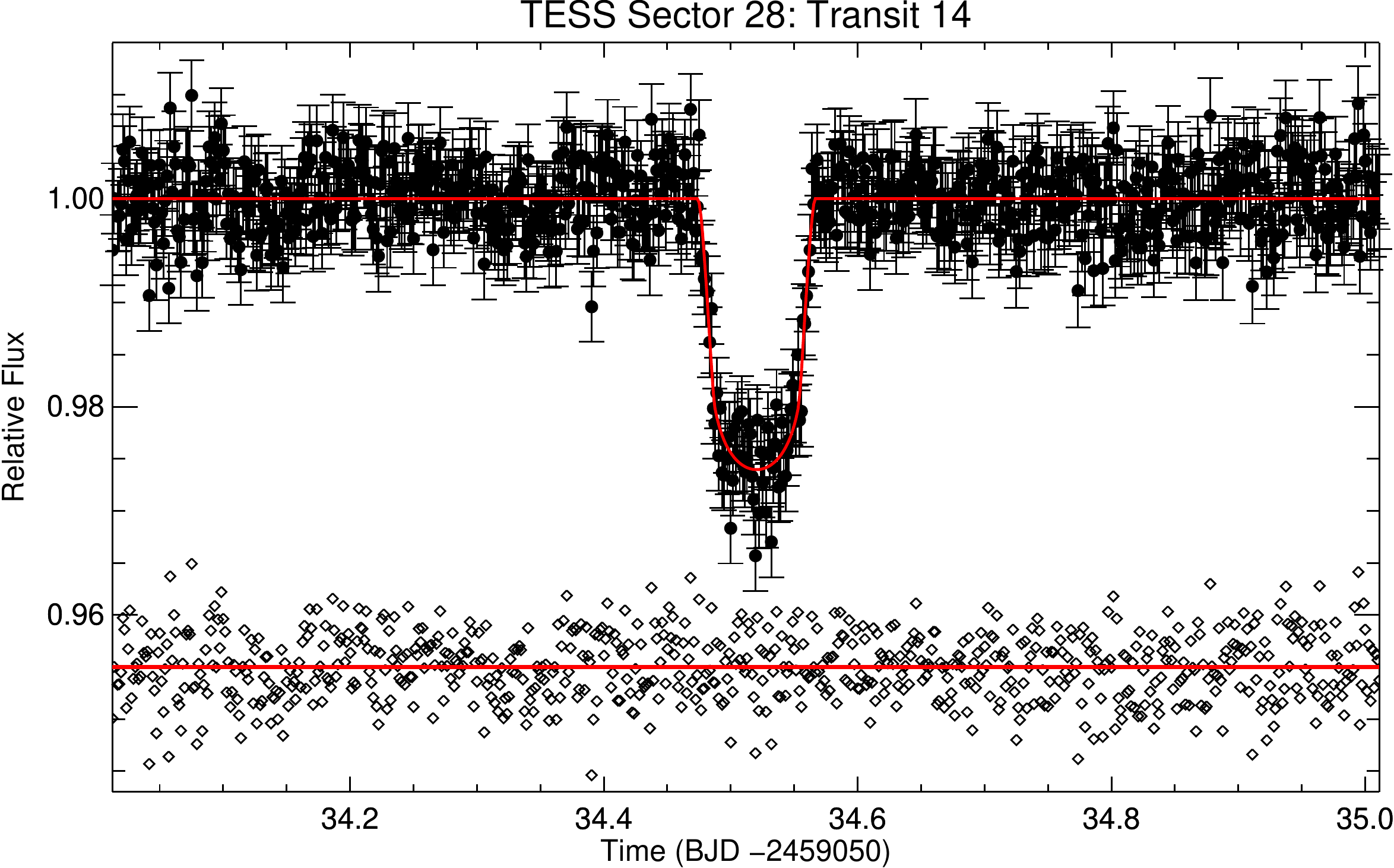}
   \end{tabular}
  \begin{tabular}{cc}
   \includegraphics[width=0.50\textwidth]{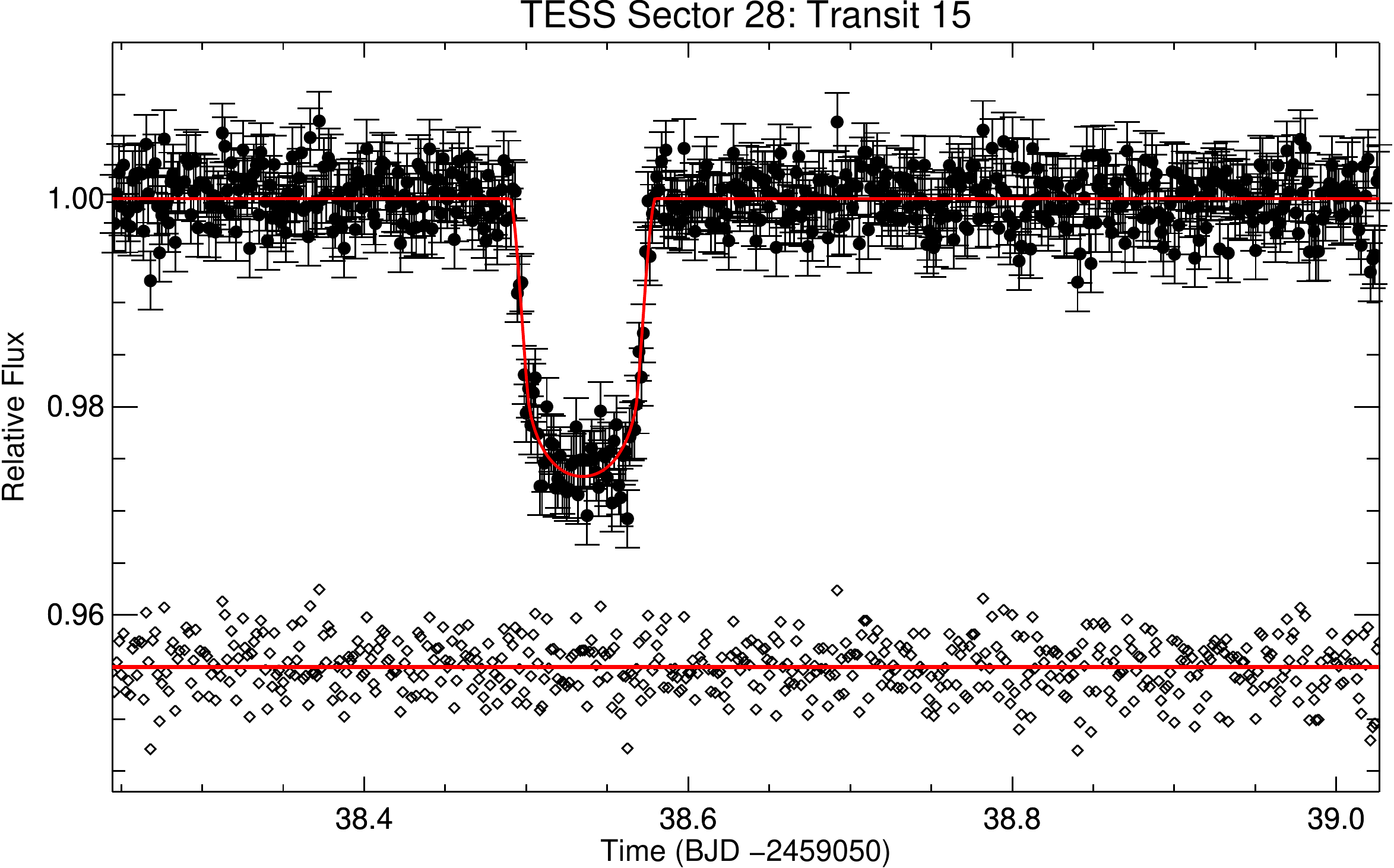}
  \end{tabular}
\caption{Individual TESS transit events (9-15) from Sector 28 of WASP-4b. Other comments are the same as Figure \ref{fig:ind_transits_sec2_1}.}
\label{fig:ind_transits_sec28_1}
\end{figure*}

\newpage
\clearpage
%%%%%%%%%%%%%%%%%%%%%%%%%%%%%%%%%%%%%%%%%%%%%%%%%%%%%%%%%%%%%%%%%%%%%%%%
%%%%%%%%%%%%%%%%%%%%%%%%%%%%%%%%%%%%%%%%%%%%%%%%%%%%%%%%%%%%%%%%%%%%%%%%
\section{Difference in mid-transit times for the Sector 2 TESS data} \label{sec:diffTESS}
\begin{figure}[!htb]
\includegraphics[width=0.85\textwidth]{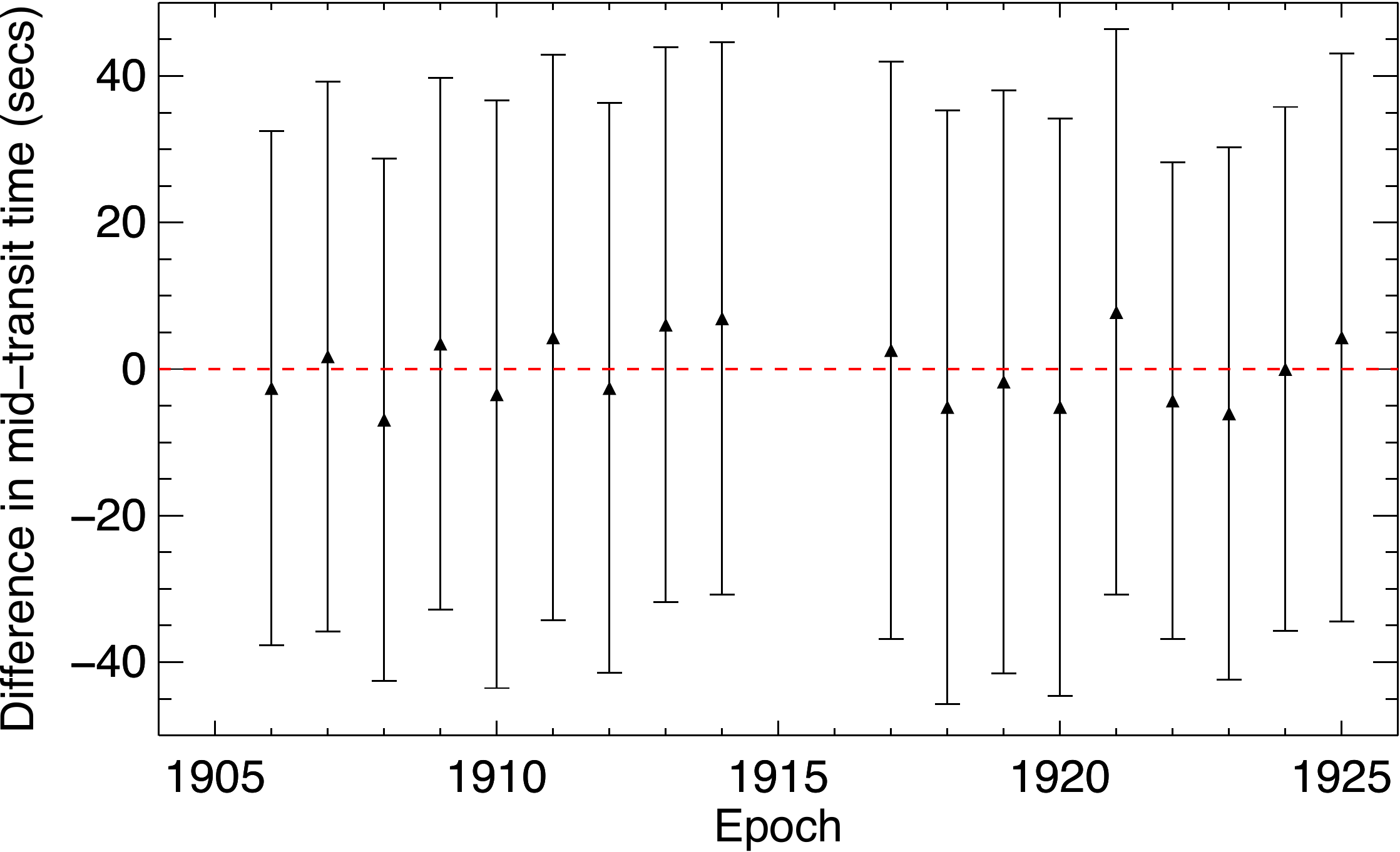}
%\plotone{TESS_Sector2_Compare.pdf}
\caption{Difference in mid-transit times between our analysis and \citet{Bouma2020} for the Sector 2 TESS data. 
\label{fig:LC_S2_comp}}
\end{figure}

The comparison in the derived mid-transit times between our analysis and \citet{Bouma2020} for the Sector 2 TESS data is found in Figure \ref{fig:LC_S2_comp}. Our results are consistent within 1$\sigma$. 

\begin{comment}
%%%%%%%%%%%%%%%%%%%%%%%%%%%%%%%%%%%%%%%%%%%%%%%%%%%%%%%%%%%%%%%%%%%%%%%%
%%%%%%%%%%%%%%%%%%%%%%%%%%%%%%%%%%%%%%%%%%%%%%%%%%%%%%%%%%%%%%%%%%%%%%%%
\begin{comment}
\section{TESS Occultation} \label{app:occ}

We created an occultation light curve by phase-folding all the data about the secondary eclipse using the first TESS transit as the reference transit time. As shown in Figure \ref{fig:tess_occ}, we do not see an occultation of WASP-4b in the TESS data. This is the case for both the PDC and DVT TESS light curves. We find the 3-$\sigma$ upper limits on the occultation depth (\delta$_{occ}$) and the geometric albedo to be 1.34 $\times 10^{-5}$ and 0.017, respectively.   

\begin{figure}[th!]
\centering
 \begin{tabular}{c}
   \sidesubfloat[\textbf{(a.)}]{   \includegraphics[width=0.85\textwidth,page=1]{WASP4b_Occ.pdf}} \\
  \sidesubfloat[\textbf{(b.)}]{    \includegraphics[width=0.85\textwidth,page=2]{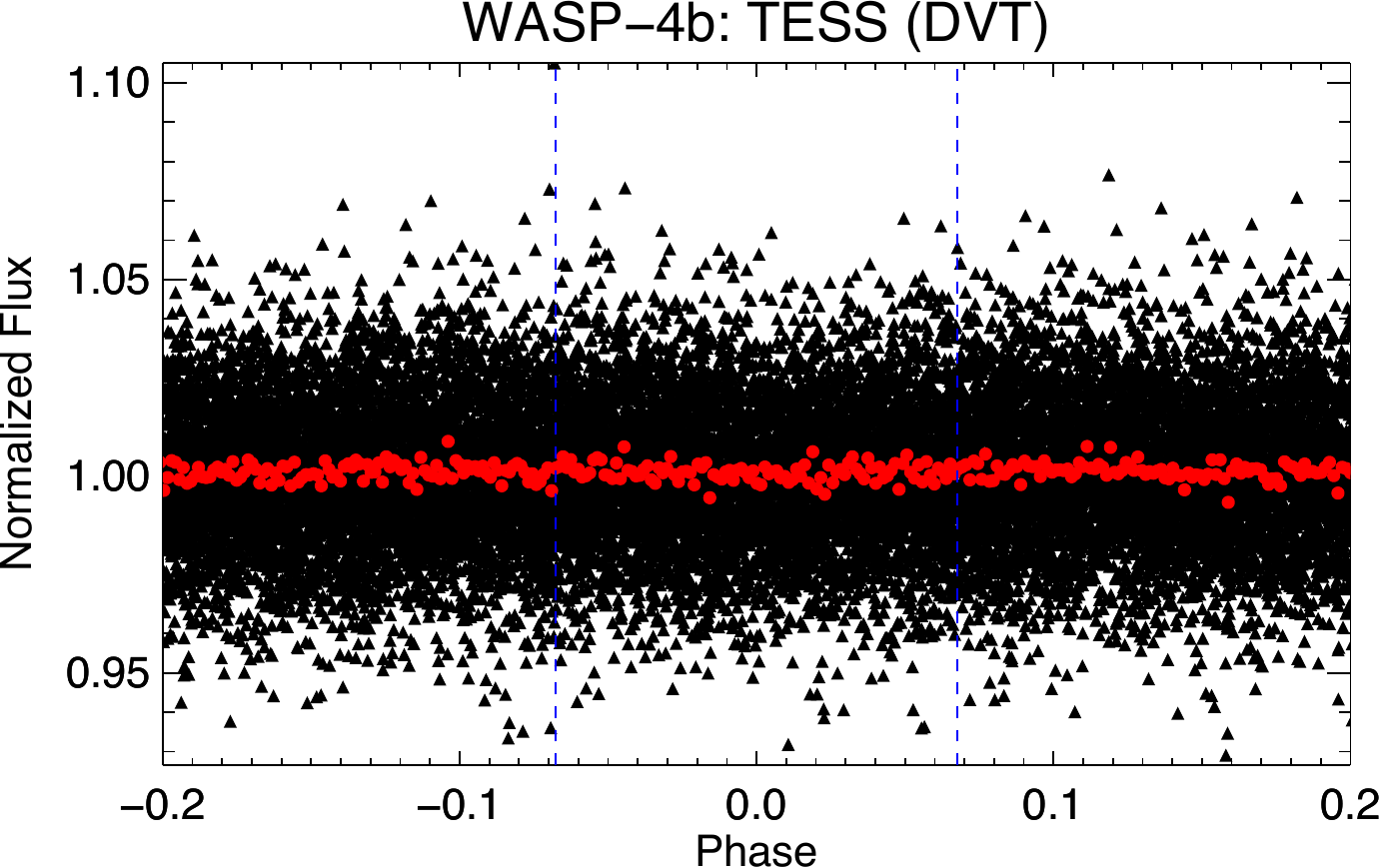}} \\
 \end{tabular}
\caption{Phased-folded occulation PDC (panel a) and DVT (panel b) light curve of WASP-4b from TESS. The unbinnned and data binned by 18 minutes are shown in blue and black, respectively.
}
\label{fig:tess_occ}
\end{figure} 
\end{comment}
 
%%%%%%%%%%%%%%%%%%%%%%%%%%%%%%%%%%%%%%%%%%%%%%%%%%%%%%%%%%%%%%%%%%%%%%%%
%%%%%%%%%%%%%%%%%%%%%%%%%%%%%%%%%%%%%%%%%%%%%%%%%%%%%%%%%%%%%%%%%%%%%%%%
\section{Radial-Velocity Models} \label{app:RV}
\begin{figure*}[thb!]
\centering
 \begin{tabular}{c}
  \includegraphics[width=0.95\textwidth,page=1]{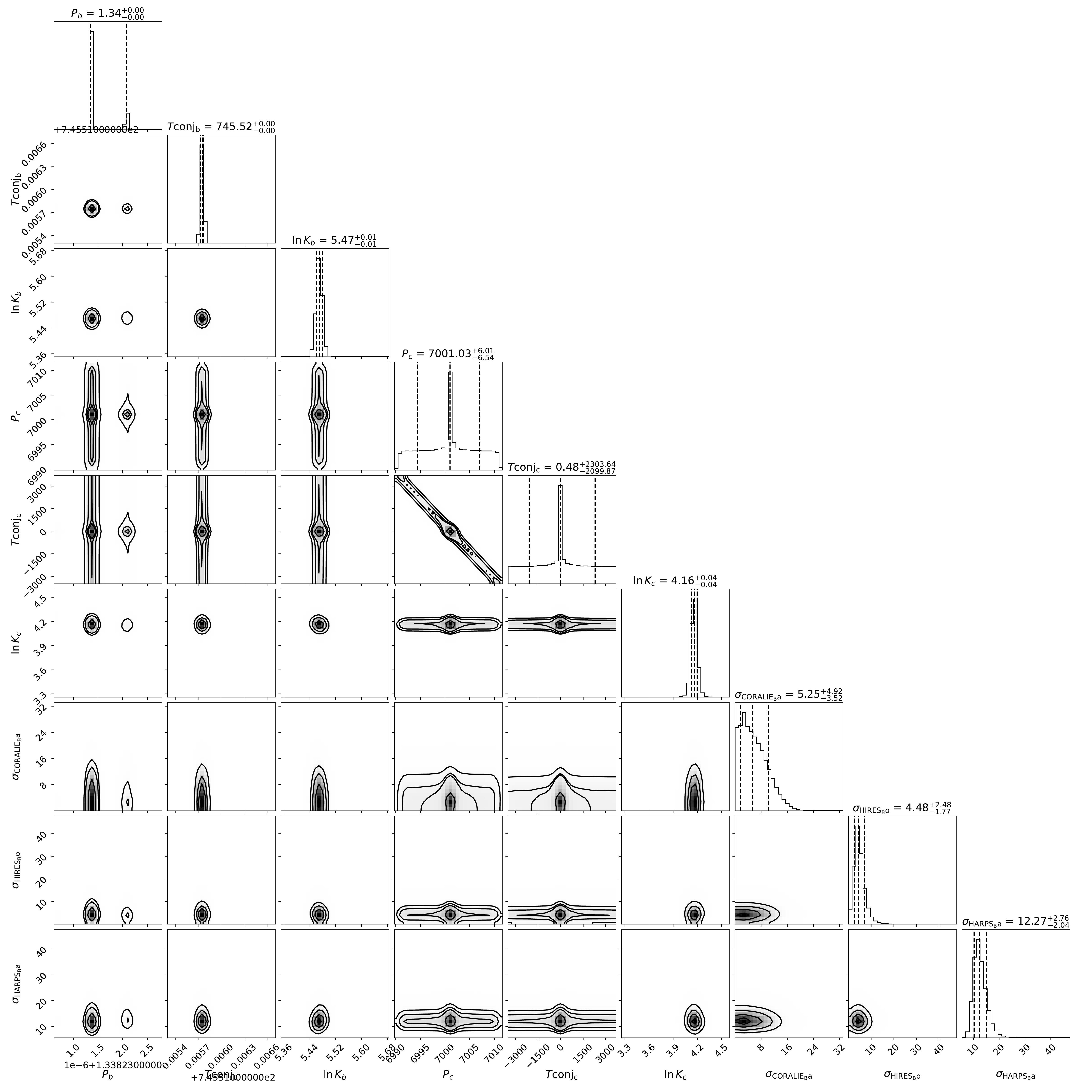}
 \end{tabular}
\caption{Posterior distributions for all free parameters for the best-fit two-planet RV analysis (Model $\#$ 3). The results of this model can be found in Table \ref{tb:RV_results_All} and Figure \ref{fig:RVModels_all}.}
\label{fig:corner_2planet}
\end{figure*}

\begin{figure*}[thb!]
\centering
 \begin{tabular}{c}
  \includegraphics[width=0.95\textwidth,page=1]{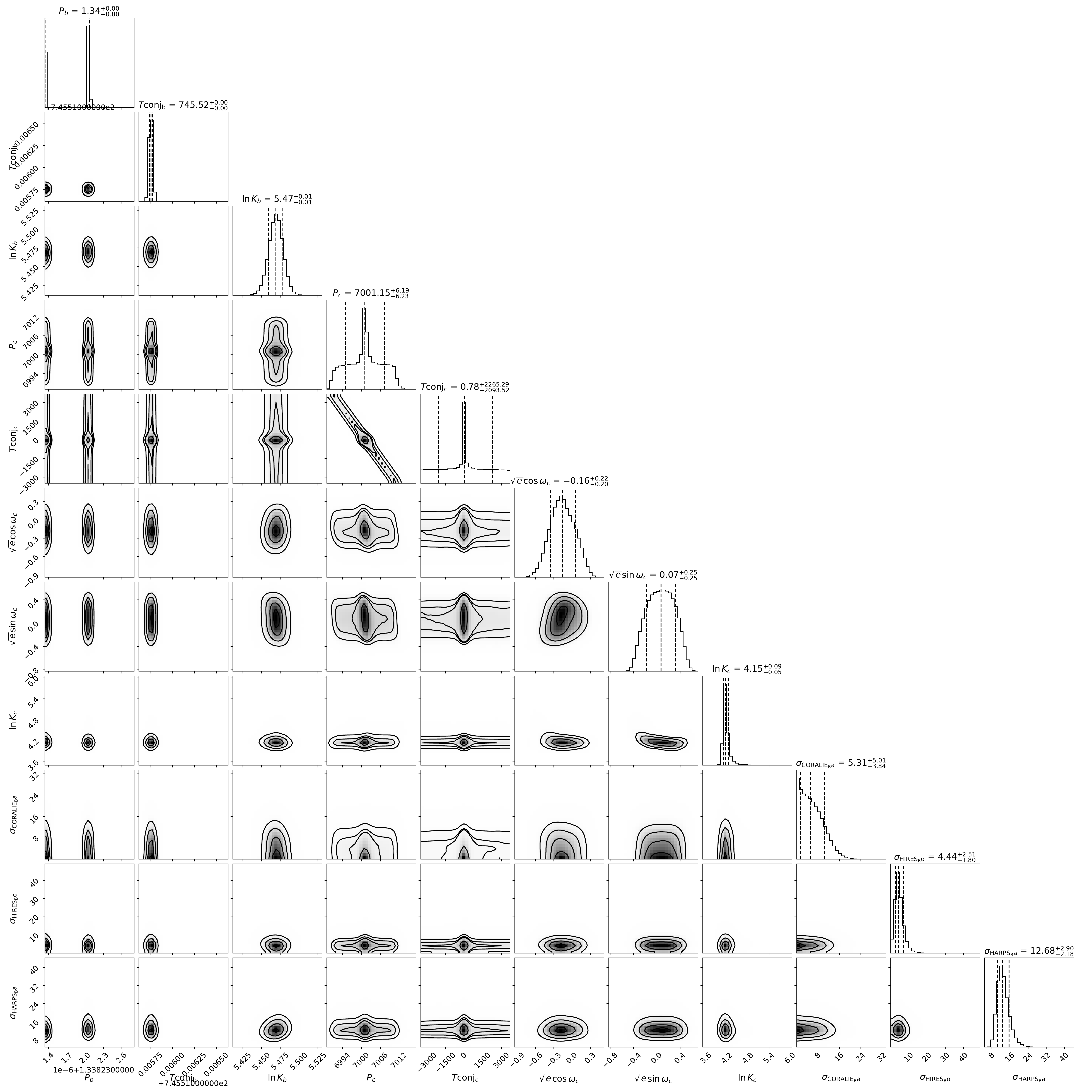}
 \end{tabular}
\caption{Posterior distributions for all free parameters for the two-planet RV analysis (Model $\#$ 4) fitting for e$_{c}$. The results of this model can be found in Table \ref{tb:RV_results_All} and Figure \ref{fig:RVModels_all}.}
\label{fig:corner_model4}
\end{figure*}

We performed an RV analysis on all the RV data in the literature with \texttt{RadVel}. Table \ref{tb:RV_compare} compares all the one-planet and two-planet RV models carried out in our analysis. The priors used in \texttt{RadVel} can be found in Table \ref{tb:priors_RV}. The best fit of the data is the two-planet model (Model $\#$ 3 in Table \ref{tb:RV_compare}) with a BIC of 628.34. In this model, the eccentricities of both bodies (e$_{b}$ and e$_{c}$) are fixed to zero and there is no linear acceleration. The planetary parameters derived of all models can be found in Table \ref{tb:RV_results_All}. The posterior distributions for all free parameters of Model $\#$ 1 - 4 can be found in Figures \ref{fig:corner_2planet}--\ref{fig:corner_model4}. The best-fit Keplarian orbital models compared to the RV data can be found in Figure \ref{fig:RVModels_all}. 

%The results of this model can be found in Table \ref{tb:RV_results_twoplanet} and Figure \ref{fig:RVfits_2planet} in the main text. The posterior distributions for all free parameters of Model $\#$ 1 - 4 can be found in Figures \ref{fig:corner_2planet}--\ref{fig:corner_model4}. 

%The results of the two-planet model fitting for e$_{c}$ is found in Table \ref{tb:2planet_ecc} and Figure \ref{fig:RVfits_2planet_ecc}. The posterior distributions for all free parameters of Model $\#$4 can be found in Figure \ref{fig:corner_model4}.  More observations are needed to full constrain the parameters of the second body in the WASP-4 system. 

\begin{comment}
\end{comment}\begin{table}[phtb]
 \caption{Comparison of all the RV models performed with \texttt{RadVel}}
    \centering
    \begin{tabular}{llllll}
    \hline
Model &  Free Parameters & $N_{\rm free}$  & RMS & $\ln{L}$ & BIC  \\
  \hline
  \hline 
  \multicolumn{6}{c}{\textbf{One-Planet Model}}\\
  1 & $K_{b}$, {$\sigma$}, {$\gamma$} & 6 & 29.85 & -325.78 & 675.08  \\
  2 &  $K_{b}$, $\sigma$, $\gamma$, $\dot{\gamma}$ & 7  & 29.78 & -651.38 & 682.76  \\
 %  $K_{b}$, $\sigma$, $\gamma$, $\dot{\gamma}$, $\ddot{\gamma}$ & 8 & 74 &19.33 & -297.58 & 626.94 & 610.72\\ 
 \multicolumn{6}{c}{\textbf{Two-Planet Model}} \\
3   & $K_{b}$, $K_{c}$, {$\sigma$}, {$\gamma$}  & 9 & 18.91& -304.40 & 628.34    \\
%4   & $K_{b}$, $K_{c}$, {$\sigma$},  {$\gamma$}, $\dot{\gamma}$  & 10  & 74 & 18.97 & -304.11 & 632.08 & 612.53 \\
4   & $K_{b}$, $K_{c}$, {$\sigma$}, {$\gamma$}, e$_{c}$  & 11 & 19.28 & -629.22 & 637.74  \\
%6   & $K_{b}$, $K_{c}$, {$\sigma$}, {$\gamma$}, $\dot{\gamma}$, e$_{c}$ & 12  & 74 & 18.99 & -304.12 & 640.72 & 618.19 \\
  \hline
    \end{tabular}
   \tablecomments{The number of data points ($N_{\rm data}$) for all the models was 74.}
    \label{tb:RV_compare}
\end{table}
\end{comment}

%The results of the one-planet RV analysis without (Model $\#$ 1) and with (Model $\#$ 2) fitting for the linear acceleration ($\dot{\gamma}$) can be found in Table \ref{tb:RV_results} and Figure \ref{fig:RVfits} in the main text. The posterior distributions of all free parameters of Model~$\#$ 1 can be found in Figure \ref{fig:corner_without}. We find that no linear acceleration is needed to explain the observations.        

\begin{figure}[pthb!]
\centering
 \begin{tabular}{c}  \includegraphics[width=0.95\textwidth,page=1]{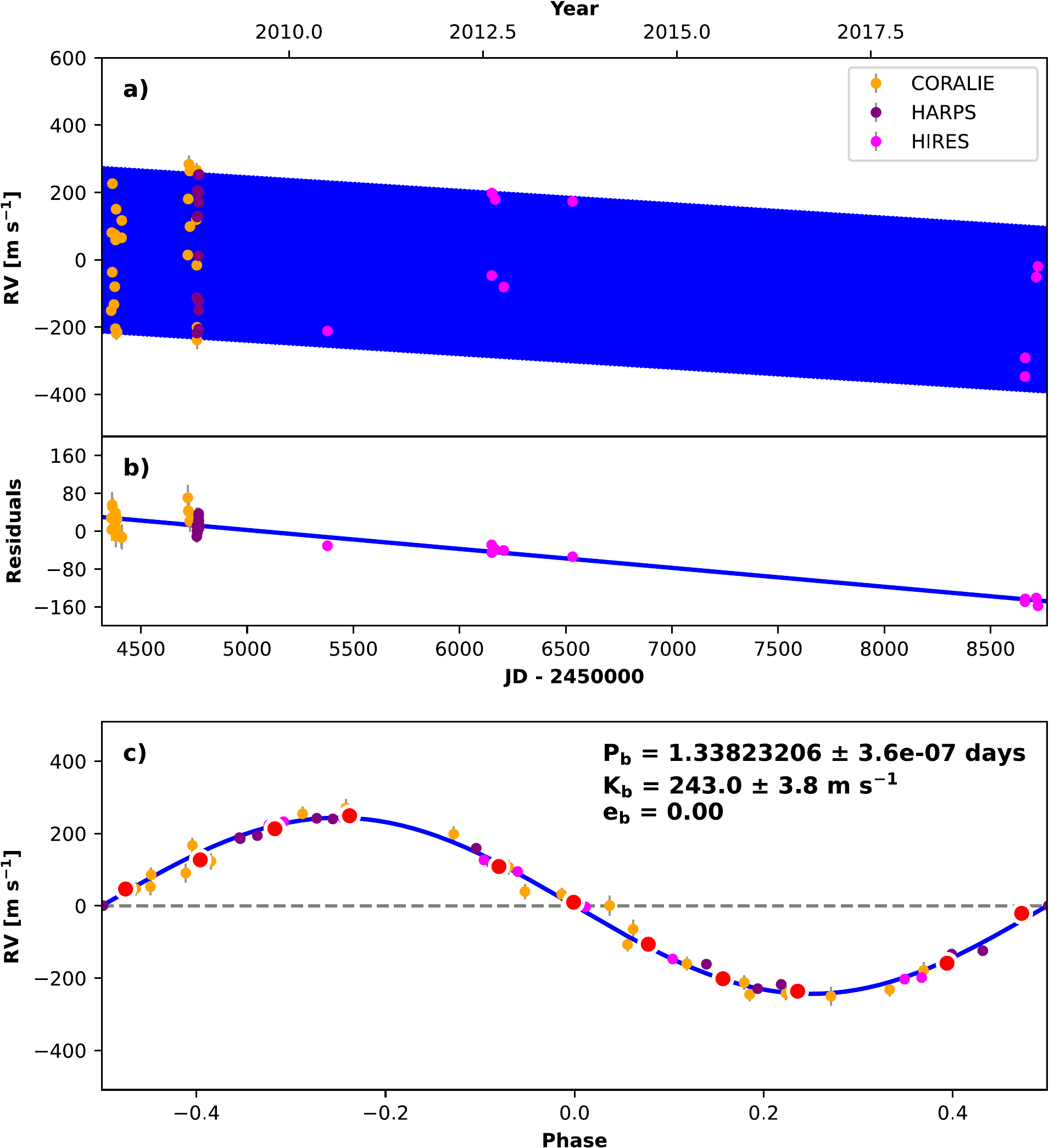}
 \end{tabular}
\caption{Best-fit 1-planet Keplerian orbital model
  for WASP-4b using \texttt{RadVel} of only the \citet{Bouma2020} RV data. {\bf a)} The maximum likelihood model is plotted while the orbital parameters listed in Table \ref{tb:BoumaRV} are the
  median values of the posterior distributions. The thin blue line is
  the best fit 1-planet model. We add in quadrature
  the RV jitter terms listed in Table \ref{tb:BoumaRV} with the
  measurement uncertainties for all RVs.  {\bf b)} Residuals to the
  best fit 1-planet model. {\bf c)} RVs phase-folded
  to the ephemeris of WASP-4b. The small point colors
  and symbols are the same as in panel {\bf a}.  Red circles are the same velocities binned in 0.08 units of orbital phase. The phase-folded model for WASP-4b is shown as the blue line. }
\label{fig:RVfit_Bouma}
\end{figure}

We also performed an RV analysis only on the data presented in \citet{Bouma2020}. We used priors listed in Table \ref{tb:priors_RV}. The results of that analysis can be found in Table \ref{tb:BoumaRV} and Figure \ref{fig:RVfit_Bouma}. We find a $\dot{\gamma}$ = $-0.0400^{+0.0037}_{-0.0032}$ m s$^{-1}$ d$^{-1}$, which is consistent the value found by \citet{Bouma2020} of $\dot{\gamma} = -0.0422^{+0.0028}_{-0.0027}$ m s$^{-1}$ day$^{-1}$.

\begin{table}[bt!]
 \caption{Priors used in the \texttt{RadVel} analysis of the RV data of WASP-4b}
    \centering
    \begin{tabular}{l}
    \hline
        e$_{b}$ constrained to be $<$ 0.99 \\
        Gaussian prior on T$_{conjb}$ (BJD-2455804): 0.515752$\pm$ 0.000019 \\ %old  2455059
        Gaussian prior on P$_{b}$ (days): 1.33823147 $\pm$0.000000023\\
        Bounded prior: 0.0 $< \sigma_{\text{HARPS,Bouma2020}} < $ 50.0 \\
        Bounded prior: 0.0 $< \sigma_{\text{HARPS,Baluev2019}} < $ 50.0   \\
        Bounded prior: 0.0 $< \sigma_{\text{CORALIE,Baluev2019}} < $ 50.0  \\
        Bounded prior: 0.0 $< \sigma_{\text{CORALIE,Bouma2020}} < $ 50.0 \\
        Bounded prior: 0.0 $< \sigma_{\text{HIRES}} < $ 100 \\
    \hline
    \end{tabular}
    \tablecomments{The Gaussian priors on T$_{conjb}$ and P$_{b}$ were taken from constant period model of \citet{Bouma2019}.}
    \label{tb:priors_RV}
\end{table}

\begin{figure*}[thb!]
\centering
 \begin{tabular}{c}
   \includegraphics[width=0.95\textwidth,page=1]{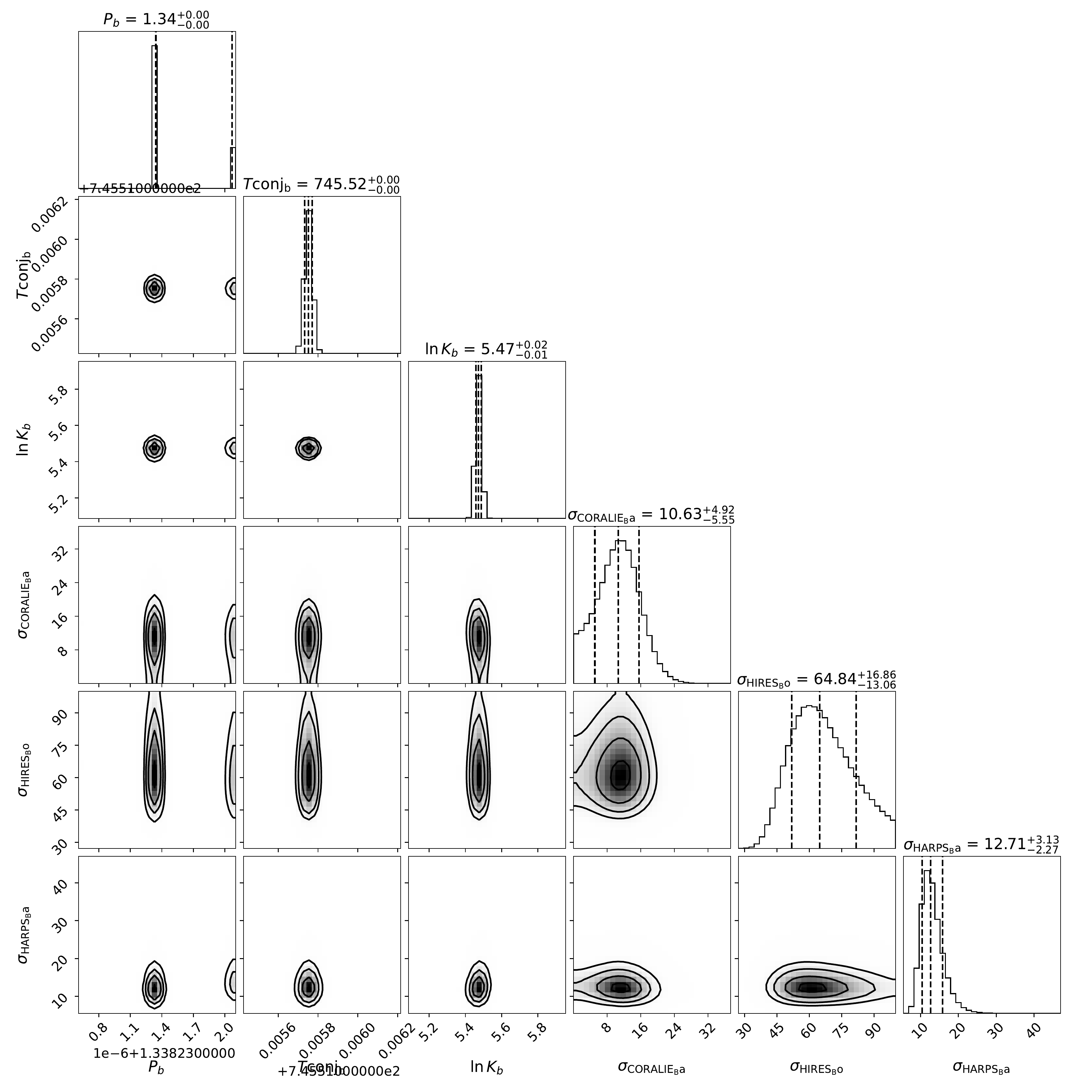}
 \end{tabular}
\caption{Posterior distributions for all free parameters for the one-planet RV analysis without (Model $\#$ 1) without fitting for the linear acceleration. The results of this model can be found in Table \ref{tb:RV_results_All} and Figure \ref{fig:RVModels_all}.}
\label{fig:corner_without}
\end{figure*}

%%%% Model $\#$ 2)
\begin{figure*}[thb!]
\centering
 \begin{tabular}{c}
   \includegraphics[width=0.95\textwidth,page=1]{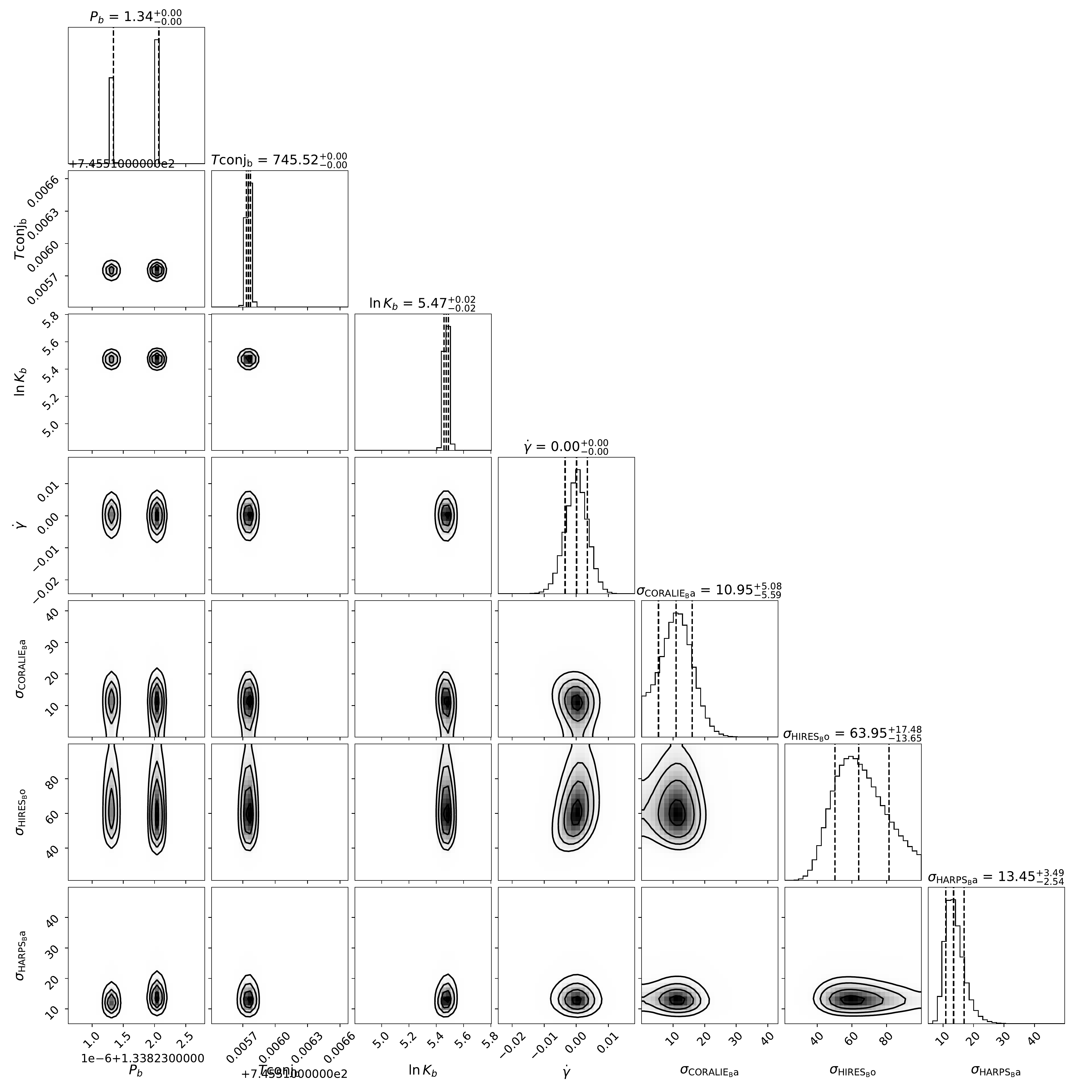}
 \end{tabular}
\caption{Posterior distributions for all free parameters for the one-planet RV analysis without (Model $\#$ 2) fitting for the linear acceleration. The results of this model can be found in Table \ref{tb:RV_results_All} and Figure \ref{fig:RVModels_all}.}
\label{fig:corner_without}
\end{figure*}

%%%Model $\#$ 3)

%%%% Model $\#$ 4)

\newpage
\clearpage 
\begin{table}[phtb]
 \caption{Results of fitting only the \citet{Bouma2020} radial-velocity data of WASP-4b with \texttt{RadVel}}
    \centering
    \begin{tabular}{lll}
    \hline
   \multicolumn{3}{c}{\bf{Modified MCMC Step Parameters}} \\
  $P_{b}$ & $1.3382320649^{+2.8e-09}_{-7.2e-07}$  & days \\

  $T\rm{conj}_{b}$ & $745.515752\pm 1.9e-05$  &JD \\

  $e_{b}$ & $\equiv0.0$ &   \\

  $\omega_{b}$ & $\equiv0.0$ &  radians \\

  $K_{b}$ & $243.0\pm 3.8$ & m s$^{-1}$ \\

\hline
\multicolumn{3}{c}{\bf{Orbital Parameters}}\\

  $P_{b}$ & $1.3382320649^{+2.8e-09}_{-7.2e-07}$ & days \\

  $T\rm{conj}_{b}$ & $745.515752\pm 1.9e-05$ & JD \\

  $e_{b}$ & $\equiv0.0$ &   \\

  $\omega_{b}$ & $\equiv0.0$ &  radians \\

  $K_{b}$ & $243.0\pm 3.8$ & m s$^{-1}$ \\

\hline
\multicolumn{3}{c}{\bf{Other Parameters}}\\

  $\gamma_{\rm HIRES,Bouma2020}$ & $\equiv48.8703$ & m s$-1$ \\

  $\gamma_{\rm HARPS_Bo}$ & $\equiv-52.8844$ & m s$-1$ \\

  $\gamma_{\rm CORALIE_Bo}$ & $\equiv-21.1286$ &m s$-1$ \\

  $\dot{\gamma}$ & $-0.0400^{+0.0037}_{-0.0032}$  m s$^{-1}$ d$^{-1}$ \\

  $\ddot{\gamma}$ & $\equiv0.0$ &  m s$^{-1}$ d$^{-2}$ \\

  $\sigma_{\rm HIRES,Bouma2020}$ & $11.6^{+5.1}_{-3.1}$ & $\rm m\ s^{-1}$ \\

  $\sigma_{\rm HARPS_Bo}$ & $14.4^{+4.4}_{-3.1}$ & $\rm m\ s^{-1}$ \\

  $\sigma_{\rm CORALIE_Bo}$ & $12.3^{+6.9}_{-7.0}$ &  $\rm m\ s^{-1}$ \\
  \hline
    \end{tabular}
    \tablecomments{  Reference epoch for $\gamma$,$\dot{\gamma}$,$\ddot{\gamma}$: 2455059 
}
    \label{tb:BoumaRV}
\end{table}

\newpage
\clearpage 
\bibliographystyle{aasjournal} % style aa.bst
\bibliography{reference_new.bib} % your references Yourfile.bib



\end{document}